\documentclass[openacc]{rstransa}

\usepackage{amsmath}
\usepackage{amsfonts}
\usepackage{bm}
\usepackage{bigints}
\usepackage{threeparttable}
\usepackage{ragged2e}
\usepackage{orcidlink}
\usepackage{twoopt}
\newcommandtwoopt{\myfrac}[4][0pt][0pt]{\genfrac{}{}{}{}{\raisebox{#1}{$#3$}}{\raisebox{-#2}{$#4$}}}

\newcommand\valpha{\ensuremath{\bm{\alpha}}}
\newcommand\vbeta{\ensuremath{\bm{\beta}}}
\newcommand\vtheta{\ensuremath{\bm{\theta}}}
\newcommand\dl{\ensuremath{D_\mathrm{L}}}
\newcommand\ds{\ensuremath{D_\mathrm{S}}}
\newcommand\dls{\ensuremath{D_\mathrm{LS}}}

\newcommand\zs{\ensuremath{z_{\rm S}}}
\newcommand\zl{\ensuremath{z_{\rm L}}}
\newcommand\zmin{\ensuremath{z_{\rm min}}}
\newcommand\zmax{\ensuremath{z_{\rm max}}}
\newcommand\mR{\ensuremath{\mathcal{R}}}

\newcommand\sigcrit{\ensuremath{\Sigma_{\rm crit}}}
\newcommand\Rlensed{\ensuremath{R_{\rm lensed}}}

\newcommand\thE{\ensuremath{\theta_{\rm E}}}

\renewcommand\sec{\ensuremath{\rm sec}}
\newcommand\eV{\ensuremath{\rm eV}}
\newcommand\zf{\ensuremath{z_{\rm F}}}
\newcommand\zh{\ensuremath{z_{\rm H}}}
\newcommand\Ntot{\ensuremath{N_{\rm tot}}}
\newcommand\Nlensed{\ensuremath{N_{\rm lensed}}}
\newcommand\Nlensdet{\ensuremath{N_{\rm lensed}^{\rm det}}}
\newcommand\Nlenspred{\ensuremath{N_{\rm lensed}^{\rm pred}}}

\newcommand\mumin{\ensuremath{\mu_{\rm min}}}
\newcommand\vlum{\ensuremath{\bm{\Lambda}}}
\newcommand\mK{\ensuremath{\mathcal{K}}}

\newcommand\mnras{MNRAS}

\begin{document}

\title{Multi-messenger Gravitational Lensing}

\author{The author list is at the end of the paper.}

\address{}

\subject{physics, astronomy, cosmology}


\corres{Graham P.\ Smith\\
\email{gps@star.sr.bham.ac.uk}}

\begin{abstract}\sloppypar
We introduce the rapidly emerging field of multi-messenger
gravitational lensing -- the discovery and science of gravitationally
lensed phenomena in the distant universe through the combination of
multiple messengers.  This is framed by gravitational lensing
phenomenology that has grown since the first discoveries in the 20th
century, messengers that span 30 orders of magnitude in energy from
high energy neutrinos to gravitational waves, and powerful ``survey
facilities'' that are capable of continually scanning the sky for
transient and variable sources.  Within this context, the main focus
is on discoveries and science that are feasible in the next 5-10 years
with current and imminent technology including the LIGO-Virgo-KAGRA
network of gravitational wave detectors, the Vera C.\ Rubin
Observatory, and contemporaneous gamma/X-ray satellites and radio
surveys.  The scientific impact of even one multi-messenger
gravitational lensing discovery will be transformational and reach
across fundamental physics, cosmology and astrophysics.  We describe
these scientific opportunities and the key challenges along the path
to achieving them. This article is the introduction to the Theme Issue
of the Philosophical Transactions of The Royal Society A on the topic
of Multi-messenger Gravitational Lensing, and describes the consensus
that emerged at the associated Theo Murphy Discussion Meeting in March
2024.
\end{abstract}




\maketitle

\section*{Executive Summary}
\label{sec:exec}

In recent years a broad consensus has developed that the
multi-messenger discovery and science of gravitationally lensed
phenomena in the distant universe is inevitable and will deliver
scientific breakthroughs across some of the biggest open questions in
fundamental physics, cosmology and astrophysics. Many of these
questions are shared across the US Decadal Review, the AstroNet
Roadmap, and the science books of major facilities including the Vera
C.\ Rubin Observatory (Rubin), Square Kilometre Array (SKA), next
generation gravitational wave (GW) detectors, and 30-m class
telescopes.

Multi-messenger gravitational lensing is well-placed to make decisive
contributions on questions relating to the nature of gravity, the
cosmological model including the expansion rate of the Universe and
the nature of dark matter (DM), the demographics and formation
channels of compact objects, the chemical enrichment of the Universe
via the r-process, the equation of state (EoS) of dense nuclear
matter, connections between and physics of diverse explosive transient
populations, and the host galaxies of GW sources. Many of these are
feasible within the next 5-10 years, i.e.\ on a timescale that is
accelerated relative to that which is feasible without assistance from
gravitational lensing.

These exciting opportunities are driven both by the discoveries of the
last decade, and by the rapid advances in detector sensitivity that
together span $\simeq30$ orders of magnitude in energy scale,
including high-energy neutrinos, gamma/X-rays, optical and infrared
(IR) photons, radio waves and GWs. In particular, the synergy between
the superb arrival time precision of neutrino, gamma-ray, radio and GW
detectors and the superb angular precision of optical/IR detectors and
upcoming radio interferometers will drive the science in the coming
decade and beyond.

Multi-messenger gravitational lensing advances in testing the nature
of gravity will benefit from both the magnifying power of
gravitational lenses to probe long travel times, and multiple
detections of the same source to boost the effective number of GW
detectors. Generically, long travel times will significantly boost the
sensitivity of searches for departures from General Relativity (GR),
because potential deviations accumulate over large cosmological
distances. Moreover, polarization constraints are the next frontier
for tests of gravity with GWs. Therefore, multiple detections of the
same chirp, due to gravitational lensing, will at least double the
effective number of GW detectors that probe whether the number of GW
polarization modes exceeds the two predicted from GR.

Multi-messenger gravitational lensing advances in cosmology will be
driven by the complementary arrival time and angular position
accuracies of the respective messengers, and the wave-like nature of
the GW signals. A multi-messenger time delay cosmography measurement
of the Hubble Constant, $H_0$, will suppress the uncertainty on the
arrival time difference measurement to a negligible level, will bring
complementary insights in to microlensing-related systematics, and for
short arrival time differences may leverage the wave nature of GWs to
break the mass-sheet degeneracy. There is also the exciting prospect
of combining time delay cosmography with standard siren cosmology in a
single multi-messenger lensing cosmology experiment. On smaller
scales, complementary constraints from different messengers, including
their sensitivity to microlensing signatures, will deliver novel
constraints on the DM sub-halo mass function, stellar mass function,
and compact DM.

Multi-messenger gravitational magnification and arrival time
differences will also open new windows on the physics of compact
binary coalescences (CBCs; also referred to as binary compact object
mergers) at high-redshift. One of the biggest unsolved mysteries
following the discovery of AT2017gfo, the kilonova counterpart to
GRB170817A / GW170817, is the physical interpretation of the early
blue kilonova emission. Very early rest-frame ultraviolet (UV)
observations are crucial to break degeneracies between competing
models. Multiple detections of the same gravitationally lensed
kilonova can access this early phase of evolution, in potentially
spectacular fashion if the second image arrives while optical target
of opportunity (ToO) observations are following up the first GW image
that arrived. Detection of lensed gamma-ray burst (GRB) counterparts
will further constrain the physics of gravitationally lensed CBCs that
emit electromagnetic (EM) radiation, including experiments to probe
the structure of GRB jets, benefiting from the different lines of
sight to the jet afforded by gravitational lensing. Progenitor compact
binary populations will also be probed, for example by testing objects
that appear to populate relatively sparse regions of parameter space,
such as the putative gap between neutron star (NS) and black hole (BH)
masses, and the transition from the most massive stellar remnant BHs
to intermediate mass BHs. This progress, coupled with rapid progress
in, e.g.\ Fast Radio Bursts (FRBs), will also drive fresh insights in
to the putative association of some FRBs with CBCs.

This article concentrates on discoveries and science that are within
reach in the next 5-10 years, with a broad focus on ground-based GW
detectors, the Vera C.\ Rubin Observatory's (Rubin's) imminent Legacy
Survey of Space and Time (LSST) and contemporaneous gamma/X-ray
satellites and radio surveys. As such, the focus is on facilities
that, in complementary ways, continually monitor the celestial sphere
and/or are capable of rapid ToO observations in response to detections
via other messengers. ``Static sky'' discoveries are also of huge
importance to multi-messenger gravitational lensing discoveries and
science, because the several order of magnitude expansion of the
census of gravitational lenses from Rubin/LSST, \emph{Euclid} and
their contemporaries will provide an unprecedented and comprehensive
view of the high-magnification lines-of-sight to the distant universe.

Exciting scientific opportunities naturally come with challenges that
the community must overcome, in this case on a timescale of 3-5
years. The most obvious cross-cutting challenge is to localize
gravitationally lensed CBCs that are discovered via messengers with
large localization uncertainties (GWs and GRBs) to $\lesssim1\,\rm
arcsec$, i.e.\ the angular scale of gravitational lensing and the
gravitationally lensed host galaxies. This requires cross-community
collaboration including to develop efficient methods to select
candidates, plan follow-up observations, and exploit synergies with
rapidly growing gravitational lens catalogues. Example outcomes of
cross-community collaboration include definitions of appropriate data
sharing requirements and protocols, and end-to-end simulations of
multi-messenger gravitational lensing source populations, signals, and
detection strategies.

Robust multi-messenger gravitational lensing discoveries and
interpretation of non-detections also requires significant advances in
our knowledge of the gravitational lens population in the observable
Universe. As alluded to above, the zeroth order cross-cutting
requirement is to build large and well-defined samples of
gravitational lenses from EM surveys, with well-calibrated selection
functions. However, robust multi-messenger gravitational lensing
discovery strategies also require the internal structure of the lenses
in these samples to be characterized as a function of the lens
mass. Specifically, the density profile slope and density of lenses at
their Einstein radii are the key parameters -- in addition to Einstein
radius -- that control the expected arrival time difference and image
separation for a given lens magnification.

Essential progress is also required in preparation for specific
science cases including those sketched above, for
example: \vspace{-3mm}
\begin{itemize}
    \item incorporation of ultra-precise arrival time difference
      measurements in to time delay cosmology inference
      pipelines;\vspace{0.5mm}
    \item detailed simulations of EM-bright CBCs with different mass
      ratios and NS equations of state;\vspace{0.5mm}
    \item model agnostic analysis pipelines for GW propagation,
      polarization and birefringence tests of GR;\vspace{0.5mm}
    \item detailed theoretical predictions for how these phenomena
      present in cosmologically motivated theories of gravity beyond
      GR;\vspace{0.5mm}
    \item the theory and computation of microlensing in the wave
      optics regime, models for data analysis and low-latency
      microlensing searches for EM follow-up;\vspace{0.5mm}
    \item development of methods to identify and follow-up candidate
      gravitationally lensed GRBs and FRBs in real-time, to enable
      discoveries before and during future GW runs;\vspace{0.5mm}
    \item detailed simulations of gravitational lensing as a probe of
      GRB jet structure, as a path to optimize gravitationally lensed
      GRB searches.
\end{itemize}

\section{Introduction}\label{sec:intro}

Multi-messenger gravitational lensing combines multiple messengers to
discover and study transient and variable phenomena in gravitationally
lensed host galaxies in the distant universe (typically redshifts of
$z\gtrsim1$) to probe a broad range of physics. The messengers span at
least 30 orders of magnitude in energy, from $\simeq10^8\,\rm GeV$ to
$\simeq10^{-14}\,\rm eV$, and include high-energy neutrinos, gamma-
and X-rays, UV/optical/IR photons, radio waves, and GWs. As they
traverse the gravitational potential of a dense foreground structure
such as a galaxy or group/cluster of galaxies, several paths of least
action through the potential may cause multiple ``images'' of the
source to arrive at an observer at different times. Each of the images
corresponds to a different trajectory that is perturbed relative to
that in the absence of lensing by up to $\simeq1\,\rm arcmin$, and
their flux can be magnified significantly.

Messengers associated with transient and variable sources typically
emanate from sources related to compact objects (black holes and
neutron stars), some of which are the end points of stellar
evolution. These sources include the collapse of stellar cores
(supernovae, GRBs, neutrinos and GWs), CBCs (GWs, kilonovae, GRBs and
their afterglows), phenomena associated with supermassive black holes
in galaxies (active galactic nuclei including blazars, and tidal
disruption of stars), plus fast-fading X-ray, optical and radio
sources of currently uncertain origin. EM messengers that emanate from
stars and dust (i.e.\ detectable as neither transient nor variable) in
the gravitationally lensed host galaxies are also of central
importance to locating transient/variable sources within them.

Interest in multi-messenger gravitational lensing has been fuelled by
the breakthrough direct detections of GWs \cite{Abbott2016,
  LIGOScientific:2016sjg, LIGOScientific:2016dsl,
  LIGOScientific:2018mvr}, the first multi-messenger discovery of a
CBC \cite{gw170817, Abbott2017e, LIGOScientific:2017ync}, and the
first discoveries of gravitationally lensed supernovae
\cite{Quimby2013, Kelly2015, Goobar2017, Goobar2023}. These
discoveries have helped to unlock a broad range of science that spans
fundamental physics, cosmology, high-energy astrophysics, nuclear
physics, the chemical enrichment of the universe, and galaxy
evolution. Multi-messenger gravitational lensing is well placed to
significantly { expand and accelerate scientific progress in these
  topics} (Section~\ref{sec:science}, and references therein).

In particular, a robust detection of a gravitationally lensed CBC via
EM and GW messengers would enable novel tests of GR, providing the
broadest-band large-scale laboratory for such experiments to
date. GRB170817A / GW170817 / AT2017gfo provided rich evidence of how
multi-messenger approaches enhance discovery science
\cite{gw170817,Margutti2021}. Similarly, EM messengers associated with
gravitationally lensed CBCs can make game-changing contributions by
localising the gravitationally lensed merger. Identification of the EM
counterpart to a candidate gravitationally lensed GW will achieve
sub-arcsecond localization in the host galaxy
\cite{Smith2023,Ryczanowski2025}. EM information about host galaxies
of gravitationally lensed binary black hole (BBH) mergers can also
place powerful constraints on these host galaxies, and potentially
achieve a similar level of precision \cite{Hannuksela2020, Wempe2024,
  Shan:2023ngi, Uronen:2024bth}. The combination of sub-arcsecond
angular resolution from EM messengers with the millisecond temporal
resolution of the GW detectors is then key to unlocking novel
science. Moreover, this exciting new lensing regime that combines
superb angular and temporal resolution is also available by combining
the timing precision of radio, gamma-ray, and / or neutrino detections
with optical/IR detection
(Sections~\ref{sec:signals}~\&~\ref{sec:science}\ref{sec:sources}\ref{sec:grb}~\&~\ref{sec:science}\ref{sec:sources}\ref{sec:frb}).

When considering direct detection of multiple messengers from
transient and variable sources, multi-messenger gravitational lensing
is multi-messenger astronomy enhanced by multiple magnified
lines-of-sight to sources at redshifts beyond those typically
accessible without lensing. Multi-messenger astronomy itself began in
the late 1980s when neutrinos were detected from a core collapse
supernova (SN1987A) in the Large Magellanic Cloud \cite{Hirata1987,
  Bionta1987, Alekseev1987} and from the Sun
\cite{Hirata1989,Hirata1990}. Three decades later in 2017 the merger
of a binary neutron star (BNS) and its aftermath at a distance of
$D=40\,\rm Mpc$ were detected via many messengers, spanning gamma-rays
to radio waves in the EM spectrum and GWs \cite{LV2017}, and
coincident neutrino and gamma-ray flares were detected from an active
galactic nucleus (AGN), the Blazar TXS\,0506$+$056, at a redshift of
$z=0.3365$ \cite{IceCube2018}. That these latter multi-messenger
discoveries were sources at cosmological distances is central to
demonstrating the feasibility of multi-messenger gravitational lensing
discoveries (Section~\ref{sec:channels}\ref{sec:relative}), echoing
the historical development of gravitational lensing.

AGN, in the form of quasars, were central to enabling the step from
early work on gravitational lensing \cite{Dyson1920, Einstein1936,
  Zwicky1937a, Zwicky1937b, Refsdal1964} to modern discoveries. The
intrinsic brightness of quasars renders them detectable out to high
redshift ($z\gtrsim1$) without requiring any gravitational
magnification. This was key to the first discovery of a
gravitationally lensed source at cosmological distances in 1979, when
the quasar pair 0957$+$561 was confirmed as a single quasar at
$z=1.405$ that is gravitationally lensed into two detectable images by
a massive foreground galaxy at $z=0.39$ \cite{Walsh1979,
  Young1980}. Gravitationally lensed quasars received significant
impetus from the Sloan Digital Sky Survey (SDSS) in the first decade
of the 21st century \cite{Inada2012}. Thanks to long-term monitoring
of these intrinsically variable sources \cite[for example]{Millon2020}
they now provide state of the art time delay cosmography measurements
of $H_0$ \cite[and references therein]{Treu2022}.

GRBs are more luminous than quasars, and therefore also prime
candidates for gravitationally lensed discoveries. Early discussion of
gravitationally lensed GRBs was contemporaneous with establishing the
extragalactic nature of most GRBs in the 1980s, when Paczynski
considered the gravitational lensing interpretation of three similar
bursts from the source B\,1900$+$14
\cite{Mazets1979,Paczynski1986}. Prospects for testing the lensing
interpretation of candidate lensed GRBs improved around a decade
later, following the discovery of afterglow emission from GRBs that
spans X-ray to radio wavelengths \cite{Djorgovski1997, Frail1997,
  Piro1998}, and the joint association of some supernovae (detected at
optical wavelengths) and GRBs with the core collapse of massive stars
\cite{Galama1998, Hjorth2003}. These breakthroughs enabled the
localization of GRBs to their host galaxies and thus also to the
angular scale of gravitational lensing. In the modern era,
\emph{Fermi's} Gamma-ray Burst Monitor (GBM) alone has detected
$>3000$ GRBs to date, with typical sky localization uncertainties of
up to $\simeq10^3\,\rm degree^2$, of which $\simeq20\%$ have arcsecond
localizations via detection of an afterglow, mostly because of their
co-discovery with \emph{Swift}
\cite{Connaughton2015,Goldstein2020}. In parallel, several studies
have searched for and discussed candidate gravitationally lensed GRBs,
with no confirmed discoveries to date \cite[for example]{Pedersen2005,
  Rapoport2012, Ahlgren2020, Paynter2021, Veres2021}.

The first discoveries of gravitationally lensed supernovae in the
mid-2010's \cite{Quimby2013, Kelly2015, Goobar2017} propelled
gravitational lensing into a new regime of lensed transients --
i.e.\ objects that subsequently fade completely and thus, unlike
lensed quasars, allow detailed studies of their host galaxies. These
and subsequent discoveries
\cite{Rodney2021,Goobar2023,Pierel2024,Frye2024} are more highly
magnified than the typical lensed quasars, because supernovae are
intrinsically fainter than quasars, and therefore at comparable
detector sensitivity they require higher magnification to be detected
at cosmological distances. Importantly, in the context of
multi-messenger gravitational lensing, these lensed supernovae
confirmed that discovery of gravitationally lensed optical transients
is feasible. Moreover, the gain in survey sensitivity from Rubin/LSST
will drive significant growth in the number of discoveries in the
coming decade \cite{Wojtak2019,Goldstein2019,Arendse2024}, and
motivates optimisation of discovery methods for gravitationally lensed
optical transients relevant to multi-messenger gravitational lensing
\cite{Ryczanowski2020, Ryczanowski2023, Magee2023, Murieta2023,
  Murieta2024, Sagues2024, Goobar2024}.

Following the first direct detection of GWs \cite{Abbott2016} by the
ground-based network that now comprises the two LIGO detectors
\cite{LIGOScientific:2014pky}, the Virgo detector
\cite{VIRGO:2014yos}, and the KAGRA detector \cite{KAGRA:2020tym},
signatures of gravitational lensing were discussed and searched for
both by the LIGO-Virgo-KAGRA (LVK) collaborations and groups external
to the LVK \cite{Smith2018, Broadhurst2018, Haris:2018vmn,
  Hannuksela2019, Dai:2020tpj, Diego2021, Liu2021, McIsaac2020,
  Basak:2021ten, LVlens2021, Kim:2022lex, LVKlens2023,
  Janquart:2023mvf, Janquart:2024ztv}. These drove the development of
several analysis methodologies \cite[for example]{Haris:2018vmn,
  McIsaac2020, Li:2019osa, Liu:2020par, Wright_2022, Seo:2021psp,
  Lo:2021nae, Janquart:2021qov, Janquart:2023osz, Goyal:2021hxv,
  Barsode:2024zwv, Ezquiaga:2023xfe, Chakraborty_2024,
  chakraborty2024muglancenoveltechniquedetect}, in addition to
forecasts of the rate of detection of lensed CBCs \cite{Ng2018,
  Li2018gw, Oguri:2018muv, Xu:2021bfn, Wierda:2021upe, Yang2022,
  Smith2023, Magare2023, Phurailatpam2024}. The first direct detection
of GW was swiftly followed by the first multi-messenger detection of a
CBC \cite[and references
  therein]{gw170817,Goldstein:2017mmi,Savchenko:2017ffs,Abbott2017e,Coulter:2017wya,LV2017,Margutti2021,Nicholl2025}. In
the intervening years, several scientific applications of
multi-messenger gravitational lensing discoveries were proposed,
including tests of GR \cite{Mukherjee_2020_gw_gal_sur,
  Mukherjee_2020_tgr_weak}, the speeds of light and GWs
\cite{Baker:2016reh, Collett:2016dey, Fan2017} and measurements of the
expansion of the Universe \cite[for example]{Liao:2017, Wei2017,
  Balaudo_2023}.

To facilitate multi-messenger discovery of gravitationally lensed
CBCs, significant attention has focused on EM follow-up observations
of GW sources with masses that are consistent with the lensing
hypothesis \cite{Smith2019, Smith2019lsst, Robertson2020,
  Ryczanowski2020, Ryczanowski2023, Smith2023, SmithJ2023,
  Bianconi2023, Magare2023, Manieri2024, Andreoni2024}. The aim of
these studies, given the proven association of GRBs and kilonovae with
BNS mergers \cite{Abbott2017e}, is to localise candidate lensed GW
sources to sub-arcsecond accuracy within their respective
gravitationally lensed host galaxies, via detection of a lensed EM
counterpart. There are also intriguing claims that some BBH mergers
might have EM counterparts in the form of AGN flares caused by a
merger occurring in an AGN accretion disk
\cite{Connaughton2016,Graham2020,Ashton2021,Palmese2021,Morton2023}. If
AGN flares are confirmed as EM counterparts to BBH mergers, this may
lead to the gravitational lensing of stellar remnant CBCs by the AGN
or galaxies/groups/clusters that intervene along the line of sight
\cite{Leong:2024nnx,Andreoni2024}. Identification of the host galaxies
of BBH mergers without EM counterparts has also been investigated, via
comparison of the properties of known gravitational lenses derived
from EM surveys with candidate gravitationally lensed BBH mergers
\cite{Hannuksela2020, Wempe2024, Shan:2023ngi, Uronen:2024bth}.

FRBs, first discovered in 2007 \cite{Lorimer_2007}, are located at
cosmological distances and are of intriguing unknown origin
\cite{Platts2019,Petroff2019,Petroff2022}. The rapidly growing numbers
of detections, already in the hundreds, and the timing and sky
localization accuracy of the detections identify them as exciting and
relevant for multi-messenger gravitational lensing
discoveries. Indeed, numerous works have explored the potential for
gravitationally lensed FRBs to probe the nature of DM, to test
fundamental physics including GR, and to elucidate the putative
connection between FRBs and sources of GWs \cite[for
  example]{Munoz2016, Wang2018, Li2018frb, Pearson2021, Leung2022,
  Gao2022, Ho2023, Kalita2023, Singh2024}.

Multi-messenger gravitational lensing discovery and science span a
diverse community and many disciplines. A significant fraction of the
community came together for the first time in Manchester on March
11-12 in 2024 at a Theo Murphy Discussion Meeting hosted by The Royal
Society. This meeting focused mainly on opportunities in the upcoming
decade with facilities that survey a large fraction of the celestial
sphere, including \emph{Fermi}, Rubin/LSST, and LVK. This article
captures the consensus that emerged in Manchester and aims to share it
with the wider community. We give an overview of the relevant
multi-messenger signals (Section~\ref{sec:signals}), outline the
essentials of gravitational lensing theory and phenomenology
(Section~\ref{sec:lensing}), describe the multi-messenger
gravitational lensing discovery channels, discovery rates and key
challenges (Section~\ref{sec:channels}), and present the main
multi-messenger gravitational lensing science cases
(Section~\ref{sec:science}).

\section{Multi-messenger signals and instruments}\label{sec:signals}

\begin{table}[]
  \begin{threeparttable}
    \scriptsize
    \caption{Summary of messengers, multi-messenger transient and
      variable sources, survey instruments (angular grasp of
      $\gtrsim2\pi\,\rm sr$), and prospects for gravitationally lensed
      discoveries in the coming decade.}
    \label{tab:messengers}
    \begin{tabular}{p{23mm}p{8mm}p{8mm}p{5mm}p{25mm}p{12mm}p{12mm}p{15mm}}
      \hline
      \hline
      \noalign{\smallskip}
      {\bf\RaggedRight Messengers and sources\tnote{a}} & \multispan3{\bf\hspace{1.5mm} Detections to date\tnote{b}\dotfill} & \multispan3{\bf\hspace{2mm}Expectations in next decade\tnote{c}\dotfill} & {\bf References}\cr
      \noalign{\vspace{1mm}}
      & {$\Ntot$} & {$\langle z\rangle$} & {\hspace{-4mm}$\Nlensdet$} & Detector/facility & {$\zh$} & $\Nlenspred$ \cr
      \noalign{\smallskip}
      \hline
      \noalign{\medskip}
      \multispan{8}{\sc \dotfill Flux-like messengers\dotfill}\cr
      \noalign{\medskip}
      \hspace{-1mm}\underline{Neutrinos} & \multispan{6}\emph{~~~[Timing accuracy\tnote{\rm d}, $\sigma_t\simeq10^{-9}\sec$; Sky localization uncertainty\tnote{\rm e}, $\Delta\Omega\lesssim5\,{\rm degree^2}$]}\hfill\smallskip\cr
      Blazar            & \hphantom{$>$}{$1$} & $0.34$    & {$0$} & IceCube-Gen2     & $1$ & \hphantom{$<$}$1$  & \cite{IceCube2018,Taak2023nu}\cr
      Ext.\ emission GRB             & \hphantom{$>$}{$0$} & $0.03$    & {$0$} & IceCube-Gen2     & $0.07$  & $<1$ & \cite{Kimura2017}\cr
      Millisec. magnetar              & \hphantom{$>$}{$0$} & $0.002$   & {$0$} & IceCube-Gen2     & $0.02$  & $<1$ & \cite{Fang2017}\cr
      Core collapse SN           & \hphantom{$>$}{$1$} & $10^{-4}$ & {$0$} & Hyper-Kamiokande & $0.001$ & $<1$ & \cite{Abe2018,Mori2022}\cr
      Binary NS merger                  & \hphantom{$>$}{$0$} & {...}     & {$0$} & Hyper-Kamiokande & $10^{-4}$ & $<1$ & \cite{Abe2021,Chen2023}\cr
      \noalign{\smallskip}
      \hspace{-1mm}\underline{Gamma- and X-rays} & \multispan{6}\emph{~~~[$\sigma_t\simeq10^{-3}\sec$; $\Delta\Omega\simeq10^{-2}{-}10^4\rm degree^2$]}\hfill\smallskip\cr
      Long GRB        & \hphantom{$>$}$10^4$   & $3$    & {$0$} & \emph{StarBurst, SVOM} & $3$ & \hphantom{$<$}$10$ &\cite{Moresco2022}\cr
      Short GRB       & \hphantom{$>$}$10^3$   & $1$  & {$0$} & \emph{StarBurst, SVOM} & $1.5$ & \hphantom{$<$}$1$ &\cite{Berger2014}\cr
      Relativistic TDE            & $<$$10$  & $1$    & {$0$} & \emph{Einstein Probe, SVOM} & $6$ & $<$$1$ &\cite{Brown2015}\cr
      Fast X-ray transients       & \hphantom{$<$}$100$  & $3$    & {$0$} & \emph{Einstein Probe, SVOM} & $3$ & $<$$1$ & \cite{Quirola2022,Quirola2023}\cr
      \noalign{\smallskip}
      \hspace{-1mm}\underline{Optical and near-IR\tnote{f}} & \multispan{6}\emph{~~~[$\sigma_t\simeq10^4{-}10^6\sec$; $\Delta\Omega\simeq10^{-8}\rm degree^2$]}\hfill\smallskip\cr
      Super-luminous SN  & \hphantom{$>$}$300$ & $0.4$ & $0$ & LSST WFD & $1.7$ & $>$$10$ & \cr
      Type Ia SN           & $>$$10^4$ & $0.3$  & $2$ & LSST WFD & $0.8$  & $>$$100$ & \cite{Goldstein2019,Wojtak2019,Arendse2024}\cr
      TDE      & \hphantom{$>$}$100$ & $0.3$ & $0$ & LSST WFD & $0.7$  & \hphantom{$>$}$10$ & \cite{Szekerczes2024,Chen2024}\cr
      Core collapse SN     & $>$$10^3$  & {$0.2$} & {$0$} & LSST WFD & $0.5$ & $>$$100$& \cite{Goldstein2019,Wojtak2019}\cr
      GRB afterglow               & $>$$10^3$  & {$0.1$} & {$0$} & LSST WFD/ToO\tnote{g} & $0.3/0.3$ & \hphantom{$>$}$10/10$ & \cite{Andreoni2024}\cr
      Kilonovae     & \hphantom{$>$}$10$  & {$0.1$}  & {$0$} & LSST WFD/ToO & $0.2/0.7$ & \hphantom{$>$}$1/1$ & \cite{Nicholl2021,Gompertz2023,Andreoni2024}\cr
      \noalign{\smallskip}
      \hline
      \noalign{\medskip}
      \multispan{8}{\sc \dotfill Amplitude-like messengers\dotfill}\cr
      \noalign{\medskip}
      \hspace{-1mm}\underline{Radio waves} & \multispan{6}\emph{~~~[$\sigma_t\simeq10^{-3}\sec$; $\Delta\Omega\simeq10^{-8}\rm degree^2$]}\hfill\smallskip\cr
      FRB            & \hphantom{$>$}$10^3$ & $1$ & $0$ & CHIME/FRB, CHORD & $3$ & \hphantom{$>$}$10$& \cite{Jahns-Schindler2023}\cr
      GRB afterglow               & $>$$400$ & 1 & $0$ & {SKA-Mid} & 5 & \hphantom{$>$}$10$ & \cite{Burlon2015,Levine2023}\cr
      \noalign{\smallskip}
      \hspace{-1mm}\underline{Gravitational waves} &  \multispan{6}\emph{~~~[$\sigma_t\simeq10^{-3}\sec$; $\Delta\Omega\simeq10{-}10^4\rm degree^2$]}\hfill\smallskip\cr
      Binary BH merger            & ${>}90$ & $0.4$  & $0$ & LVK A$^+$/A$^\sharp$\,\,(XG) & $2/5(40)$ & $>1/5(50)$ & \cite{LVKgwtc2023,OGCcatalogue2023,ETscience2020,CEwhitepaper2023}\cr
      NS-BH merger                & \hphantom{$>$}$3$   & $0.1$  & $0$ & LVK A$^+$/A$^\sharp$\,\,(XG) & $0.3/0.6(20)$ & & \cite{LVKgwtc2023,OGCcatalogue2023,ETscience2020,CEwhitepaper2023,GW230529}\cr
      Binary NS merger            & \hphantom{$>$}$2$   & $0.04$ & $0$ & LVK A$^+$/A$^\sharp$\,\,(XG) & $0.2/0.4(8)$ & $<1/1(50)$ & \cite{LVKgwtc2023,OGCcatalogue2023,ETscience2020,CEwhitepaper2023}\cr
      Core collapse SN     & \hphantom{$>$}{$0$} & $10^{-5}$  & $0$ & LVK A$^+$/A$^\sharp$\,\,(XG) & $(10^{-4})$ & & \cite{CEwhitepaper2023}\cr
      \noalign{\smallskip}
      \hline
      \noalign{\smallskip}
    \end{tabular}
    \begin{tablenotes}
      \scriptsize
      \item[a] Messengers (underlined) are listed in order of
        decreasing energy scale, and under each messenger the sources
        are listed in order of decreasing intrinsic brightness.
      \item[b] Summary of the detections to date by wide-angle survey
        facilities with sustained operations that span years and at
        least half of the celestial sphere: $\Ntot$ is the total
        number of sources detected to date, $\langle z\rangle$ is the
        typical redshift of the detected sources (approximate peak of
        the redshift distribution of a signal-to-noise ratio limited
        sample; not a formally computed mean), and $\Nlensed$ is the
        number of confirmed gravitationally lensed sources detected to
        date by these wide-angle surveys.
      \item[c] Wide-angle surveys and detectors that have come online
        recently, or will do so in the next decade, with their
        sensitivity summarised by the expected redshift horizon out to
        which they can detect sources without assistance from
        gravitational magnification ($\zh$), and order of magnitude
        expected number of lensed detections in the next ten
        years. For GWs, we quote the expected number of events per
        year with three different detector sensitivities: 2 expected
        upgrades of LVK detectors ($A^+$/$A^\sharp$) and 1 for
        next-generation (XG) ground-based detectors such as Einstein
        Telescope and Cosmic Explorer.
      \item[d] The accuracy with which the arrival time of the
        transient signals can measured, $\sigma_{\rm t}$. This is set
        by the properties of the detectors for all messengers except
        optical/near-IR messengers, that are limited by the shape of
        their lightcurve. For example, faster transients
        (e.g.\ kilonovae) have $\sigma_{\rm t}\simeq10^4\sec$, the
        slowest transients (e.g.\ super-luminous supernovae) have
        $\sigma_{\rm t}\simeq10^6\sec$, and GRB afterglow light curves
        do not constrain arrival time.
      \item[e] The uncertainty on the sky localization of the
        messenger. For surveys that use reflecting optics, this is
        given as the solid angle subtended by a circle of diameter
        comparable with the full width at half maximum of point
        sources. For all other surveys it is given as the solid angle
        of the typical 90\% confidence interval on the sky.
      \item[f] The following peak absolute magnitudes have been
        adopted: SLSN, $-21.5$, Type Ia Supernova (SNIa), $-19.4$; TDE
        $-19.3$; Core Collapse SN (CCSN), $-18$; GRB afterglow, $-17$;
        AT2017gfo-like kilonova, $-15.7$; Conservative KN, $-14.5$;
        NS-BH KN, $-13.0$. The assumed sensitivity of ongoing optical
        surveys is an apparent magnitude of $m=20$,
        i.e.\ approximately matching the depth of PanSTARRS, ATLAS,
        ZTF, GOTO, LS4, and BlackGEM.
      \item[g] The assumed sensitivity of Rubin/LSST ToO observations
        is $m=24$ for lensed GRB afterglows counterparts to candidate
        lensed GRBs and $m=27$ for lensed kilonova counterparts to
        candidate lensed BNS mergers, respectively
        \cite{Andreoni2024}.
    \end{tablenotes}
  \end{threeparttable}
\end{table}

\noindent
The messengers span at least 30 orders of magnitude in energy
(Table~\ref{tab:messengers}), from high-energy neutrinos
($E_\nu\gtrsim10^{15}\eV$) through to low-frequency GWs ($E_{\rm
  GW}=hf\lesssim10^{-15}\eV$, where $f$ is the GW frequency).  This
vast range of energy is mirrored by differences in the technology
required to detect the messengers, the relative sensitivities of
instruments across the energy scale, and how the messengers complement
each other in the context of gravitational lensing.  We refer the
interested reader to review articles in this volume and elsewhere, for
further details of the physics of each messenger, how they are
detected, and the science questions that each messenger is well-suited
to probing \cite{Margutti2021, pastor-marazuela_fast_submitted,
  Levan2025, Collins2025, Goobar2024, Kurahashi2022}.

A key distinction between different messengers is whether they are
detected via flux or amplitude (see ``Flux-like'' and
``Amplitude-like'' sections of Table~\ref{tab:messengers}), and among
those detected via flux whether or not the individual
particles/photons energies are measured.  This is important in the
context of gravitational lensing because gravitational magnification
($\mu$) describes the transformation of solid angle
(Section~\ref{sec:lensing}\ref{sec:delayetal}), therefore flux scales
with $\mu$, and wave amplitude scales with $\sqrt{\mu}$.  Neutrino and
gamma-ray instruments (e.g.\ IceCube, Kamiokande, \emph{Fermi} and
\emph{Swift}) count and measure the energy of individual particles and
photons, whilst { most} optical and IR { instruments}
(e.g.\ PanSTARRS, ZTF, Rubin/LSST) count photons without measuring
their individual energies.  These messengers are therefore detected
primarily via the flux of energy that arrives at the respective
instruments.  Lower energy messengers are detected via their wave
amplitude.  Radio instruments measure radio wave amplitude via the
time varying voltages that they detect (e.g.\ CHIME/FRB).  GW
instruments detect the amplitude of GWs that arrive at Earth via the
strain signal that is measured with interferometers.  Whilst detection
does not rely on whether messengers are polarized, all of them can in
principle be polarized, and this can lead to important science
applications
(e.g.\ Section~\ref{sec:science}\ref{sec:gravity}\ref{sec:polarisation}).

The timing accuracy and sky localization uncertainty of the
instruments differ dramatically between the messengers
(Table~\ref{tab:messengers}). Broadly speaking, superb timing accuracy
($\sigma_{\rm t}<1\,\rm sec$) is associated with poor sky localization
uncertainties ($\Delta\Omega>10\,\rm degree^2$), and vice versa. This
is important because combining multiple gravitationally lensed
messengers that have complementary strengths in timing and sky
localization has great potential to unlock discoveries
(Section~\ref{sec:channels}) and novel science
(Section~\ref{sec:science}). Discovery and science are enhanced by
direct detection of different messengers from an EM-bright
gravitationally lensed transient/variable source
(Section~\ref{sec:channels}\ref{sec:pathways}\ref{sec:bns}). A
complementary approach uses optical information about galaxies located
behind known gravitational lenses to search for lenses responsible
pairs of EM-dark GW detections that are lensed images of the same
source (Section~\ref{sec:channels}\ref{sec:pathways}\ref{sec:bbh}).

To be more specific, the superb timing accuracy of gamma-ray, radio,
and GW detection complements the superb angular resolution (sky
localization uncertainties) of optical/IR transient surveys via which
lensed optical counterparts can be identified. The latter can be
further significantly enhanced by the astrometric precision that can
be achieved with \emph{Hubble Space Telescope} and \emph{James Webb
Space Telescope} follow-up observations.  Measurements of the arrival
times of optical/IR signals stands out in Table~\ref{tab:messengers}
as the least accurate among the messengers. The accuracy of optical
measurements is currently set by the measurement uncertainties on when
the respective lightcurves peaks -- typically of order days.
Interferometric radio detection also stands out in
Table~\ref{tab:messengers}, as the only messenger for which accurate
sky localisation \emph{and} arrival time difference measurements are
feasible. The points to exciting prospects for scientific exploitation
of gravitationally lensed FRBs, especially in combination with other
messengers (Section~\ref{sec:science}\ref{sec:sources}\ref{sec:frb})
\cite{pastor-marazuela_fast_submitted}.

The redshift horizons, $\zh$, listed in Table~\ref{tab:messengers}
indicate the maximum redshifts at which sources are detectable in the
coming decade via each messenger without assistance from gravitational
magnification. The numbers of gravitationally lensed sources that are
forecast to be detected in the coming decade roughly scale with $\zh$
because, for reasonable assumptions on the comoving rate density of
sources, a larger value of $\zh$ indicates a larger comoving volume
within which detectable sources may be located. This has important
consequences for the focus and balance of this article, and is
explained in detail in Section~\ref{sec:channels}. In summary, we
focus on messengers for which $\zh\gtrsim0.1$, as these offer the
strongest potential for discovery of lensed sources in the coming
decade.

\section{Gravitational lensing theory and phenomenology}\label{sec:lensing}

We give an overview for the non-expert of gravitational lensing theory
and phenomenology in the context of multi-messenger astronomy, and
refer readers to other works, \cite[for
  example]{Blandford1986,Schneider1992}, and others cited below for
further theoretical details.

\subsection{Arrival time, deflection and magnification}\label{sec:delayetal}

The travel time, $t$, from a source at a redshift of $\zs$ along a
null geodesic through a gravitational field at a redshift of $\zl$
depends on distances and the Fermat potential of the lens, $\tau$:
\begin{equation}
  c\,t=\left(1+\zl\right)\frac{\dl\ds}{\dls}\,\,\tau(\vtheta,\vbeta),
  \label{eqn:ct}
\end{equation}
where $\dl$, $\dls$ and $\ds$ are the angular diameter distances from
the observer to the lens, lens to source, and observer to source
respectively, $\vbeta$ is the true position of the source on the
celestial sphere, and $\vtheta$ is the position of the gravitationally
lensed image of that source.

The Fermat potential comprises the geometrical path length difference
between the unperturbed observer-source path and the actual path
(first term on the right hand side of Equation~\ref{eqn:fermat}), and
a relativistic term \cite{Shapiro:1964} that is described by the
deflection potential of the lens, $\psi$ (second term):
\begin{equation}
  \tau(\vtheta,\vbeta)=\frac{(\vtheta-\vbeta)^2}{2}-\psi(\vtheta).
  \label{eqn:fermat}
\end{equation}
The deflection potential satisfies the two-dimensional Poisson
equation, $\nabla^2\psi=2\kappa$, where $\kappa\equiv\Sigma/\sigcrit$
is the dimensionless projected matter density of the lens, and
$\sigcrit$ is the critical density given by
\begin{equation}
  \sigcrit=\frac{c^2}{4\pi G}\,\frac{\ds}{\dl\dls}.
\end{equation}

In practice, the arrival time \emph{difference} between two
gravitationally lensed images ($\Delta t_{\rm AB}=t_{\rm A}-t_{\rm
  B}$) and the positions of the images ($\vtheta_{\rm A},\vtheta_{\rm
  B}$) are measurable if the measurement uncertainties are
sufficiently small, whilst the difference between the unperturbed
travel time and either $t_{\rm A}$ or $t_{\rm B}$, and also $\vbeta$,
are intrinsically not measurable.  The arrival time difference is
therefore conventionally written as
\begin{equation}
  \Delta t_{\rm AB} = \frac{D_{\rm \Delta t}}{c} \left[\tau(\vtheta_{\rm A},\vbeta) -\tau(\vtheta_{\rm B},\vbeta) \right],
\end{equation}
where $D_{\Delta t}$ is the so-called time-delay distance that is
defined as
\begin{equation}
  D_{\Delta t} \equiv (1+\zl)\frac{\dl\ds}{\dls}.
\end{equation}
By construction, the time-delay distance is therefore inversely
proportional to $H_0$, and is central to time-delay cosmography
(Section~\ref{sec:science}\ref{sec:cosmology}\ref{sec:cosmography}).

Applying Fermat's Principle to Equation~\ref{eqn:fermat}
(i.e.\ requiring $\nabla\tau=0$) yields the locations of the image(s)
of gravitationally lensed sources, i.e.\ the lens equation:
\begin{equation}
  \vtheta=\vbeta+\nabla\psi(\vtheta)=\vbeta+\valpha(\vtheta),
  \label{eqn:lensequation}
\end{equation}
where $\valpha=\nabla\psi$ is the deflection angle. Strong lensing --
the formation of multiple images -- corresponds to multiple solutions,
$\vtheta_k$, of Equation~\ref{eqn:lensequation} for a given source
position, $\vbeta$.

Most gravitational lenses are approximately axially symmetric with
$\kappa$ decreasing as a function of angular offset from the lens
centre, $\theta=|\vtheta|$, and produce multiple images of distant
sources at lens-centric angles that satisfy
$\langle\kappa(<\theta)\rangle=1$ \cite{Subramanian1986}. These images
form at or close to the so-called Einstein radius, $\thE$, which is
defined as follows for an axially symmetric lens:
\begin{equation}
  \thE=\left(\frac{4GM}{c^2}\,\frac{\dls}{\dl\ds}\right)^{1/2}\,,
  \label{eqn:thE}
\end{equation}
where $M=M(<\thE)$, i.e.\ the projected mass interior to the Einstein
radius.

The flux that arrives at Earth from a gravitationally lensed source
differs from the flux that would arrive in the absence of
gravitational lensing by a factor $|\mu|$, where $\mu$ is the lens
magnification:
\begin{equation}
  \mu=\left[(1-\kappa)^2-\gamma^2\right]^{-1}. 
\end{equation}
where $\kappa$ and $\gamma$ are the convergence and shear respectively
-- i.e.\ the isotropic and anisotropic contributions to the
magnification. They are related to the second order partial
derivatives of the deflection field:
$\kappa=(\psi_{,11}+\psi_{,22})/2=\nabla\psi^2/2$ (as above),
$\gamma_1=(\psi_{\rm ,11}-\psi_{\rm ,22})/2$, and $\gamma_2=\psi_{\rm
  ,12}=\psi_{\rm ,21}$. The subscripts on $\psi$ denote partial
differentiation with respect to the two components of $\vtheta$.

The amplitude of a gravitationally lensed wave-like signal therefore
differs from the intrinsic amplitude of the signal by a factor
$\sqrt{|\mu|}$. As a consequence, both flux- and wave-like signals
that are gravitationally magnified are apparently brighter/louder than
the underlying source if $|\mu|>1$.

\subsection{Critical curves, caustics{ , and image parity}}\label{sec:critical}

The multiple images that are created by strong gravitational lenses
form adjacent to so-called ``critical curves'' in the image plane that
is accessible to our detectors. These curves are closed and are
analogous to the perfect Einstein rings of radius $\thE$ associated
with a hypothetical axially symmetric lens. A critical curve is the
boundary between regions of positive and negative gravitational
magnification, where the sign indicates the parity of the
gravitational image formed in that region. In this context parity
refers to the handedness of the image, as seen in the mirror symmetry
of a pair of optical images, or phase of the GW signal associated with
images that form on either side of a critical curve.

Critical curves map to caustics in the source plane that demarcate the
regions of different image multiplicity, such that if a source is
exterior to all caustics it produces one image of positive parity, and
it produces additional pairs of images (of no net parity per pair) for
every caustic within which it is located. Images formed at minima and
maxima of the arrival time surface have positive parity and are called
Type I and III images respectively, whilst images formed at saddle
points have negative parity and are called Type II images.

\subsection{Achromaticity}\label{sec:achromatic}

Gravitational lensing is achromatic in the geometrical optics limit
(Section~\ref{sec:lensing}\ref{sec:waveoptics}), i.e.\ the spectrum of
a source is unaltered by gravitational lensing. At optical
wavelengths, this enables multiple images of gravitationally lensed
galaxies and explosive transients to be identified via similarity in
their broad-band colours and/or spectra, in addition to mirror
symmetries due to parity conservation
(Section~\ref{sec:lensing}\ref{sec:critical}). Achromaticity is also
key to identifying strongly gravitationally lensed GWs. The multiple
images of a GW source are broadly identical in frequency evolution,
which enables multiple images of a GW source to be identified in the
time domain, and the arrival time difference between these images to
be measured. Each image, does, however, experience its own
frequency-independent phase shift which can alter waveform morphology
in some cases, depending on the image type \cite{Dai:2017huk,
  Ezquiaga_2021}. This caveat can also be exploited to identify Type
II images in cases where higher order modes are present in the GW
signal \cite{Wang:2021kzt,Janquart:2021nus,Vijaykumar:2022dlp}.

Achromaticity may not hold in several scenarios that are relevant to
multi-messenger gravitational lensing. First, the spectral similarity
of multiple gravitational images of a source assumes that emission
from the source is isotropic on angular scales probed by the multiple
sight-lines to the source afforded by lensing. For compact sources
such as supernovae or CBCs, achromaticity therefore assumes isotropy
of emission on the scale of the Einstein radius of the gravitational
lens, i.e.\ $\theta\lesssim1\,\rm arcmin$. This is discussed in the
context of the least isotropic messenger considered in this article --
gravitationally lensed GRBs -- in
Section~\ref{sec:science}\ref{sec:sources}\ref{sec:grb}. Second,
gravitational magnification can be frequency dependent in scenarios
where the geometrical optics limit breaks down, for example GWs that
are lensed by compact lenses such as stars and compact objects
\cite{Cheung_2021, Yeung_2023, Diego:2019lcd, Mishra:2021xzz,
  Meena:2022unp}. In such cases, a ``wave optics'' treatment is
necessary (Section~\ref{sec:lensing}\ref{sec:waveoptics}). Note also
that an additional frequency-dependent modulation can arise in
strongly lensed images when those encounter smaller objects present in
the lens \cite{Seo:2021psp,
  Cheung_2021,Yeung_2023,Mishra:2021xzz,Meena:2022unp}. Third,
micro-lensing may also affect the lightcurves of lensed optical
counterparts to lensed GW sources, causing systematic differences
between the photometric evolution of different lensed images of the
same source. This has, for example, been investigated in the context
of measuring arrival time differences from gravitationally lensed SNe
\cite[and references therein]{Suyu2024}.

\subsection{Physical scales of lensing, and geometric, Eikonal and
  wave optics}\label{sec:waveoptics}

\begin{figure}
  \vspace{-10pt} \centering
  \includegraphics[width=0.8\textwidth]{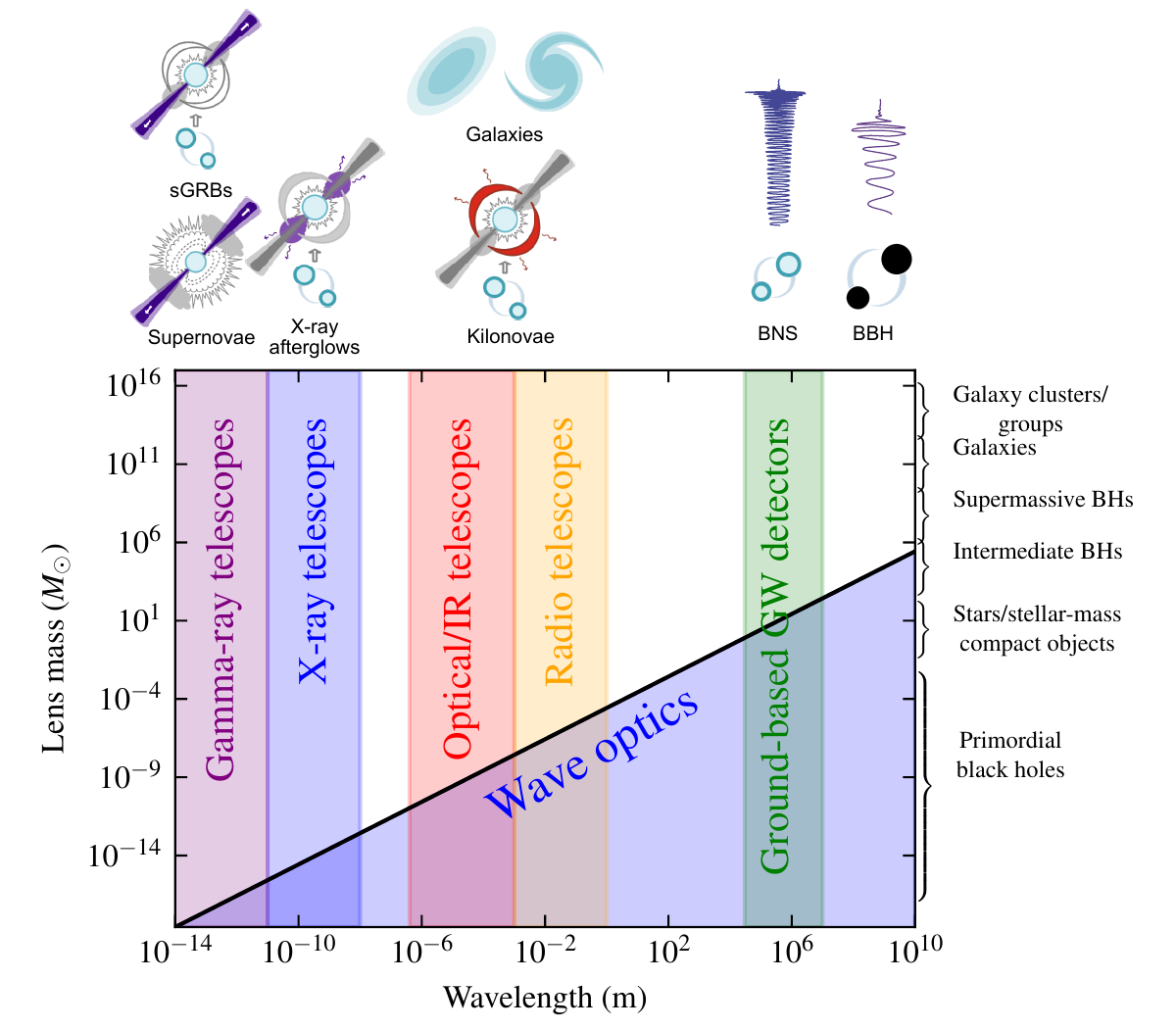}
  \vspace{-10pt}
  \caption{Illustration of the mass scales at which wave optics
    effects become relevant for gravitationally lensed signals.  The
    geometric optics regime is valid when the wavelength of the
    radiation is much smaller than the scale of the lensing potential.
    The wave optics regime is valid when the wavelength is comparable
    to the scale of the lensing potential.  Because the wavelength of
    GWs detected by the current ground-based detectors is typically
    much larger than the wavelength of most light sources, wave optics
    effects can become relevant for lenses below $\lesssim100\,\rm
    M_\odot$. Since other GW detectors like \emph{LISA} will be
    sensitive to even longer wavelengths, wave optics effects will be
    even more important.  The precise mass scale depends also on the
    lensing configuration, such as the distance from the caustic,
    where wave optics effects can become more prominent close to a
    caustic when the magnification is large.  }
  \label{fig:wave_optics_illustration}
  \vspace{-10pt}
\end{figure}

Gravitational lensing presents several phenomenological differences
that depend on the physical scale of the lens and the messenger being
considered. Firstly, lensing by massive extended objects such as
galaxies or galaxy clusters -- in which the source is inside the
caustic of the lens -- is referred to as strong lensing. Such objects
are typically hosted by DM halos that span at least
$M_{200}\simeq10^{12}-10^{15.5}\,\rm M_\odot$
\cite{Shajib2024,Natarajan2024}, where $M_{200}$ is the mass in the
spherical region within which the mean density is $200\times$ the
critical density of the universe. This produces multiple images that
are potentially both spatially and temporally resolvable. As is common
in the literature, we use the term ``images'' to denote the repeated
signals for gravitationally lensed sources regardless of the nature of
the messenger.

Moving down in physical scale whilst keeping the source within the
caustic of the lens, the angular-temporal separation between the
images shrinks, eventually leading to images that overlap and then
images that are no longer individually resolvable
\cite{Wambsganss2006}. This is the domain of millilensing and
microlensing, typified by angular separations of milli- and
micro-arcseconds respectively. Milli/micro-lensing and strong lensing
are not mutually exclusive, because small-scale lenses can be present
in large extended lenses, and thus perturb the strong lensing signal
\cite[and references therein]{Vernardos2024}. Moreover, the magnifying
effect of a strong lens can boost the detectability of
milli/micro-lensing \cite{Diego:2019lcd}.

Finally, weak lensing refers to the imprint of the gravitational field
of a lens on the signals from distant sources that are located well
outside the caustics discussed in
Section~\ref{sec:lensing}\ref{sec:critical}
\cite{Bartelmann2001,Schneider2005}. { Weak lensing therefore does not
  produce multiple images of distant sources}. At optical wavelengths,
weak lensing causes subtle distortions in the measured shapes of
distant galaxies. Measurement and analysis of weak lensing signals
requires careful statistical analysis of a large number of sources due
to the subtle effects being unmeasurable for individual
sources. Whilst we do not focus on weak lensing here, it is also a
source of bias in standard siren cosmology \cite[for
  example]{Mpetha2024}. Efforts to control this bias can benefit from
the enhanced knowledge of the host galaxies of GW sources that can be
obtained from multi-messenger gravitational lensing detections
(Section~\ref{sec:science}\ref{sec:sources}\ref{sec:hosts}).

Turning to the physical treatments, three regimes are relevant:
geometric, Eikonal, and wave optics. Geometrical optics is relevant if
the wavelength of the messenger is much smaller than the physical
scale of the lensing potential. It applies to both EM and GW
messengers (Figure~\ref{fig:wave_optics_illustration}) and is
described in
Sections~\ref{sec:lensing}\ref{sec:delayetal}-\ref{sec:achromatic}.
Wave optics is relevant if the wavelength of the messenger is larger
than or comparable with the physical scale of the lensing
potential. This applies to GWs that are detectable by ground-based
detectors, specifically for which the GW wavelength, $\lambda_{\rm
  GW}$, is comparable with the Schwarzschild radius, $R_{\rm
  S}=2GM/c^{2}$, of the lens. In such cases, effects such as
diffraction must be included in the treatment \cite{Nakamura1998,
  Takahashi:2003ix, Caliskan:2022hbu, Leung2023}, and the total
magnification ofthe GW waveforms must be fully computed as
\cite{Takahashi:2003ix}:
\begin{equation}
  F(f) = 
  \frac{1+\zl}{c}\frac{\dl\ds}{\dls} \frac{f}{i} 
  \int d^2 \vtheta 
  \exp [ 2 \pi i f \Delta t(\vtheta, \vbeta ) ]\,.
\end{equation}

For the current GW detector network, wave-optics effects are relevant
for compact lenses with masses $\lesssim100\,\rm M_\odot$. Wave-optics
effects will be relevant at much higher lens masses in the future when
space-based detectors such as \emph{LISA}
\cite{babak2021lisasensitivitysnrcalculations} probe longer
wavelengths, for example from binary supermassive black holes
\cite{Caliskan:2022hbu, Fairbairn:2022xln, Tambalo:2022wlm,
  Savastano:2023spl, Caliskan:2023zqm}.

Eikonal optics refers to the regime in which the arrival time
differences between the multiple lensed GW signals are less than the
duration of the signal. In this regime interference between the
multiple signals must also be considered. This can occur either due
directly to the mass of the object \cite{Liu_2023_eikonal} or in the
case of highly magnified images \cite[for
  example]{Creighton_Anderson_book_2011, Cremonese:2021,
  Bulashenko2022, Leung2023, Smith2023, Lo2024}. In the latter case,
for a representative galaxy-scale lens with $\thE\simeq1\,{\rm
  arcsec}$ this corresponds to gravitational magnifications of
$\mu\gtrsim50$ for a quad image configuration. For fold image pairs
produced by a representative galaxy cluster lens ($\thE\simeq5\,{\rm
  arcsec}$) this corresponds to $\mu\gtrsim200$
(Section~\ref{sec:lensing}\ref{sec:cluster}, \cite{Smith2023}).

\subsection{Mass sheet degeneracy}\label{sec:msd}

Robust physical interpretation of gravitationally lensed signals
requires the so-called ``mass sheet degeneracy'' to be broken
\cite{Falco:1985,Gorenstein1988}. This degeneracy affects inference of
the properties of the lens mass distribution, the size and luminosity
of sources and cosmological parameters, including $H_0$. Put simply,
if a projected mass distribution $\kappa(\vtheta)$ gives an adequate
fit to the measured image positions, flux ratios, and any measured
arrival time differences, then so too do the rescaled mass
distributions
\begin{equation}
  \kappa_\lambda=(1-\lambda)+\lambda\,\kappa(\vtheta),
\end{equation}
where $\lambda$ is an arbitrary scalar. 

The three gravitational lensing phenomena of arrival time, deflection
and magnification are affected differently by the mass sheet
degeneracy. The arrival time difference between two transient or time
varying signals is altered by $\Delta t_\lambda=\Delta t\,\lambda$
and, as implied above, the image positions are unaltered but the
source positions (and by association also the deflection angles) are
altered by $\vbeta_\lambda=\vbeta\,\lambda^{-1}$, while the
magnification is altered by $\mu_\lambda=\mu\,\lambda^{-2}$, where
subscript $\lambda$ denotes quantities related to the rescaled density
field. The mass sheet degeneracy can be broken if independent
information is available about the mass of the lens, e.g.\ from
stellar dynamics, the size or luminosity of the source, the
characteristic interference patterns of GW waveforms, or for multiple
source planes behind the same lens \cite[for example]{Arendse2024,
  Poon_gw_lens_degeneracies_overview_2024, Birrer2020, Khadka2024,
  Cremonese:2021, Chen_MSD_2024, Meena_wave_optics_MSD_2024}.

\subsection{Galaxy-scale strong lenses}\label{sec:galaxy}

Galaxy-scale strong lenses discovered to date in optical imaging
surveys are typically early-type galaxies with Einstein radii of
$\thE\simeq1\,{\rm arcsec}$ \cite[and references
  therein]{Shajib2024}. These discoveries are based on recognising
gravitationally lensed sources as multiply-imaged quasars and/or
gravitational arcs in the absence of time domain information. Time
domain surveys offer complementary selection methods that can exploit
the arrival time difference between images of lensed explosive
transients \cite{Goldstein:2017mmi,Wojtak2019,Arendse2024}. For
example recent discoveries of strongly lensed supernovae probe a
population of lenses with sub-arcsec Einstein radii \cite{Goobar2017,
  Goobar2024}.

Galaxy-scale lenses typically form two or four detectable images --
so-called double and quad lenses, respectively. A third or fifth
image, respectively, is strongly demagnified and located close to the
centre of the lens. The basic properties of galaxy-scale lenses are
well described by an isothermal density profile,
\begin{equation}
  \kappa(x)=\frac{1}{2x},
  \label{eqn:isothermal}
\end{equation}
where $x\equiv\theta/\theta_{\rm E}$ \cite{Koopmans2006}. Whilst
galaxy-scale lenses are approximately axially symmetric, formally the
typical model of a galaxy-scale lens is an ellipsoidal power law
\cite{Shajib2024}.

Double images arise from sources at $y\equiv\beta/\thE<1$ from the
centre of a galaxy-scale lens. The arrival time difference between,
angular separation of, and total magnification of the image pair are
given by:
\begin{equation}
  \begin{array}{rcl}
    \myfrac[3pt][4pt]{\Delta t_{\rm double}}{92\,\rm days}&=&\left[\myfrac[3pt][4pt]{\thE}{1''}\right]^2\,\bigg[\myfrac[3pt][4pt]{y}{0.5}\bigg]\,\left[\myfrac[3pt][4pt]{D_{\Delta t}}{3.3\,\rm Gpc}\right]\,,\medskip\cr
    \Delta x_{\rm double}&=&|x_+-x_-|=2\,,\cr
    \mu_{\rm double}&=&|\mu_+|+|\mu_-|\,\,=\,\,\left(1+\myfrac[3pt][4pt]{1}{y}\right)+\left(\myfrac[3pt][4pt]{1}{y}-1\right)\,\,=\,\,\myfrac[3pt][4pt]{2}{y}\,,\cr
    \label{eqn:tdpair}
  \end{array}
\end{equation}
where $x_\pm$ and $\mu_\pm$ denote the positions and magnification of
each image, respectively. Formally, the threshold for multiple image
formation is $\mu_{\rm double}=2$, for a source at $y=1$; however in
this case one image is not detectable because $\mu_-=0$. Sources that
are more closely aligned with the centre of the lens produce more
highly magnified double images with shorter arrival time differences,
for example at $y<0.5$ both images are brighter than the source, with
$\mu_+>\mu_->1$.

Quad images arise because galaxy-scale strong lenses are typically
elliptical, creating a caustic that can produce an additional image
pair if the source lies inside it. This caustic is typically located
at $y<1$, and has a characteristic astroid shape that comprises four
cusps connected by smooth curves that are called fold
caustics. Following the formalism introduced by \cite{Schneider1992},
the arrival time difference between, angular separation of, and
magnification of each image of a fold image pair can be expressed as:
\begin{equation}
  \begin{array}{rcl}
    \myfrac[3pt][4pt]{\Delta t_{\rm fold}}{0.25\,\rm days}&=&\left[\myfrac[3pt][4pt]{\Upsilon_t}{1}\right]\,\left[\myfrac[3pt][4pt]{\Delta y_0}{0.01}\right]^{1.5}\left[\myfrac[3pt][4pt]{\thE}{1''}\right]^2\left[\myfrac[3pt][4pt]{D_{\Delta t}}{\rm 3.3\,Gpc}\right],\medskip\cr
    \myfrac[3pt][4pt]{\Delta x_{\rm fold}}{0.4}&=&\left[\myfrac[3pt][4pt]{\Upsilon_x}{1}\right]\,\left[\myfrac[3pt][4pt]{\Delta y_0}{0.01}\right]^{0.5},\medskip\cr
    \myfrac[3pt][4pt]{\mu_{\rm fold}}{10}&=&\left[\myfrac[3pt][4pt]{\Upsilon_\mu}{1}\right]\,\left[\myfrac[3pt][4pt]{\Delta y_0}{0.01}\right]^{-0.5},\cr
    \label{eqn:fold}
  \end{array}
\end{equation}
where $\Delta y_0=\Delta\beta_0/\thE$ is the length of the shortest
arc that connects the source position with the fold caustic, the
density profile of the lens local to the image plane position that
corresponds to $\Delta y_0=0$ is given by $\kappa=\kappa_0
x^{\eta_0}$, and $\Upsilon_t$, $\Upsilon_x$, and $\Upsilon_\mu$
describe the density and structure of the lens at the mid-point of the
shortest arc that connects the image pair:
\begin{equation}
  \Upsilon_t=\Upsilon_x=\Upsilon_\mu\,|\eta_0|^{0.5}=\big[|\eta_0|\,(2+\eta_0)\big]^{-0.5},
\end{equation}
where $-1<\eta_0<0$, and Equation~\ref{eqn:fold} relies on the
relation $\kappa_0=1+\eta_0/2$ for approximately axially symmetric
lenses to eliminate $\kappa_0$. For an isothermal galaxy-scale lens
$\eta_0=-1$ and $\kappa_0=0.5$ (Equation~\ref{eqn:isothermal}),
yielding $\Upsilon_t=\Upsilon_x=\Upsilon_\mu=1$.

Quad images are a higher magnification regime than double images
because the source needs to be more closely aligned with the high
magnification central region of the lens to access the fold
caustic. Quads are therefore also associated with shorter arrival time
differences than doubles, because arrival time difference scales
inversely with magnification, with stronger scaling for folds than for
doubles: $\Delta t_{\rm fold}\propto\mu_{\rm fold}^{-3}$, $\Delta
t_{\rm double}\propto\mu_{\rm double}^{-1}$.

\subsection{Group/cluster-scale strong lenses}\label{sec:cluster}

Galaxy groups and clusters have typical Einstein radii in the range
$\thE\simeq3-60\,\rm arcsec$, i.e.\ larger than individual early-type
galaxy-scale lenses, due to the enhanced projected density in group
and cluster cores that is attributable to the massive DM halo
($M_{200}\simeq10^{13}-10^{15.5}\,\rm M_\odot$) in which they are
embedded \cite{Natarajan2024}. The DM contribution causes the density
profiles of group- and cluster-scale lenses to be denser
($\kappa_0>0.5$) and flatter ($\eta_0>-1$) than isothermal at their
Einstein radius \cite{More2012,Fox2022}. This reduces the efficiency
of clusters in producing multiple images in the low magnification
regime that is accessible to galaxy-scale doubles,
i.e.\ $\mu\lesssim10$ \cite{Smith2023}. A similar effect is expected
for group-scale lenses, however this has not yet been studied in
detail. The cores of group- and cluster-scale lenses tend to be
strongly asymmetric, and thus fold caustics tend to dominate the
multiple images that they form. This is particularly true for massive
galaxy clusters due to the prevalence of substructure due to the
hierarchical nature of large scale structure \cite[for
  example]{Kneib1996, Smith2005, Smith2009, Jauzac2016, Mahler2023}.

The arrival time difference, image separation, and individual image
magnifications of fold image pairs formed by group and cluster lenses
are also given by Equations~\ref{eqn:fold}.  Given the density and
structure of group- and cluster-scale lenses discussed above,
$\Upsilon_t$, $\Upsilon_x$, and $\Upsilon_\mu$ all tend to exceed
unity and thus the arrival time difference, image separation, and
magnification of fold image pairs are all larger for
group/cluster-scale lenses than for galaxy-scale quad lenses. In
summary, the phenomenology of group and cluster lenses relative to
galaxy lenses can be understood broadly in terms of how their density
profiles and substructures shape their fold caustics and efficiency of
multiple image formation at $\mu\lesssim10$.

\subsection{Compact lenses}\label{sec:compact}

Compact lenses also form images on the angular scale of their Einstein
radius (Equation~\ref{eqn:thE}). { The total magnification of these
  images}, formed by an isolated compact lens is given by
\begin{equation}
 \mu = \frac{y^2 + 2}{y\sqrt{y^2 + 4}},
\end{equation}
where $y$ is the dimensionless impact parameter, and has the same
definition as in Section~\ref{sec:lensing}\ref{sec:galaxy}.

When compact lenses are embedded in a dense environment such as
stellar fields or galaxy/group/cluster-scale lenses, the lensing
effects of compact lenses can be enhanced, leading to caustic networks
and complex lensing patterns \cite{Diego:2019lcd, Cheung_2021,
  Mishra:2021xzz, Meena:2022unp, Shan:2022xfx, Yeung_2023}.  Compact
lenses are therefore crucial in studying DM distributions, distant
stars and black holes. By tracing their gravitational signatures they
offer insights into the unseen mass in the universe, such as
primordial black holes or other forms of DM.

\subsection{Optical depth} \label{sec:opticaldepth}

The optical depth to gravitational lensing, $\tau$, is defined as the
fraction of the celestial sphere that is gravitationally lensed. It is
usually defined in the source plane, i.e.\ the fraction of the
intrinsic celestial sphere, because this is well-suited to predicting
and interpreting the number of gravitational lens discoveries. The
source plane optical depth can be defined in terms of the number of
gravitationally lensed sources that will be detected
\cite{Schneider1992}, or in terms of the number of images that will be
detected \cite{Hilbert2008,Robertson2020}. Formally, our overview in
this Section considers the latter because it is arguably better
aligned with the focus of this article, with each detectable
gravitationally lensed image representing an opportunity to make the
first multi-messenger gravitational lensing discovery.

Defined in this way, the optical depth as a function of the mass ($M$)
and redshift of the lenses ($\zl$) can be written as follows:
\begin{equation}
  \myfrac[1pt][1.5pt]{\partial^2\tau}{\partial M\,\partial\zl}=\myfrac[1pt][1.5pt]{1}{\Omega}\,\myfrac[1pt][1.5pt]{\partial^2\sigma_{\rm tot}}{\partial V\,\partial M}\,\myfrac[1pt][1.5pt]{dV}{d\zl},
\end{equation}
where $\Omega=4\pi$ is the solid angle of the celestial sphere,
$\sigma_{\rm tot}$ is the sum of the cross-sections of all the
gravitational lenses in the mass interval $dM$, in the comoving volume
element $dV$. Using this equation requires the cross-section to
gravitational lensing to be defined across the relevant mass and
redshift range.

Gravitational lenses span a wide range of mass and internal structure
that affect the arrival time differences, image separations and
magnifications of the gravitational images that they produce
(Sections~\ref{sec:lensing}\ref{sec:galaxy}-\ref{sec:compact}). One
approach is to define the cross-section in terms of multiple-image
formation and assume that all lenses are early-type galaxies, with the
number and masses of the lenses normalized to the SDSS galaxy velocity
dispersion function, \cite[for
  example]{Oguri2010,Haris:2018vmn,Goldstein2019,Arendse2024}. The
main advantage of this approach is that by concentrating exclusively
on isothermal lenses it enables self-consistent predictions of arrival
time differences, image separations, and magnifications. The main
disadvantage is that it ignores the impact of shallower than
isothermal group/cluster-scale density profile slopes on the
efficiency of multiple-image formation, arrival time differences,
image separations, and magnifications (Equations~\ref{eqn:fold}).

An alternative is to build the optical depth on cosmological n-body
simulations. Such approaches include ray tracing through DM halos from
the Millennium simulation in to which analytic galaxies have been
pasted \cite{Hilbert2008}, deriving the optical depth to gravitational
magnification from cosmological hydrodynamical simulations
\cite{Robertson2020}, and using halo models calibrated to cosmological
simulations to extend galaxy-scale lens approach to account for their
host DM halos \cite{Abe2024}. The main advantage of these methods is
that in principle they incorporate the full range of lens mass and
structure, and thus address the disadvantage of the galaxy-scale
methods outlined above.

\begin{figure}
  \vspace{-10pt}
  \centering
  \includegraphics[width=0.6\textwidth]{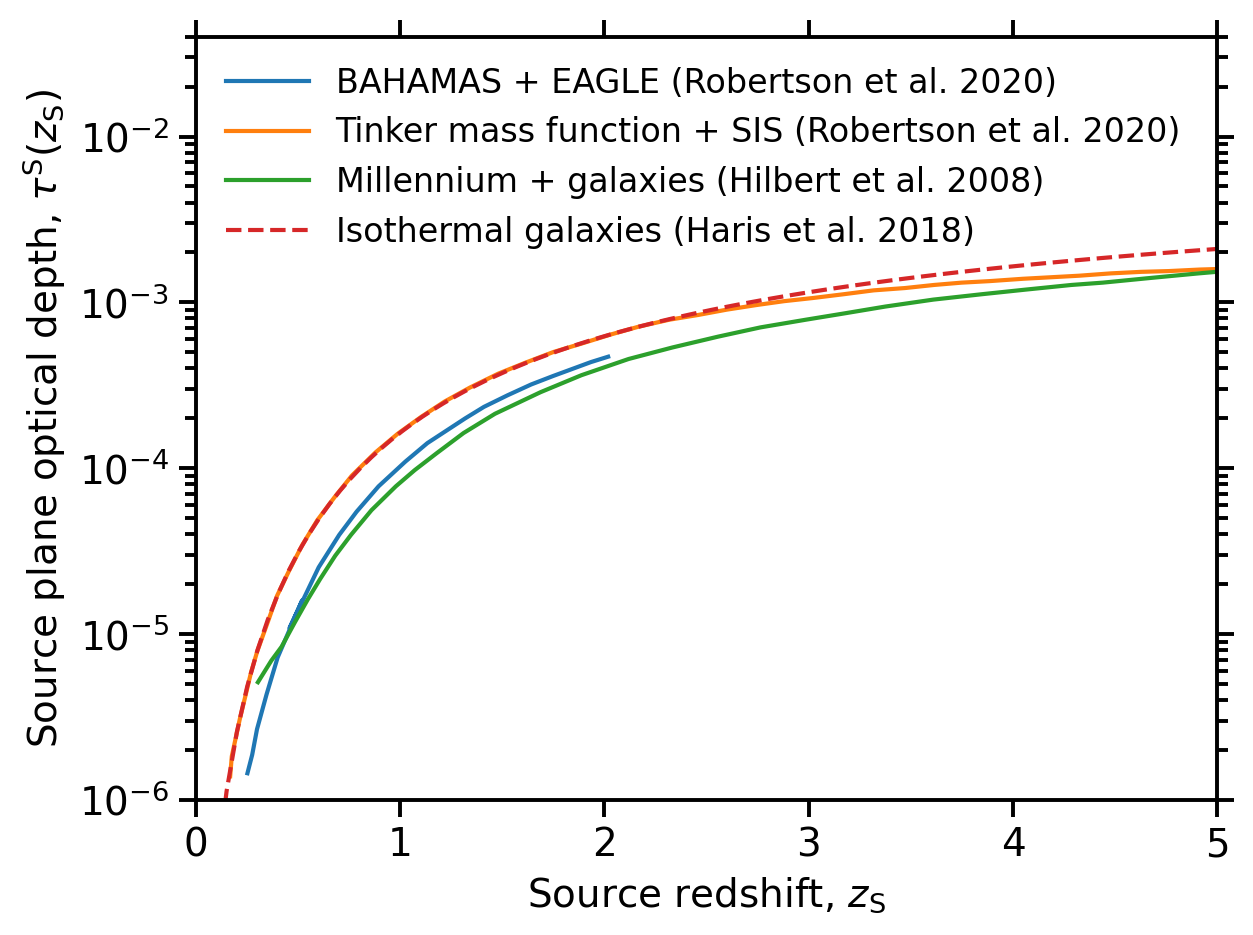} 
  \vspace{-5pt}
  \caption{Different models of the source plane optical depth to
    gravitational lensing agree within a factor $\simeq2$. This
    indicates that the on the integral of the optical depth across the
    mass function of lenses is converged. As discussed in the text,
    the distribution of the optical depth across the mass function is
    less well converged, and is a key area for theoretical and
    observational progress. Figure reproduced from \cite{Smith2023}. }
  \label{fig:taus} 
  \vspace{-10pt}
\end{figure}

The approaches outlined above tend to agree within a factor of
$\simeq2$ on the integral over the mass function of the optical depth
to gravitational magnification as a function of source redshift
(Figure~\ref{fig:taus}). This encourages confidence in the following
expression adapted from \cite{Haris:2018vmn}:
\begin{equation}
  \myfrac[2pt][1pt]{d\tau(\zs)}{d\mu}=\left(\myfrac[2pt][1pt]{D_{\rm S}(1+\zs)}{62.2\,\rm Gpc}\,\,\myfrac[2pt][1pt]{2}{\mu}\right)^3,
  \label{eqn:tau}
\end{equation}
where the first term in parenthesis on the right hand side describes
the optical depth to magnification $\mu=2$, and the second term scales
that to higher magnifications. However, the approaches outlined above
tend to disagree on how the optical depth is distributed with respect
to mass and therefore with respect to lens structure. This is either
by construction in the case of the galaxy-only models, or likely due
to differences in implementation of baryons in the simulation-based
methods, as discussed for example by \cite{Robertson2020}.

\section{Discovery channels for multi-messenger gravitational lensing}
\label{sec:channels}

The path to the first multi-messenger gravitational lensing
discoveries depends on synergies between the messengers that go beyond
the detectability of gravitationally lensed transient sources via
different messengers (Section~\ref{sec:signals}). In this Section we
describe how these synergies shape several complementary channels
through which the first discoveries will be made.

We briefly review the formalism for forecasting the rates of
gravitationally lensed transient detections, and phrase it as a model
for the relative rate of gravitational lensing detections that
conveniently side-steps messenger-specific technical details
(Section~\ref{sec:channels}\ref{sec:model}). We then apply this model
simultaneously to all messengers and present an integrated view of
relative detection rates across gravitationally lensed transient
sources and the different messengers
(Section~\ref{sec:channels}\ref{sec:relative}). This integrated view
motivates the focus of the rest of this article on gravitationally
lensed CBCs, beginning with a review of the available discovery
channels (Section~\ref{sec:channels}\ref{sec:pathways}).

We also introduce the term \emph{golden object}, echoing how several
breakthrough discoveries of individual objects have driven very
significant scientific progress, in some cases over many decades, for
example GW170817 and the Hulse Taylor pulsar. For the purpose of
convenience in this article, we define \emph{golden objects} as
gravitationally lensed sources for which multiple gravitational images
are detected directly via many messengers, at least one of which is
not electromagnetic. In the context of lensed CBCs -- the main focus
of this article -- this would include a lensed BNS for which multiple
images of the lensed merger are detected directly in GWs and more than
one EM messenger.

\subsection{A model for multi-messenger gravitational lensing rates}
\label{sec:model}

Previous works on the rates of gravitationally lensed transients
\cite{Oguri2010,Smith2018,Li2018gw,Ng2018,Haris:2018vmn,Oguri:2018muv,Wojtak2019,Goldstein2019,Wierda:2021upe,Abbott2021lens,Magare2023,Smith2023,Arendse2024,Phurailatpam2024}
are based on an underlying framework that can be summarised as:
\begin{equation}
  \Rlensed=\int d\vlum\int dz\int d\mu\,\,\frac{d\tau}{d\mu}\,\frac{dV}{dz}\,\frac{\mR(\vlum,z)}{1+z}\,\mK(z)\,p_{\rm det}(\vlum,\mu,z)\,,  \label{eqn:Rlensed}
\end{equation}
where $\Rlensed$ is the number of gravitationally lensed object
detections per unit time in the observer's frame, $\vlum$ are the
intrinsic source properties (e.g.\ luminosity, mass), $z$ is the
redshift of the sources, $dV$ is the comoving volume element, $\mR$ is
the comoving rate density of the sources, $\mK(z)$ describes how
cosmological redshifting alters detectability (analogous to optical
$k$-correction \cite{Hogg2002}), and $p_{\rm det}$ is the detection
probability for a given messenger. For definiteness,
Equation~\ref{eqn:Rlensed} expresses the optical depth to
gravitational lensing, $\tau$, in terms of gravitational
magnification, $\mu$, for the reasons outlined in
Section~\ref{sec:lensing}\ref{sec:opticaldepth}.

The comoving rate density of sources is conveniently phrased as a
separable function of redshift and intrinsic source properties,
$\mR=\mR_0\,g(z)\,\phi(\vlum|z)$, where $\mR_0$ is the local comoving
rate density, $g(z)$ describes redshift evolution, and $\phi(\vlum|z)$
represents the probability density function of CBCs, or luminosity
function of optical sources. Uncertainties in these terms are the
dominant sources of uncertainty in the \emph{absolute} number of
detectable lensed sources. In particular, the local comoving rate
density and the redshift evolution of the sources are often not
accurately known. Unknown redshift evolution is important because
gravitational magnification enables sources to be detected at
redshifts beyond those upon which models for $\mR$ are
based. Conversely, discovering gravitationally lensed sources at high
redshift and/or well-defined non-detections can constrain the redshift
evolution of the respective source populations.

Recent work on the detection rates of gravitationally lensed GW
signals tends to focus on the \emph{relative} rate of detection,
i.e.\ the ratio of lensed detections to detections that are not
lensed. This approach has the benefit that $\mR_0$ cancels, and
uncertainties on the functional form of $g$ and $\phi$ mainly impact
on the redshift and magnification distribution of the detectable
lensed populations (see below, and Figure~\ref{fig:zmupeak}). In what
follows we adopt typical (and benign) assumptions for $g(z)$, namely
non-evolving or evolution that tracks the evolution of the cosmic star
formation rate density (SFRD) \cite{Madau2014}. In the latter
scenario, the SFRD peaks at a so-called pivot redshift of $z_{\rm
  pivot}=1.9$ and declines as a power law at lower and higher
redshifts from that peak. While this model is not strictly relevant to
the details of all source populations, it serves as a useful baseline
for the overview presented here.

We write the relative rate of detection, $\varrho$, and the associated
rates of detection for events which are lensed and not lensed as
follows:
\begin{equation}
  \varrho(\zf)=\myfrac[2pt][2pt]{\Rlensed}{R_{\rm not}}=\myfrac[7pt][7pt]{\bigintsss_{\zmin}^{\zmax}dz\,\bigintsss_{\mumin(z)}^{\infty}d\mu\,\myfrac[2pt][3pt]{d\tau}{d\mu}\,\myfrac[2pt][3pt]{dV}{dz}\,\myfrac[2pt][3pt]{g(z)}{1+z}\,\mK(z)}{\bigintsss_{\zmin}^{\zf}dz\,\myfrac[2pt][3pt]{dV}{dz}\,\myfrac[2pt][3pt]{g(z)}{1+z}\,\mK(z)},
  \label{eqn:relative}
\end{equation}  
where $\mumin(z)$ is the minimum gravitational magnification required
to produce a detectable signal from a source/messenger combination,
and (\zmin,\,\zmax) denotes the redshift range over which the
respective detectors are sensitive to the different messengers. Also,
$\zf$ are the redshift frontiers for representative sources
(Section~\ref{sec:signals}, Table~\ref{tab:messengers}) out to which
they are detectable without assistance from gravitational lensing, at
the signal-to-noise ratio (SNR) limit required for detection by the
respective communities. In effect, for each source/messenger
combination, we collapse $\vlum$ to a single parameter $\Lambda$ and
assign it a single value that corresponds to the minimum signal
strength that is detectable at the relevant value of $\zf$. Finally,
we assume $\mK=(1+z)$, because this is relevant to band-limited
detections \cite{Hogg2002}. However, the take away messages from this
Section are unchanged if we were to assume $\mK=1$.

To illustrate the sensitivity of the predicted lensed populations to
the redshift horizon (as a combination of intrinsic source strength
and detector sensitivity) and assumed redshift evolution of the
source, we numerically integrate Equation~\ref{eqn:relative} and
compute the peak of the predicted redshift and and magnification
distributions. The redshift distributions for lensed detections of
evolving and non-evolving populations differ strongly
(Figure~\ref{fig:zmupeak}). For the evolving population, $z_{\rm
  peak}$ is tugged towards $z_{\rm pivot}$, and thus the detectable
lensed population is dominated by sources at $z\simeq1-3$. In
contrast, the $z_{\rm peak}$ for the non-evolving source population is
always at $z_{\rm peak}>\zh$. For $\zh\lesssim1$, an evolving
population therefore peaks at higher redshift than a non-evolving
population, and thus is more highly magnified. This behaviour reverses
at $\zh\gtrsim1$, however it is important to note that for these
redshift horizons detections of gravitationally lensed evolving and
non-evolving source populations are dominated by low magnification
lensing events, i.e.\ $\mu\lesssim10$, and $\mu_{\rm peak}$ is
dominated by the minimum magnification required for strong
lensing. Therefore, for $\zh\gtrsim0.5$, multiply-imaged detections
will be dominated by galaxy-scale lenses, whilst for $\zh\lesssim0.5$
multiply-imaged detections will be distributed across the mass
function of lenses including groups and clusters of galaxies.
\begin{figure}
  \vspace{-10pt}
  \centering
  \includegraphics[height=45mm]{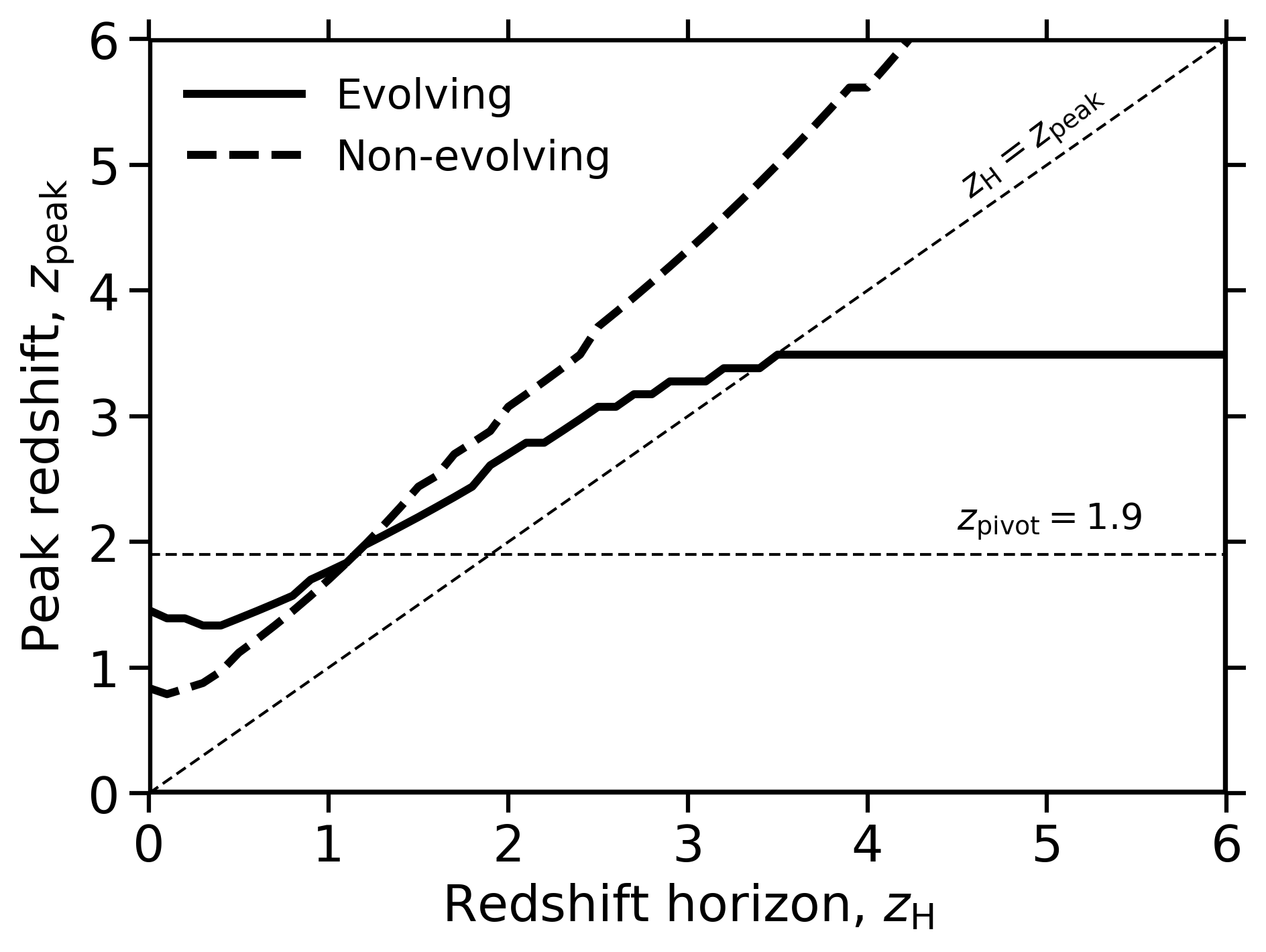} 
  \hspace{3mm}
  \includegraphics[height=45mm]{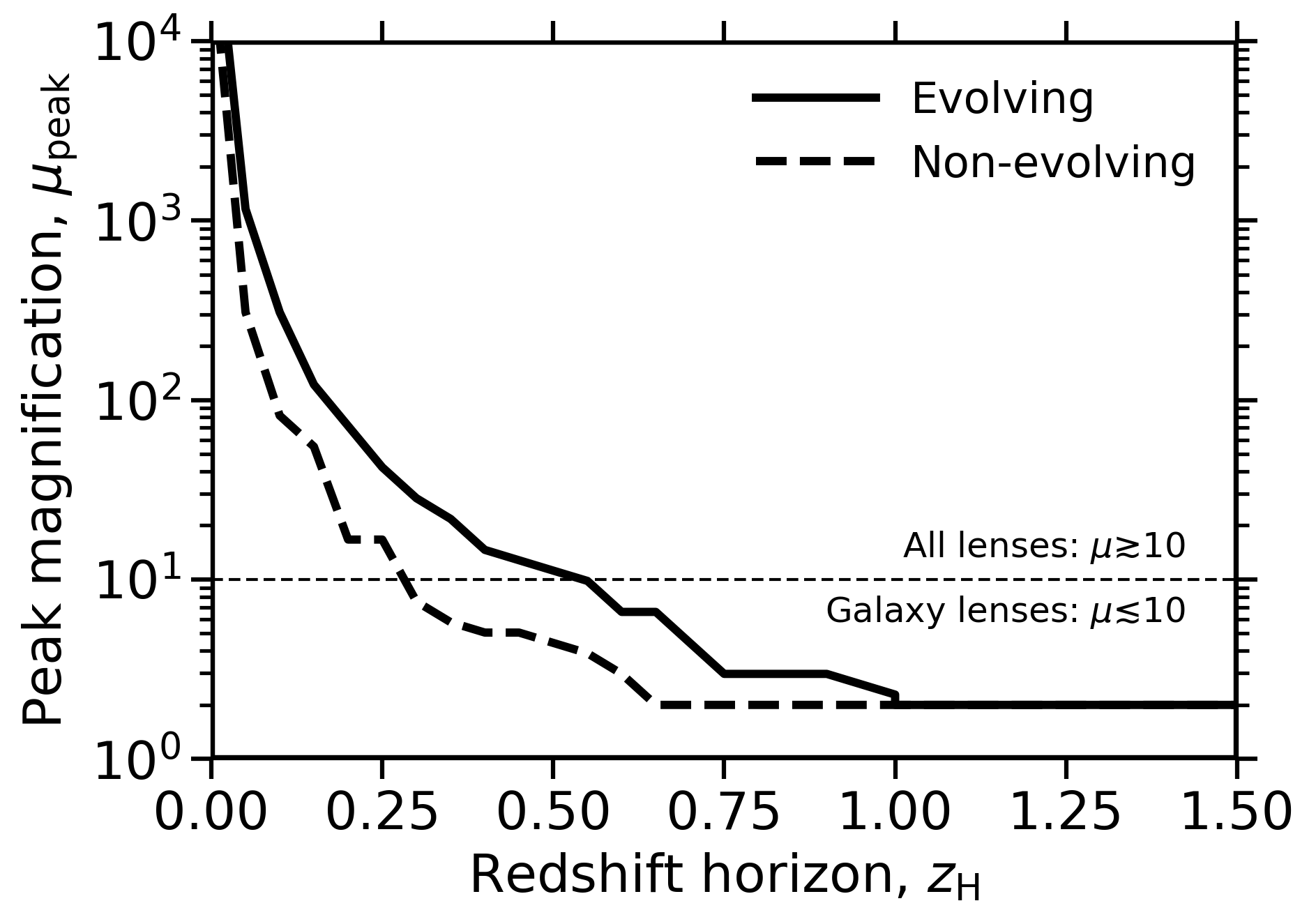} 
  \vspace{-5pt}
  \caption{The peak of the redshift (left) and magnification (right)
    distributions of detectable lensed sources, as a function of
    redshift horizon, $\zh$, based on the model and assumptions
    described in Section~\ref{sec:channels}\ref{sec:model}. The
    comoving rate density evolution of the ``evolving'' population
    tracks the SFRD history of the universe, as described in the
    text. For the evolving (more commonly used for forecasting)
    scenario, $\zh\simeq0.5$ is the approximate transition from
    detectable lensed sources being dominated by high magnification
    lensing ($\mu\gtrsim10$) to being dominated by low magnification
    lensing ($\mu\lesssim10$).}
  \label{fig:zmupeak} 
  \vspace{-10pt}
\end{figure}

\subsection{Relative detection rates by messenger and source population}
\label{sec:relative}

\begin{figure}
  \vspace{-10pt}
  \centering
  \includegraphics[width=\textwidth]{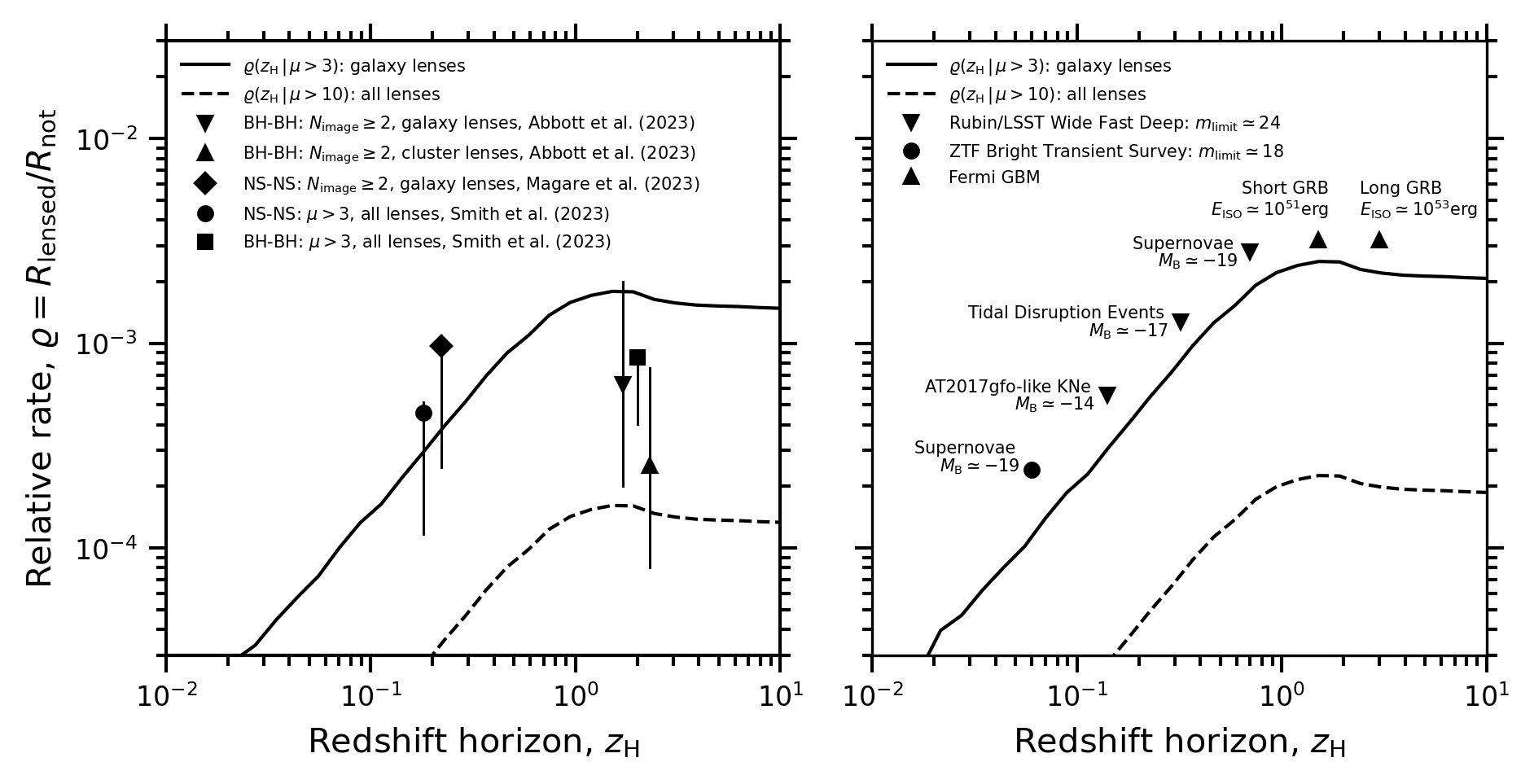} 
  \vspace{-10pt}
  \caption{{\sc Left} -- The predicted relative rates of discovery of
    gravitationally lensed GW sources from a number of detailed
    studies (data points) overlaid on curves based on the
    multi-messenger model discussed in
    Section~\ref{sec:channels}\ref{sec:model}. The upper (solid) and
    lower (dashed) curves bracket the range of threshold gravitational
    magnifications above which galaxy-scale lenses and all lenses
    (i.e.\ including massive galaxy clusters) are efficient at forming
    multiple images. The curves assume $\mathcal{K}=\,\rm {constant}$,
    as is relevant to GW detectors.  {\sc Right} -- The typical
    redshift horizons out to which different EM sources can be
    detected without assistance from gravitational magnification
    ($\zh$, points), shown at an arbitrary offset above the curves,
    for clarity. The curves assume $\mathcal{K}\propto(1+z)$ for
    simplicity, i.e.\ the $k$-correction relevant to an EM source that
    has a flat $S_\nu$ spectrum and detected photometrically. The
    difference between solid and dashed curves is the same as in the
    left panel. The expected relative rate of detection of
    gravitationally lensed images by the respective surveys can be
    read off from curves at the redshifts that correspond to each of
    the points.
  \label{fig:relative_rate} 
  \vspace{-10pt}
}
\end{figure}

The relative detection rate, $\varrho$, of gravitationally lensed
images increases rapidly at $\zh<1$ before plateauing at one lensed
detection per $\simeq10^3$ detections that are not lensed at
$\zh\gtrsim1$ (Figure~\ref{fig:relative_rate}). In the left panel, the
model for $\varrho(\zh)$ is consistent with several independent
predictions for gravitationally lensed GW sources that are based on
detailed calculations \cite{Li2018gw, Ng2018,
  Haris:2018vmn,Oguri:2018muv, Wierda:2021upe,Xu:2021bfn,
  Abbott2021lens,Mukherjee:2021qam, Magare2023, Smith2023}. In
summary, roughly one per thousand GW sources detected in the 2020s and
2030s is expected to be gravitationally lensed, independent of
detector sensitivity. In the coming decade, the improving sensitivity
of GW detectors combined with the relative lensing rate of
$\varrho\simeq10^{-3}$ will enable significant numbers of detections
gravitationally lensed GWs from LVK's O5 onwards
(Table~\ref{tab:messengers}).  The prospects are even stronger for
future detectors, such as ET or CE, that are expected to detect
$\simeq10^5$ GW sources per year \cite{Maggiore_2020,
  CEwhitepaper2023}, including lensing of other types of CBCs.

Redshift horizons lower than those shown in
Figure~\ref{fig:relative_rate}, i.e.\ $\zh<0.01$, correspond to
source/messenger combinations that are not detectable beyond
$D(\zh)\simeq40\,\rm Mpc$ unless the signal is boosted by
gravitational lensing. At $\zh<0.01$ the relative rate is very low,
$\varrho<10^{-4}$, driven by the extreme gravitational magnification
($\mu\gtrsim10^6$, i.e.\ beyond the upper limit on magnification that
is typical of finite source effects \cite{Diego2019extreme}) required
to detect a source at a typical redshift of $z\simeq1-2$ if the
redshift horizon is $\zh<0.01$. Note, the local group is at lower
redshifts still. Moreover, the cosmological volume interior to these
redshifts is very small, rendering the number of detections of sources
that are not lensed to be very small at $R_{\rm not}\ll1\,\rm
year^{-1}$. For example, even next generation GW detectors will only
be sensitive to core collapse supernovae within our own galaxy, and
next generation neutrino detectors are expected to be sensitive within
the local group of galaxies. As alluded to in
Section~\ref{sec:signals}, this is the motivation for focusing this
article on messengers from gravitationally lensed CBCs.

Turning to the right panel of Figure~\ref{fig:relative_rate}, the
relative detection rate of gravitationally lensed images is one per
$\simeq10^3-10^4$ across EM messengers from sources discussed in this
review. Therefore, as the number of detections that are not lensed
approaches $R_{\rm not}\simeq10^3-10^4$, the detection of
gravitationally lensed images becomes more likely. This is consistent
with the detection of a few gravitationally lensed SNIa by the
combination of iPTF and ZTF \cite{Goobar2017,Goobar2023}, the
expectation that some of the $\simeq10^4$ GRBs that have been detected
to date are in fact gravitationally lensed \cite{Levan2025}, and
preparations to discover hundreds of gravitationally lensed supernovae
with Rubin/LSST \cite[for
  example]{Goldstein2019,Wojtak2019,Arendse2024,Abe2024}.

\subsection{Pathways to multi-messenger gravitational lensing discovery}
\label{sec:pathways}

In general, discovery requires candidate gravitationally lensed
signals to be selected from the many signals that are detected, as the
trigger for follow-up analysis and observations. Efficient selection
requires the lensed signal to be distinctive in some way relative to
signals that are not lensed. Typically, this relies on gravitational
magnification to make lensed sources appear to be brighter and closer
than their true brightness and distance, and/or the detection of two
or more signals that are consistent with being lensed images of a
single source.

\begin{figure}
  \vspace{-10pt}
  \centerline{\includegraphics[width=0.7\textwidth]{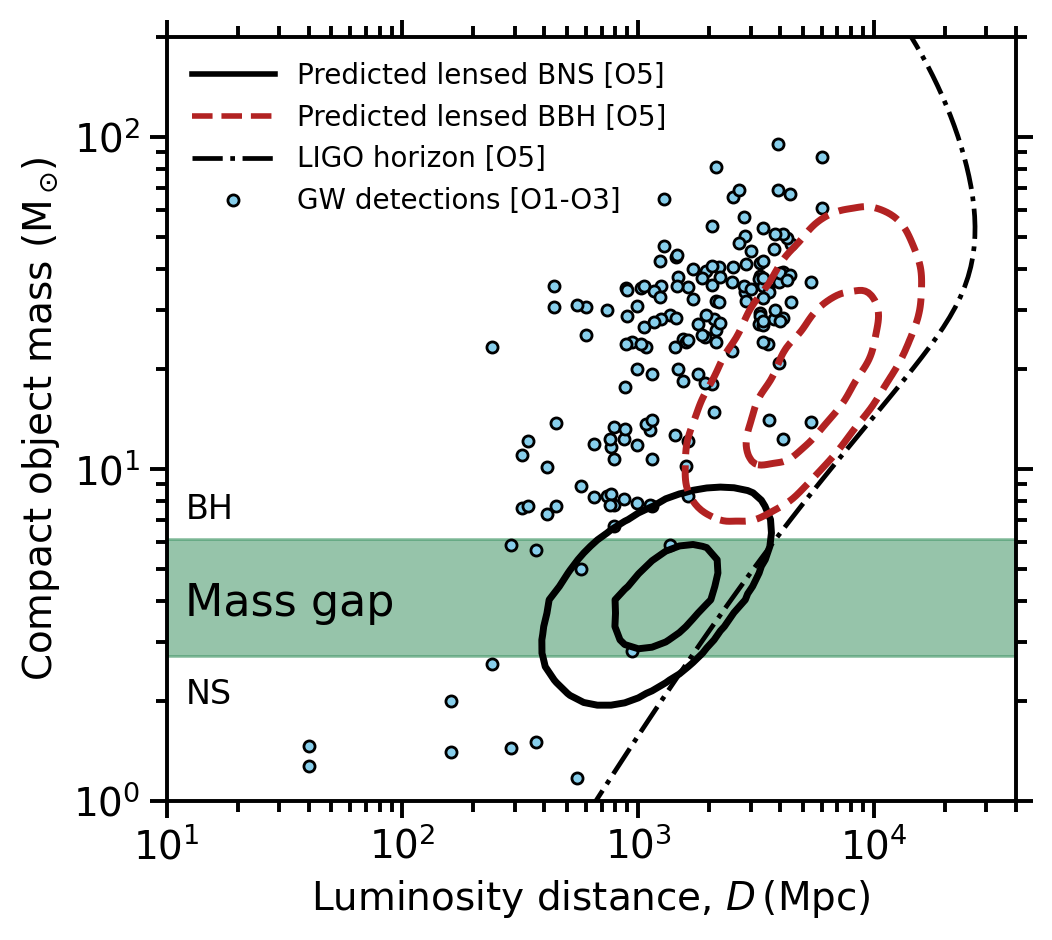}}
  \vspace{-10pt}
  \caption{GW signals from gravitationally lensed BBHs during the
    fifth LVK run (as inferred in low latency, assuming $\mu=1$) are
    predicted to overlap in mass with the bulk of the GW signals --
    compare red dashed contours with the detections from the first
    three runs. In contrast, GW signals from gravitationally lensed
    BNSs are predicted to be dominated by sources that appear in low
    latency to be located in the so-called ``mass gap'' between
    neutron stars and stellar remnant BHs. This allows a more
    efficient selection of candidate lensed BNS based using
    magnification-based methods, than for candidate lensed BBHs.  This
    figure is based on work published in
    \cite{Smith2023,Andreoni2024}.
  \label{fig:massdistance} 
  \vspace{-10pt}
}
\end{figure}

To bring the focus to gravitationally lensed CBCs, we summarise some
of the challenges involved in selecting candidate gravitationally
lensed GW sources. First, the relative detection rate of
$\varrho\simeq10^{-3}$ (Section~\ref{sec:channels}\ref{sec:relative})
motivates assuming that GW detections are not gravitationally lensed
unless strong evidence emerges to the contrary. The mass and distance
of GW detections are both degenerate with lens magnification, and
therefore they appear brighter and closer than they really are,
analogous with EM detections. However, the predicted masses that the
LVK collaboration would infer in low latency (i.e.\ assuming $\mu=1$)
for gravitationally lensed GW sources have significant overlap with
the range of masses of GW sources that (given that
$\varrho\simeq10^{-3}$) are unlikely to be lensed. For example, in
LVK's fifth run, based solely on the mass axis in
Figure~\ref{fig:massdistance}, essentially every GW detection at
$M\gtrsim2\,\rm M_\odot$ could be regarded as a candidate
gravitationally lensed GW source based on a magnification argument,
and should in principle be included in dedicated lensing analyses to
be confirmed/ruled out such as was done in previous observing runs
e.g.~\cite{LIGOScientific:2021izm, LIGOScientific:2023bwz}. Note that
the data points in this Figure relate to GW sources detected in
previous GW runs and are therefore subject to a horizon a factor
$\simeq5\times$ leftward (lower distance) than the O5 horizon that is
shown.

Rapid \emph{and} efficient identification of candidate gravitationally
lensed GW sources is most critical for those sources that have
transient EM counterparts, because both speed of EM ToO follow-up
observations and suppression of false positives among the candidate
lensed GW sources are essential. The need for rapid follow-up,
including the science case for detection of the first lensed kilonova
image to arrive
(Section~\ref{sec:science}\ref{sec:sources}\ref{sec:kilonovae}),
motivates a magnification-based selection of candidates, if false
positives can be adequately controlled. The putative ``mass gap''
between the most massive NSs and the least massive BHs
\cite{Bailyn1998, Oezel2010, Farr2011, deSa2022, GW230529, LVK03pop}
is a promising region of parameter space in which to select candidate
lensed BNS mergers. The appeal is mainly empirical, in that this
region of parameter space is sparsely populated, and not based on
asserting that this region is empty of sources that are not
lensed. The main strength of this discovery channel is the proven
association of GWs, kilonovae and GRBs with BNS mergers, and thus the
potential to discover \emph{golden objects}. The challenges include
the diversity of intrinsic properties of kilonovae, large GW sky
localization uncertainties, and the relative rarity of BNS mergers.

Most, and potentially all, BBH mergers are EM-dark, and hence the
emphasis on rapid identification of candidate lensed BBH mergers among
GW detections is less severe than for candidate lensed BNS mergers.
This, coupled with the significant overlap in the mass distributions
of BBH mergers that are lensed and not lensed
(Figure~\ref{fig:massdistance}), motivates a greater focus on
selecting candidate lensed BBH mergers for further investigation via
image multiplicity. The main strength of this discovery channel is
that BBH mergers are more numerous than BNS mergers among GW
detections, and thus detection of the relevant GW signals by LVK is
more likely. The challenges include the large GW sky localization
uncertainties that will contain many gravitational lenses even after
significant improvements in the sky localization derived from the
joint posteriors of two GW detections. Nevertheless,
magnification-based selection of candidate lensed BBH sources is
possible, for example in association with the follow-up ToO
observations of massive BBH detections to search for AGN flare
counterparts, following the candidate counterpart to GW190521
discussed by \cite{Graham2020,Ashton2021,Palmese2021,Morton2023}.  It
is, however, noted that this focus is not exclusive; signatures of
gravitational lensing may also be detected in individual GW
detections, through waveform distortions resulting from microlensing
or millilensing, or Type II images due to their negative parity.

\begin{figure}
  \vspace{-10pt}
  \centerline{\includegraphics[width=\textwidth]{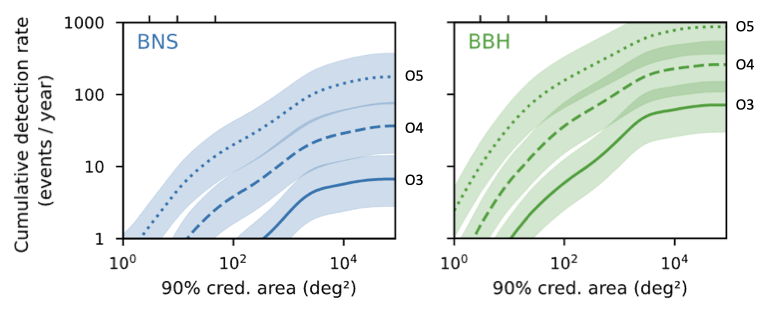}}
  \vspace{-10pt}
  \caption{This figure is adapted from the figure available at
    https://emfollow.docs.ligo.org/userguide/capabilities.html, based
    on \cite{Petrov2022}. It shows the predicted cumulative
    distributions of sky localization uncertainties of GW detections
    by LVK through to their fifth run. Independent of run or source
    type, $\simeq10\%$ of detections will be localized to better than
    $\Omega\simeq100\,\rm degree^2$ precision. Improvements on this
    await extension of the GW detector network, via LIGO India
    \cite{Saleem2022}. We also note that in the case where several
    lensed images are detected, major improvement in the sky
    localisation uncertainties are possible\cite{Uronen:2024bth}.
  \label{fig:localization} 
  \vspace{-10pt}
}
\end{figure}

Before moving on to discuss the channels introduced above in more
detail, we provide further context on GW sky localization
uncertainties in Figure~\ref{fig:localization}. As the sensitivity of
the current GW detector network improves towards O5, the fraction of
GW detections with sky localizations of $\Omega_{90}\lesssim100\,\rm
{degree}^2$ remains at around ten per cent. This fraction will
increase significantly when the planned LIGO-India detector comes
online \cite{Saleem2022}. The size of the GW sky localization
uncertainties are key to the synergies between GW and EM messengers,
both for efficient use of telescope time to follow-up GW sources and
for efficient comparison with EM-based catalogues of known
gravitational lenses. It is also important to note that detection of
multiple GW signals from a gravitationally lensed CBC merger helps to
reduce the sky localisation uncertainties considerably.

\subsubsection{Gravitationally lensed binary neutron star mergers}
\label{sec:bns}

\begin{figure}
\vspace{-10pt}
\centerline{\includegraphics[width=0.7\textwidth]{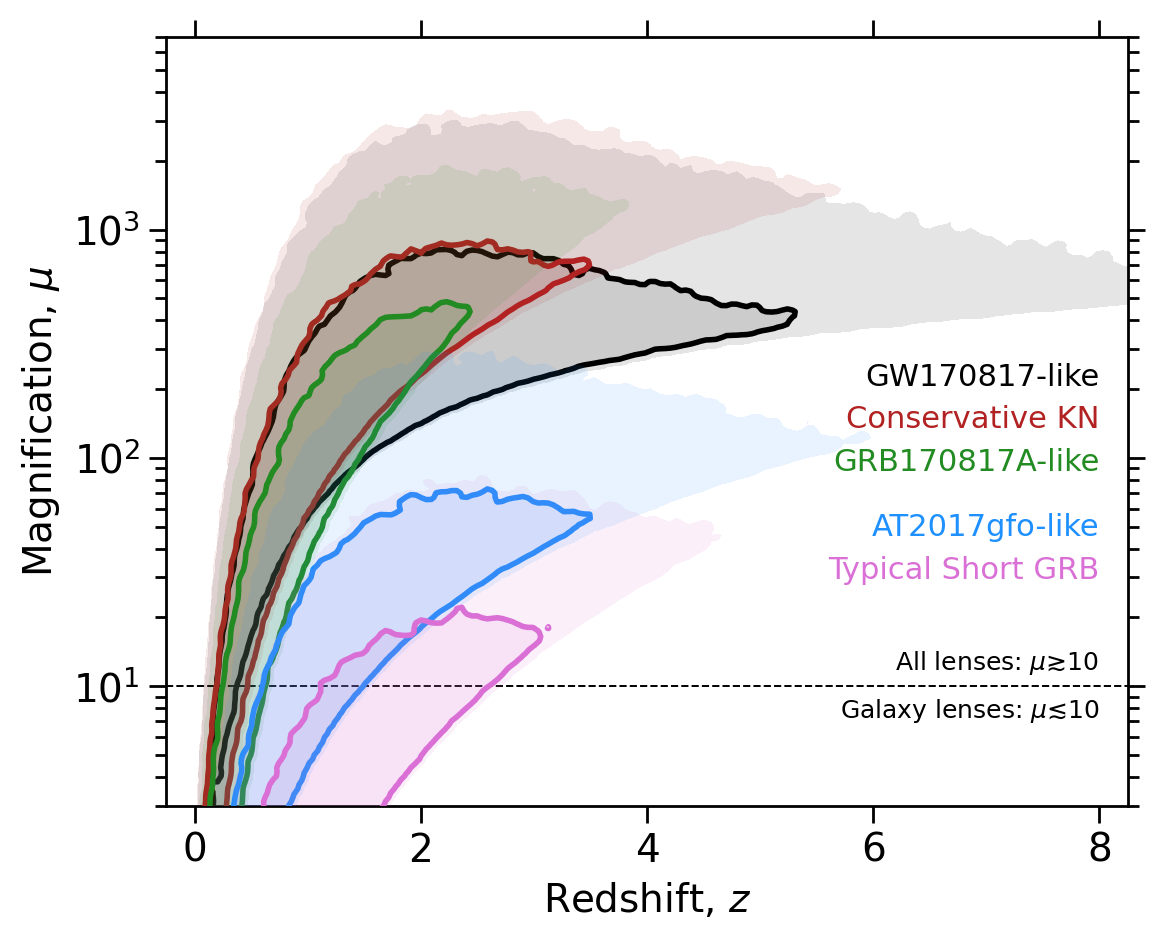}}
  \vspace{-10pt}
  \caption{Magnification-redshift distributions of messengers from
    gravitationally lensed BNS mergers, based on
    Table~\ref{tab:messengers}, Equation~\ref{eqn:relative}, and
    Section~\ref{sec:channels}\ref{sec:model}. Contours encloses 90\%
    of the predicted lensed detections, and the shaded areas extend to
    99\% to visualise the tails of the respective distributions, as
    explained in
    Section~\ref{sec:channels}\ref{sec:pathways}\ref{sec:bns}.
  \label{fig:muz} 
  \vspace{-10pt}
}
\end{figure}

The detection of multiple messengers from a BNS merger in 2017
(GRB170817A, GW170817, AT2017gfo), combined with current / imminent
detector sensitivities has opened up the exciting prospect of
detecting a gravitationally lensed CBC via multiple messengers. To
give a concrete example, in Figure~\ref{fig:muz} we show the location
of detectable messengers from a gravitationally lensed BNS merger in
LVK's fifth run, based on the multi-messenger lensing model described
in Section~\ref{sec:channels}\ref{sec:model}, and assuming the LVK
A$+$, LSST ToO and \emph{Fermi}/GBM sensitivities listed in
Table~\ref{tab:messengers}. For each messenger the lower edge of the
respective contour represents a hard detection limit based on the
respective horizons and the implied gravitational magnification
required for detection. The detectable messengers have a tail to high
magnification as is apparent from the extension of pale shaded regions
beyond their respective contours.

In the context of initial discovery via GWs (black contour), the key
takeaway from Figure~\ref{fig:muz} is that none of the other
distributions are peaking / extending to higher magnifications than
the black contour and pale grey shading. The EM instrument
sensitivities are therefore well-matched to detecting the EM
counterpart to an LVK detection of a gravitationally lensed BNS
signal. Importantly, this is not strongly dependent on the details of
the EM signals because the red and green contours that overlap well
with the black GW contour are based on conservative assumptions about
the brightness of EM signals. The red contour assumes that the
kilonova counterpart is redder and fainter than AT2017gfo, following
the ``conservative'' model discussed by
\cite{Nicholl2021,Smith2023,Andreoni2024}. Equally, the green contour
assumes that the GRB counterpart is fainter than a typical short GRB,
for example due to being viewed off-axis, as was GRB20170817A. The
blue (AT2017gfo-like kilonova) and pink contours (typical short GRB)
correspond to brighter EM scenarios in which a lensed BNS merger that
is detected by LVK in GWs would be detectable as a kilonova and short
GRB, albeit in the respective high-magnification tails.

\begin{figure}
  \vspace{-10pt}
  \centerline{\includegraphics[width=\textwidth]{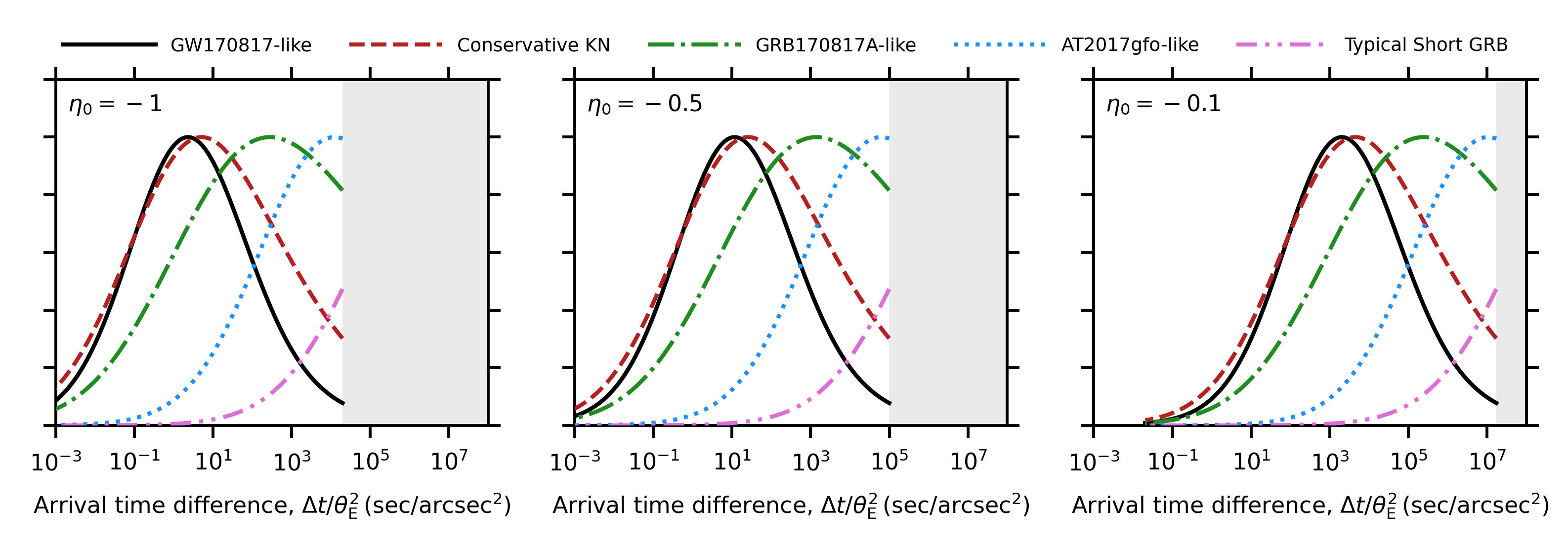}}
  \vspace{-10pt}
  \caption{ Arrival time difference distributions for the five
    messenger/instrument combinations shown in Figure~\ref{fig:muz},
    normalised to an Einstein radius of $\thE=1\,\rm arcsec$, based on
    combining the magnification distributions shown in that Figure
    with Equation~\ref{eqn:fold}, for lens density profiles that are
    steep ($\eta_0=-1$), intermediate ($\eta_0=-0.5$), and flat
    ($\eta_0=-1$) at the mid-point between fold image pairs
    (Sections~\ref{sec:lensing}\ref{sec:galaxy}~\&~\ref{sec:lensing}\ref{sec:cluster}). The
    grey shaded region in each panel indicates the region in which
    $\mu<10$, i.e.\ where lenses with flatter density profiles tend to
    be less efficient at forming multiple images
    (Section~\ref{sec:lensing}\ref{sec:cluster}). The distributions
    are all normalised to the same arbitrary peak value. The overlaps
    of the arrival time distributions shown in this Figure reflect the
    overlapping distributions in Figure~\ref{fig:muz}.
  \label{fig:dt} 
  \vspace{-10pt}
}
\end{figure}

Identification of GW signals from candidate gravitationally lensed BNS
can be based on identifying sources that have a high probability of
comprising one of more compact objects with mass consistent with
$3<M<5\,\rm M_\odot$
\cite{Smith2023,Bianconi2023,Ryczanowski2025}. This selection, based
on the information released with low latency by LVK, is also the
baseline for the current planning of Rubin/LSST ToO follow-up of
candidate gravitationally lensed BNS \cite{Andreoni2024}. Clearly, a
joint magnification plus multiplicity selection would be extremely
powerful if the arrival time difference between two lensed GW signals
is $\Delta t\lesssim1\,\rm hr$, as highlighted by
\cite{Smith2023,Magare2023}. The short arrival time differences
associated with lensed BNS mergers are a direct consequence of the
relatively large magnifications required to detect them. For example,
the arrival time difference between a fold image pair formed by a
galaxy-scale lens (likely part of a quad image configuration) can be
typically as short as a second, and typically reach a day for a very
flat cluster-scale lens (Figure~\ref{fig:dt}). The shorter arrival
time differences for lensed BNS mergers therefore have potential to
probe the Eikonal optics regime
(Section~\ref{sec:lensing}\ref{sec:waveoptics}~\&~Section~\ref{sec:science}\ref{sec:cosmology}\ref{sec:cosmography}).

Additional information from the GW data also has significant potential
to suppress false positives when selecting candidate lensed GW
signals. For example, the mass ratio of the CBC and its detector-frame
chirp mass are both invariant to gravitational lensing, and therefore
are well-suited to improving the selection of candidates, if
available. Detection of GW signals that appear to emanate from the
mass gap and that also contain signatures of tidal deformability of
the compact objects involved in the merger that are consistent with
them being NSs would also add further weight to a magnification-based
mass gap selection \cite{Pang:2020qow}. This enhanced method likely
awaits next-generation GW detectors because measurements of tidal
deformabilities of GW sources are rather poorly constrained with
current GW detectors \cite[for example]{PhysRevX.9.011001,GW230529}.

We also consider the scenario of EM-led detection of gravitationally
lensed BNSs. Again to give concrete examples, if this was based on
multiple detections of a typical short GRB and/or a gravitationally
lensed AT2017gfo-like kilonova, then the corresponding GW signals are
unlikely to be above the detection threshold of the LVK data. This can
be seen in the blue and pink contours being below the black contour in
Figure~\ref{fig:muz}. Therefore, if LVK was operating at the time of
such a detection, then a sub-threshold search of the LVK data would
probably be required to search for the GW signals, following similar
approaches to sub-threshold GW searches on GRB detections
\cite{LIGOScientific:2022jpr}. A GRB-led approach also highlights that
the initial GRB sky localization uncertainties can span thousands of
$\rm degree^2$. To succeed, GRB-led discovery would therefore require
progress on rapid identification of candidate lensed GRBs (via
multiplicity), and then rapid localization via their afterglow and/or
kilonova emission \cite{Andreoni2024,Levan2025}.

\subsubsection{Gravitationally lensed dark binaries}\label{sec:bbh}

To date the number of GW signals detected from BBH mergers outnumbers
those from BNS mergers by a factor of
$\simeq50$\cite{LIGOScientific:2018mvr,
  theligoscientificcollaboration2022gwtc21deepextendedcatalog,
  LVKgwtc2023}, and the rate of lensed GW detections is also expected
to follow this pattern assuming the two types of mergers follow the
relative lensing rates (valid only for next-generation detectors). The
detection rate of BBH mergers continues to grow and indicates that LVK
will be capable of detecting a few lensed BBH mergers per year during
their fifth run in the late-2020s
\cite{Ng2018,Yang2022,Li:2018prc,Smith2023,Wierda:2021upe,Xu:2021bfn,Phurailatpam2024}. Various
tools and pipelines have been developed in recent years to analyse and
identify lensed candidates in LVK GW data, though no conclusive
evidence for lensing has been found so far
\cite{Hannuksela2019,LVlens2021,LVKlens2023,Janquart:2023mvf}. For
single, standalone GW events, searches for lensing signatures such as
Type II images \cite{Ezquiaga_2021, Wang:2021kzt, Janquart:2021nus,
  Janquart:2023osz}, micro- and millilensing
\cite{Wright_Janquart_Hendry,Liu_2023_eikonal} are conducted. For
multiple images occupying similar regions of the GW parameter space,
we can analyse the pairs, triplets, or quadruplets against the chance
of coincidental parameter match
\cite{Liu:2020par,Janquart:2021nus,Janquart:2023osz,Lo:2021nae}. This
is particularly effective for regions of the parameter space that are
less densely populated, such as very high-mass events. While the risk
of coincidental association between multiple candidate images
increases with the number of detected GW events
\cite{Wierda:2021upe,Caliskan2023lensingorluck}, introducing
`time-delay windows’ (i.e.\ limiting the time window within which
events are paired up together in lensing searches), according to
predicted time-delay distributions from gravitational lenses
significantly reduces the false-alarm probability \cite{Haris:2018vmn,
  Wierda:2021upe, More_2022, Janquart:2022zdd}.

\begin{figure}
  \centering
  \includegraphics[width=1\linewidth]{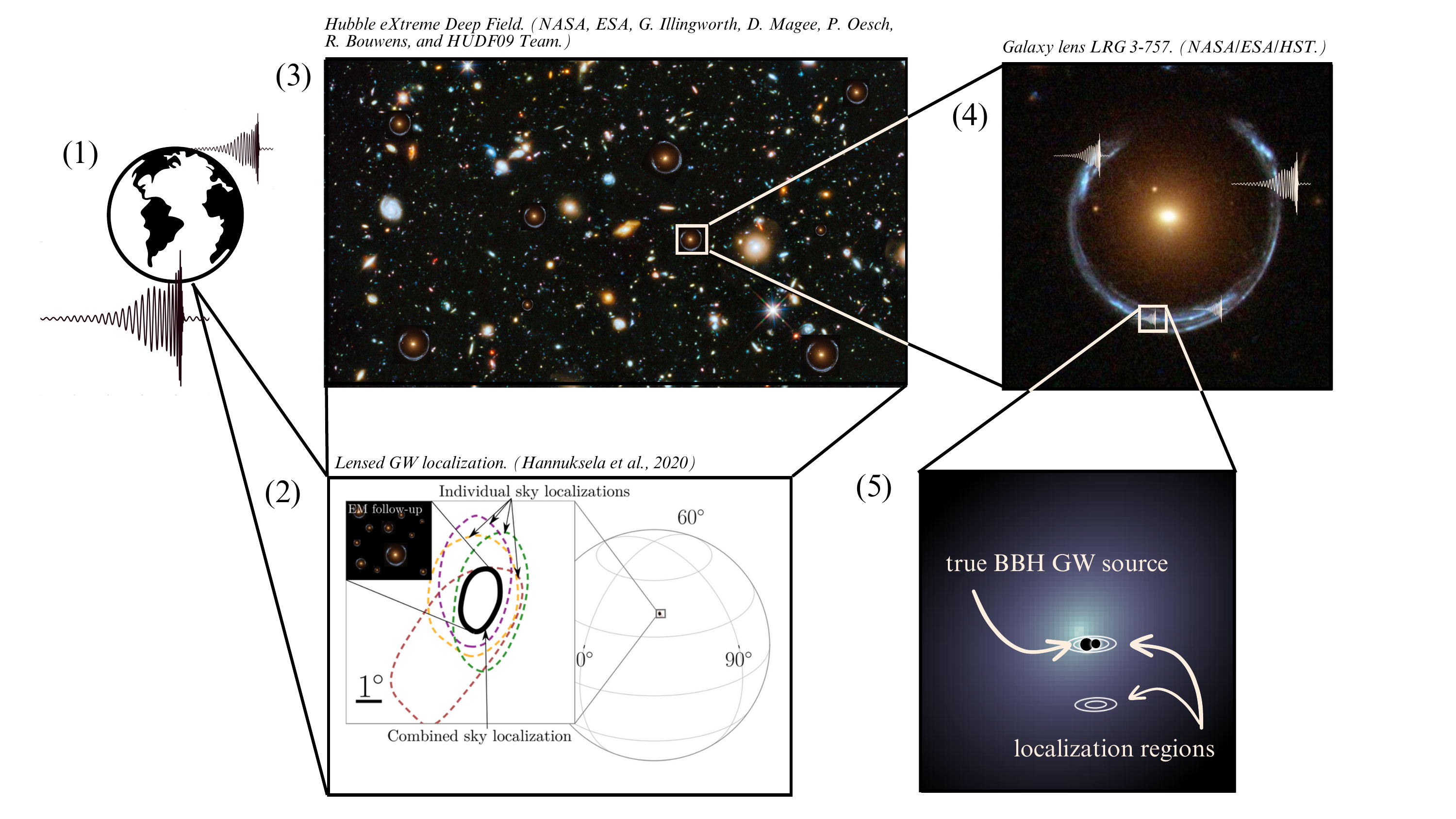}
  \caption{A schematic for the steps to localize a dark lensed binary
    merger. (1) Lensed GW images are detected by the ground-based LVK
    observatories. (2) The sky localizations from the multiple
    identified images can be analysed jointly to reduce the final sky
    localization region \cite{Hannuksela2020}. (3) { The joint sky
      region can be cross-matched with gravitational lens catalogues
      from, LSST, \textit{Euclid} and their contemporaries}. (Edited
    from: NASA, ESA, Illingworth, Magee, Oesch, Bouwens, and HUDF09
    team.) (4) The candidate lenses are individually analysed and
    reconstructed to test their match to the GW images. (Edited from:
    NASA/ESA/HST.) (5) If a gravitationally lensed galaxy from the EM
    lens catalogues stands out as a distinctly high-ranked candidate
    host of the dark lensed CBC merger, the CBC can then be localized
    accurately in the source plane.}
  \label{fig:schematic}
\end{figure}

BBH mergers are not typically expected to be accompanied by direct EM
counterparts -- although see
\cite{Connaughton2016,Graham2020,Ashton2021,Palmese2021,Morton2023}
for intriguing candidates, of which we discuss the AGN disk scenario
in more detail below. The EM counterparts to NSBH mergers are expected
to be fainter than counterparts to BNS mergers \cite[for
  example]{Gompertz2023}.

Host identification is a challenge for all CBCs without an identified
EM counterpart (Figure~\ref{fig:schematic}). However, with strong
lensing we obtain multiple images of the same GW event. If each of
these images is strong enough to be initially detected as if they were
independent GW observations, they each come with their own
$\mathcal{O}(10-1000)$ degree$^2$ sky localization
\cite{abbott2020prospects} that can be jointly analyzed to reduce the
localization to $\mathcal{O}(10)$ degree$^2$ for double- and
triple-lensed GWs, and $\mathcal{O}(1)$ degree$^2$ for
quadruplets\cite{Hannuksela:2019kle,Janquart:2021qov,Uronen:2024bth}. We
can also first do ``dark lens reconstruction’’ by using the properties
of the lensed GW signals themselves to narrow down the parameter
spaces of the lens directly, though this remains subject to
degeneracies in particular for axially asymmetric lens models
\cite{seo2024inferring, Poon_gw_lens_degeneracies_overview_2024}. This
lets us narrow down the list of candidate lenses and hosts in the sky
region, and the full lens reconstruction of the candidate lens
profiles can then test if a particular lens model created the observed
GW event \cite{Hannuksela:2019kle, Wempe2024, Wright_Janquart_Hendry,
  Shan:2023ngi, Uronen:2024bth,seo2024inferring}. When the sky
localization region is sufficiently constrained, the lens uniquely
identifiable and the lensed host galaxy bright enough, the host galaxy
can be identified in up to about 30\% of cases for quadruply-lensed
GWs\cite{Wempe2024}. With upcoming detectors the rate of lensed GWs is
forecasted to increase, and we will be able to observe multiple lensed
events each year, giving us information about the broader population
of GW hosts.

From the GW side, the principal challenges remain around instrument
sensitivity. The ability to detect as many of the lensed GW images as
possible also carries a dependence on GW detector run length and duty
cycle, therefore reducing the efficiency of discovery
\cite{Smith2023}. Higher detector sensitivities can reduce the number
of lensed images missed, which will assist in better constraining the
lens parameters and the sky region, which is crucial for EM follow-up
observations and cross-matching with EM-based lens catalogues. More
detectors operating will also significantly improve sky localizations
\cite{Fairhurst:2012tf,Saleem2022}. Dedicated methods to find weaker
images, which would separately fall below the usual detection
threshold, by leveraging information from one or more already detected
images (referred to in the literature as ``targeted sub-threshold
searches'') \cite{McIsaac2020, Li:2019osa ,Dai:2020tpj, Li:2023zdl,
  Ng:2024ooy} can also help to increase the multiplicity of detected
lensed systems. The sensitivity of such searches can, in turn, be
improved by obtaining better constraints from lens models or lists of
candidate lenses.

Improvements in the success rate of host galaxy identification will
also come from deeper and more complete catalogues of gravitational
lenses from EM surveys such as Rubin/LSST and \emph{Euclid}
\cite{Collett:2015,Shajib2024rubin} and improved empirical
understanding of the covariance of lens density, structure, mass and
image multiplicity (Section~\ref{sec:lensing}\ref{sec:cluster}). Both
survey sensitivity and sub-arcsec second angular resolution are
critical for GW host identification. The former lets us maximise the
number of lenses identified, while the second provides enough detail
about lenses for initial reconstructions to narrow down the candidate
lists as much as possible. Should the initial resolution not be high
enough, or the lens not unique enough, to identify a single host
candidate distinctly, the top-ranked candidates would need
higher-resolution dedicated follow-up observations, for example with
the \emph{Hubble Space Telescope}, \emph{James Webb Space Telescope},
and 30-m class telescopes.

Another possible avenue comes directly from the example of GW190521, a
high-mass BBH GW source
\cite{Abbott2020GW190521,Abbott2020Astrophysical}. This event prompted
a great deal of interest not only due to its high mass, but also
because of its possible association with an AGN flare
\cite{Graham2020,Palmese2021,Ashton2021,Morton2023}. When a BBH merger
occurs in an AGN disk, the merger can cause a shock inside the gas
disk that results in an observable flare
\cite{Kimura2021,Rodriguez2023,tagawa2023}. In this particular case,
the AGN flare’s association with the BBH merger remains uncertain
\cite{Palmese2021,Ashton2021,Morton2023}. Detecting a lensed AGN flare
associated with an unusually high-mass BBH event could thus be
considered a direct observable counterpart to a lensed BBH
\cite{Andreoni2024}. Furthermore, since a substantial fraction of
AGN-disk BBHs are expected to be strongly lensed by the AGN
super-massive BH, the non-detection of strong lensing can place
constraints on the fraction of BBHs formed in AGN disks
\cite{Leong:2024nnx}.

\section{Multi-messenger gravitational lensing science}\label{sec:science}

This Section describes many of the science cases for multi-messenger
gravitational lensing, organised under those relating to the nature of
gravity (Section~\ref{sec:science}\ref{sec:gravity}), cosmology
(Section~\ref{sec:science}\ref{sec:cosmology}), and the physics of the
source popluations (Section~\ref{sec:science}\ref{sec:sources}). Each
science case includes a summary of the key challenges and progress
that is required in the next 3-5 years.

\subsection{The nature of gravity}\label{sec:gravity}

As direct manifestations of the space-time metric, it is no surprise
that GWs offer new tools with which to probe directly the nature of
gravity \cite{Berti_2015, Barausse:2020rsu}. Gravitationally lensed
GWs and EM counterparts expand and enhance these tools, thanks to
detection of multiple magnified copies of multi-messenger signals
offset in time from each other, and the greater distances over which
lensed signals typically travel relative to typical signals that are
not lensed.

Deviations from GR that affect large cosmological scales are also a
highly-studied probe of the nature of dark energy. If such deviations
exist, GWs should pass through the modified gravitational regime on
their way from the source to our detectors, resulting in changes to
the amplitude and phase evolution of GWs. Gravitational lensing can
play a key role in revealing some of these changes; hence, the
detection of a multi-messenger lensing event could offer new
opportunities to pin down or rule out causes of cosmic
acceleration. However, it is important to note that both lensing and
departures from GR share some common phenomena. This could lead to the
possibility of searches for either effect having false positives
caused by the other. For example, see
\cite{gupta2024possiblecausesfalsegeneral} for a broader review of
possible sources of false positives in searches for deviations from GR
and \cite{mishra2023unveilingmicrolensingbiasestesting,
  wright2024effectdeviationsgeneralrelativity} for specific
investigations for lensing and deviations from GR. Such false positive
systematics must be carefully modelled.

Some common effects of cosmological modified gravity theories on GW
propagation, that are discussed below, can be represented
schematically as follows \cite{Saltas:2014dha}:
\begin{equation}
  \label{eq:MG_prop}
  h^{\prime\prime}_{ij}+\Big[2+\nu(z)\Big]{\cal H}h^\prime_{ij}+\left[c_T^2(z) k^2+a^2m_g^2\right]h_{ij} = a^2\Gamma(z) \gamma_{ij},
\end{equation}
where primes denote derivatives with respect to conformal time, ${\cal
  H}=a^\prime/a$ is the conformal Hubble factor, $h_{ij}$ represents
either the plus or cross GW polarisation, and $\gamma_{ij}$ is a
transverse-traceless tensor.  $c_T^2(z)$ encodes the speed of
propagation of GWs. The terms $\nu(z)$, $m_g^2$ and $\Gamma(z)$ all
represent new phenomenology \cite{Saltas:2014dha}; the standard GR
propagation equation on a Friedmann-Robertson-Walker metric is
recovered in the limit $\nu(z)$,\;$m_g^2,\;\Gamma=0$ and $c_T^2=1$.

Regarding the physical interpretation of Equation~\ref{eq:MG_prop},
$\nu(z)$ is sometimes referred to as a `GW friction' term, as it
affects the rate of change of the GW amplitude as it propagates. In
scalar-tensor gravity theories -- the largest, simplest class of
models -- $\nu(z)$ is related to the time derivative of the
gravitational coupling. This is equivalent to the rate of change of
the effective Planck mass or gravitational constant. Meanwhile,
$m_g^2$ represents the mass of the graviton; in GR gravitons are
massless, but they can become massive in other theories
\cite{deRham:2014zqa}. Finally, $\Gamma(z)$ can be thought of as a
source term for GWs; in GR, GWs are unsourced once they leave the
region of their parent CBC. However, in some bimetric gravity
theories, there can be interactions between the `normal' metric
$g_{\mu\nu}$ and a second tensor field, which acts to source the GWs
as they propagate \cite{Schmidt-May:2015vnx}.

Apart from the graviton mass, a constant, the non-standard terms in
Equation~\ref{eq:MG_prop} are all functions of redshift. For a
specific modified gravity model, this redshift dependence can be
computed directly from the gravitational Lagrangian. Whilst in
principle any functional form is possible, in practice the
time-dependence of $c_T^2(z)$, $\nu(z)$ and $\Gamma(z)$ is often
shaped by the hypothesis that deviations from GR should be responsible
for late-time cosmic acceleration. That is, most modified gravity
models are designed to leave the early universe unaffected, and only
deviate from GR at late times (say $z\lesssim 2$). That behaviour will
be carried over to deviations from GR in
Equation~\ref{eq:MG_prop}. With this expectation in hand,
phenomenological parameterisations of $c_T^2(z)$ and $\nu(z)$ have
been widely investigated \cite{Chen:2023wpj, Colangeli:2025bnb}.

In reality, Equation~\ref{eq:MG_prop} must be corrected to account for
perturbations in the gravitational field sourced by matter density
fluctuations. These will source lensing and gravitational redshift
effects for the propagating GW. The amplitude of these perturbations
themselves can depend on the theory of gravity; indeed this is one
place that ``screening'' effects may show up. Screening is a set of
mechanisms by which modified gravity theories reduce to GR in
particular environments: typically highly dense perturbations will be
screened (behave like GR), and linear perturbations will be unscreened
(deviate from GR). The discussion of screening effects goes beyond the
scope of the present work; see \cite{Joyce:2014kja} for a
comprehensive review.

\subsubsection{The first detection of gravitational lensing of
  gravitational waves as a test of GR}

The first convincing detection of gravitational lensing of GW signals
will itself be a first-of-a-kind test of GR since this theory predicts
that GWs travel along geodesics and hence are gravitationally lensed
as they traverse a gravitational field \cite{Einstein1936}. The
detection of a lensed GW will confirm this property. From a
multi-messenger perspective, in the absence of a direct counterpart,
EM information such as comprehensive catalogues of gravitational
lenses from surveys such as LSST and \emph{Euclid} could provide
additional support for candidate lensed GW signals that are found by
GW lensing searches by matching them with their lensed host galaxy
(see \cite{Hannuksela2019, LVlens2021, LVKlens2023}, for example). An
initial proof-of-concept of such catalogue matching was, for instance,
performed in \cite{Janquart:2023mvf} for some of the ultimately
discarded GW lensing candidates from the third GW run. Should a lensed
GW detection be accompanied by lensed EM counterparts (\emph{golden
objects}) we can go a step further and test whether GWs travel on
\textbf{null} geodesics, i.e.\ whether they propagate at the speed of
light.

\subsubsection{Constraining the relative speed of messengers beyond
  GRB170817A/GW170817}
\label{sec:speed}

In GR, both EM radiation and GWs are massless and should propagate at
the same speed ($c_T^2=1$ in Equation~\ref{eq:MG_prop}). However, in
some theories, the graviton can have a mass and thus GWs can travel at
a speed that is different from EM radiation. Any difference in speed
can be measured when GW and EM signals are detected from the same
source, known as \textit{bright sirens}. However, a potentially
confounding factor is that in a wave optics regime, GWs can sometimes
\textit{appear} to be travelling superluminally, due to distortions of
the waveform \cite{Ezquiaga:2020spg}.

The tightest constraints on $c_T$ are obtained from the most
accurately measured arrival times, namely for GW and GRB signals. For
GRB170817A/GW170817 the relative difference was constrained to be
between $-3\times10^{-15}$ and $7\times10^{-16}$
\cite{Abbott2017e,Baker:2017hug,Creminelli:2017sry,Ezquiaga:2017ekz},
thus ruling out many alternative gravity theories. The dominant
uncertainty in analysis of GRB170817A/GW170817 is the GRB physics,
i.e.\ the unknown details of the physics of GRB jet launching that can
introduce a physical delay between GW and GRB emission that is not
related to their speed \cite{Abbott2017e}.

Multi-messenger gravitational lensing offers a complementary method
that side-steps the systematic uncertainty relating to the GRB
physics. For each image of a lensed CBC detected in GWs and GRBs, one
can measure the delay between the GRB and GW signals. Now, rather than
analyzing these delays individually, one can take the difference
between them, thus eliminating the dependency on the GRB physics --
namely the delay between GW and GRB emission -- because it is the same
for both lensed images. In this way, multi-messenger constraints on
the relative speed of messengers, and in turn on the mass of the
graviton, and possibly on the total neutrino mass, can be pushed to a
new level \cite{Baker:2016reh,Collett:2016dey,Fan2017}.

An additional feature is that a joint GRB/GW detection will place a
bound on the propagation speed of GWs at a much higher redshift
($z\simeq1-2$) than GW170817 (located at $z\simeq 0.01$). Whilst in GR
$c_T$ is a constant, in a modified gravity theory its value can vary
according to the cosmological evolution of (say) a scalar field or
dark energy EoS. As such, bounding $c_T$ at higher redshifts has
additional importance when constraining deviations from GR.

\subsubsection{Probing GW propagation with gravitationally magnified sources}
\label{sec:propagation}

The new phenomenology introduced by modified gravity effects may be a
function of redshift, as outlined in
Equation~\ref{eq:MG_prop}. Moreover, to be good dark energy
candidates, most extensions of GR are constructed to reduce to GR at
high redshifts ($z\gtrsim 2$). Thus, as we noted above when discussing
bounds on the GW propagation speed, probes that are capable of
reaching the distant universe are particularly constraining. Detecting
modified GW propagation as a function of redshift is challenging with
solely GW data, because redshift is not constrained directly by the GW
data. Information about the redshift of a GW event, is obtained from
the set of siren techniques that were originally developed to measure
$H_0$ from GW data
\cite{Schutz:1986ix,Baker:2017hug,Ezquiaga:2018btd,Lagos:2019kds,Leyde2022,Chen:2023wpj}. We
have already introduced bright sirens above\cite{Abbott2017H0}, but
let us note here that there exist two further techniques appropriate
in the absence of EM counterparts, known as spectral sirens
\cite{Ezquiaga:2022zkx} and dark sirens
\cite{Gair:2022zsa,Gray:2023wgj,Mastrogiovanni:2023emh}.

Whether bright or dark, gravitationally lensed GWs are powerful probes
of GW propagation \cite{Finke:2021znb, Iacovelli2022,
  Narola2024}. Detecting multiple lensed copies of a GW event will
enable tighter constraints on both the source parameters and the
functions that describe modifications of GR,
$\nu(z),\;m_g^2,\;\Gamma(z)$ and $c_T^2(z)$, that appear in
Equation~\ref{eq:MG_prop}. Moreover, the lens magnification will allow
detection of distant systems which would otherwise not be detected,
boosting our distance reach as motivated above. This will be most
pronounced for gravitationally lensed BNS because they are expected to
be more highly magnified than gravitationally lensed BBH
\cite{Smith2023,Magare2023}. However, even in the absence of an EM
counterpart, a convincing gravitationally lensed GW source can benefit
from other lensing information: identification of a plausible
gravitationally lensed host galaxy consistent with the lensed GW
signal will give direct access the the redshift of the GW source
\cite{Hannuksela2020,Wempe2024}.

Maximal exploitation of a multimessenger lensed event and forecast
constraints with upcoming data is an ongoing area of study. Further
work is needed to understand exactly how GWs propagate around a lens
outside of GR (e.g.\ if the graviton has a mass), and how this would
affect observables such as time delays and magnification ratios. All
the subtleties of lens modelling in GR must be folded in on top of
this (for example, the precise location of the source within the host
galaxy), and their degeneracy with modified gravity parameters
investigated.

\subsubsection{A step change in gravitationally lensed GW polarisation
  constraints}\label{sec:polarisation}

GR predicts two GW polarisation modes $(+,\times)$, in contrast to
alternative theories of gravity that may predict up to six
modes. Detecting polarisation modes individually depends sensitively
on the number of GW detectors, because the GW signal at each detector
is a linear combination of the GW polarisations, which depends on the
sky location of the source relative to the detector.  Due to the
limited sensitivity of the present detector network, the current
state-of-the-art employs simplified hypotheses as alternatives to
GR. That is, the alternative hypothesis assumes that the polarizations
contain only scalar modes or only vector modes (no tensor
modes). These analyses have concluded that the tensor-only hypothesis
is preferred over scalar-only or vector-only hypotheses
\cite{Isi:2017fbj, gw170814, GWTC1TestofGR, GW17O817:tgr}.

Robust detection of multiple images from a gravitationally lensed GW
source would at least double the number of GW signals available for
polarisation measurements of that source, with each lensed image of
that GW source containing a different linear combination of the
polarisations. This is because the lensed GW signals arrive at Earth
at different times that are independent of the rotation of the Earth,
and thus independent of the orientation of the detectors to the
respective signals. This $\ge2$-fold increase in the number of signals
therefore dramatically improves our ability to distinguish between the
polarisations, simply by boosting the number of GW detectors from the
four currently available to at least eight
\cite{Smith2019,Goyal:2020bkm}.

Methodologies need to be developed to extract the individual
polarisations from lensed GW signals in a model-agnostic way to test
GR efficiently. In the context of multi-messenger gravitational
lensing, it is important to recognise that currently the best GW
polarisation constraints come from the multi-messenger detection of
GRB170817A/GW170817/AT2017gfo \cite{GW17O817:tgr}. This is because the
sub-arcsecond localization of the GW source derived from the EM
detection delivers precise and accurate constraints on the GW the
detector responses (antenna pattern functions) to the GW
polarisations. A \emph{golden object} discovery would therefore
facilitate constraining the additional polarisation modes to the next
level.

Some theories of gravity also predict novel birefringence phenomena
for lensed GWs, whereby each GW polarisation mode is deflected
differently by a lens, leading to a net time delay between them
\cite{Ezquiaga:2020dao,Dalang:2020eaj,Streibert:2024cuf}.  Such
modifications could be tightly constrained with a multi-messenger
lensing event, although they could also imprint deviations that
distort the waveforms themselves.  On the one hand, detection of
birefringence would violate GR, and on the other hand strict limits on
birefringence would constrain beyond GR theories. Treating this effect
phenomenologically, \cite{Goyal:2023uvm} found no significant
deviation from GR using the latest catalog of GW events (GWTC-3), and
in turn constrained the birefringence probability and parameters of
alternative theories of gravity. These constraints will get better as
the number of detections increase. In addition to birefringence, other
wave-optics phenomena provide a smoking gun for deviations from
GR. Novel gravitational interactions also produce GW dispersion
(frequency-dependent phase corrections) on the $+,\times$ and
additional fields \cite{Menadeo:2024uoq}, and apparent polarizations
distinct from GR \cite{Dalang:2021qhu}. Diffraction can provide even
further tests of GR through frequency-dependent modulations of the
amplitude \cite{udeepta, Takeda:2024ghe}.

\subsection{Cosmology}\label{sec:cosmology}

Gravitational lensing is a powerful and well-established probe of
cosmology, including the expansion of the universe \cite{Treu:2022,
  Birrer:2022} and both the nature of DM and the structure and content
of DM halos \cite{Vegetti2023,Natarajan2024,Buckley2018}. Measuring
cosmic expansion with gravitationally lensed transient and variable
sources -- so called time delay cosmography -- is central to resolving
the tension between measurements of $H_0$ that are based on the
distance ladder and on the cosmic microwave background
\cite{Abdalla2022}. The sensitivity of gravitational lensing signals
to all matter, regardless of its nature bestows upon lensing a central
role in the quest to constrain DM. GWs and their EM counterparts -- in
the absence of lensing -- are also emerging as valuable tools for
cosmology, including as dark and bright standard sirens
\cite{Schutz1986,Abbott2017H0,Abbott2021H0, Gair:2022zsa,
  Gray:2023wgj, Mukherjee_2021, Cigarr_n_D_az_2022, M_ller_2024}.

Multi-messenger gravitational lensing expands and enhances the
cosmological applications of gravitational lensing in several
important ways. Firstly, the timing accuracy of GW, GRB and FRB
instruments promises to push time delay cosmography in to a new regime
of ultra-precise arrival time difference measurements. Secondly, a
\emph{golden object} will enable joint constraints on $H_0$ from
multiple detections of the same bright standard siren and
multi-messenger time-delay cosmography. Third, the ultra-precise
timing of GW instruments in the wave optics limit, and ultra-precise
localization of optical detectors in the geometric optics limit, are
highly complementary for probing DM and the structure of DM halos.

\subsubsection{Multi-messenger time-delay cosmography}\label{sec:cosmography}

Multi-messenger gravitational lensing is an exciting new channel for
time-delay cosmography, with the GW signal replacing the EM signal for
the purpose of measuring the arrival time difference \cite{Liao:2017,
  Wei:2017, Li:2019, Hou:2020, Jana:2023}. GW instruments have a
timing accuracy of $\sim10^{-3}\,\sec$, which together with the
well-understood GW waveforms for CBCs enables an arrival time
difference measurement with an uncertainty of $\sim10^{-3}\,\sec$. For
comparison, the most precise EM-based arrival time difference
measurements to date have an uncertainty of $\simeq1\,\rm day$,
reflecting the cadence of optical observations and optical brightness
fluctuations \cite{Millon2020}.

Gravitationally lensed GRBs
(Section~\ref{sec:science}\ref{sec:sources}\ref{sec:grb}) and FRBs
(whether or not associated with a merger;
Section~\ref{sec:science}\ref{sec:sources}\ref{sec:frb}) would also
yield a dramatic gain in the precision of the arrival time difference
measurement relative to lensed quasars and supernovae, thanks to the
sub-second timing accuracy of gamma-ray and radio instruments.
Therefore, gravitationally lensed GW, GRB and FRB signals are all in
the regime of ultra-precise arrival time difference measurements. To
give a concrete example, discovery of a multiply-imaged GRB that is
localised to its gravitationally lensed host galaxy via its lensed
afterglow emission would unlock ultra-precise time-delay cosmography.

In this ultra-precise arrival time difference regime, other
uncertainties will dominate. Statistically, the relative astrometric
precision of the arriving images will likely dominate
\cite{Birrer:2019}. For example, an arrival time difference precision
of $<1\,\sec$, corresponds to a displacement of an image of order
$<10^{-4}$ mas, which is several orders of magnitude below the best
possible astrometric constraints that would be achievable with
space-based optical/IR follow-up observations of an EM-bright
gravitationally lensed CBC such as a gravitationally lensed BNS
merger. However, the ultra-precise time-delay measurements, in
particular when measured in a quadruply lensed system, can add
significant constraints also on the lens model and the expected
position of the images, thus mitigating, at least in part, the
astrometric uncertainties \cite{Birrer:2025}.  For EM-dark
gravitationally lensed mergers, such as BBH, accurate astrometry for
time-delay cosmography will again rely on EM observations, for example
via identifying the plausible handful of EM-detected gravitationally
lensed host galaxies located within the joint GW-based sky
localization of a candidate gravitationally lensed BBH
\cite{Hannuksela2020,Wempe2024}.

A major systematic in time delay cosmography is the modelling of the
lens.  The mass-sheet degeneracy (MSD)
\cite{Falco:1985,Gorenstein1988} is an important systematic
uncertainty in time-delay cosmography, and is relevant to all
messengers. In the geometrical optics limit, the MSD cannot be broken
with solely the lensing constraints upon which the $H_0$ inference
relies. It can however be broken with measurements of the velocity
dispersion of stars in the lens \cite[for example]{Shajib:2023} or
with weak lensing measurements \cite{Khadka2024}. The same or similar
approaches are likely to be relevant to time-delay cosmography based
on gravitationally lensed GW, GRB and FRB signals. In addition, the
possibility of breaking the MSD in the Eikonal optics regime, using
the beat pattern of two gravitationally lensed GW signals that overlap
temporally has been explored
\cite{Cremonese:2021,Meena_wave_optics_MSD_2024,Chen_MSD_2024}.  This
is because the frequency-dependent distortions in the GW waveform
encode more information about the lens model than just the time delay
between the images.  In brief, the MSD can potentially be broken with
the GW data themselves if the arrival time difference is comparable
with the duration of the GW signals ($\sim1\,\rm minute$ in the case
of BNS signals). The MSD may therefore be suppressed for
multi-messenger gravitational lensing time-delay cosmography in the
high-magnification regime that is typical for gravitationally lensed
BNS mergers \cite{Smith2023}.

Further work in this area includes exploring and implementing optimal
search strategies to identify multi-messenger lensing events that are
well-suited to time-delay cosmography, given the predicted region of
parameter space in which such lensed events will occur. Work is also
required to investigate and develop optimal analysis methods to
combine all/some of the messengers, and to break the MSD. This work
will also shape the requirements on follow-up observations of
multi-messenger lensing events ahead of detections.

\subsubsection{Gravitationally lensed standard sirens}\label{sec:sirens}

The standard siren method of measuring $H_0$ combines GW-based
luminosity distance measurements to CBCs with estimates/measurements
of the redshift of CBC host galaxies to constrain the
redshift-distance relation \cite{Schutz1986}. The first bright
standard siren measurement was enabled by the multi-messenger
discovery of GRB170817A / GW170817 / AT2017gfo
\cite{Abbott2017H0}. The multi-messenger gravitational lensing
analogue of this measurement would involve a standard siren
measurement of $H_0$ for each of the images of a gravitationally
lensed EM-bright CBC. The distance measurement for each image would be
derived from the respective GW strain signal, and corrected for
gravitational magnification, while the redshift measurement would come
from follow-up EM observations of the images of the lensed EM
counterpart and/or lensed host galaxy. Such measurements of $H_0$
would extend the redshift reach of bright standard siren constraints
from redshifts of $z\lesssim0.1$ with LVK detections of EM-bright
standard sirens that are not lensed to $z\simeq1-2$ for their
counterparts that are gravitationally lensed. Whilst this is science
is relevant to \emph{golden objects}, it is mainly reliant on GW and
optical detections for the distance and redshift measurements
respectively.

EM-dark standard siren measurements have also been made using the
growing catalogue of BBH mergers that LVK have detected
\cite{Soares2019,Abbott2021H0,Bom2024,Afradique2024}. It has also been
demonstrated that gravitationally lensed EM-dark standard sirens can
yield interesting constraints on $H_0$, by probabilistic ranking of
plausible lensed host galaxies within the joint sky localization of
pairs of candidate lensed EM-dark GW signals
\cite{Hannuksela2020,Wempe2024,PhysRevD.109.043017,Uronen:2024bth}. A
key advantage of this method over EM-dark standard sirens that are not
lensed is that in the lensing case the number of plausible host
galaxies is \emph{significantly} suppressed by the joint sky
localization of two GW detections, the requirement that the host
galaxy candidates must themselves be gravitationally lensed, and the
galaxy lens must be able to reproduce the gravitational-wave lensing
observables.

Further work is required to build the Rubin/LSST and \emph{Euclid}
strong lens samples, as these will play a key role in this science for
both EM-bright and EM-dark standard sirens. It is also important to
develop methods to combine multi-messenger time delay cosmography and
gravitationally lensed standard sirens to optimise the synergy between
these novel constraints on $H_0$.

For completeness, we also note that weak lensing of GWs can be a
source of bias for the measurement of the Universe's
expansion. Gravitational potentials present on the GW travel path from
source to observer will lead to additional magnification which will
bias the measured luminosity distance
\cite{Canevarolo:2023dkh}. Moreover, magnified events are more likely
to be detected, meaning it will worsen the bias
\cite{Cusin:2019eyv,Cusin:2020ezb,Shan:2020esq}. Such biases can be
accounted for as an additional source of noise in distance
measurements \cite{Hirata:2010ba}. For bright sirens, it can be shown
that, in some cases, lensing can lead to bias larger than statistical
uncertainty \cite{Canevarolo:2023dkh}, also showing the importance to
properly modify model the lensing magnification probability density
function \cite{Mpetha2024}. For dark events, magnification can bias
the source-frame chirp mass estimate as it would lead to a biased
luminosity distance, and consequently the redshift if no external
observables can be used to alleviate the degeneracy
\cite{Dai:2016igl,Smith2023,Canevarolo:2024muf}.

\subsubsection{The dark matter subhalo mass function}
\label{sec:subhalo}

Numerous astronomical observations point to the matter content of the
Universe being dominated by cold (non-relativistic) DM \cite[for
  example]{Zwicky1937c,Rubin1970,Treu2004,Clowe2006,Okabe2016,Planck2020,Shajib2021}. However,
on small length and mass scales the cold dark matter (CDM) faces a
number of challenges, including the so-called ``missing satellite''
problem \cite[and references therein]{Bullock2017}.  It has therefore
been proposed that CDM may not be the correct picture, and there could
be other kinds of DM such as fuzzy \cite{Hu_2000}, interacting
\cite{Spergel_2000}, or warm \cite{Lovell_2014} DM.

Gravitational lensing is a well-established probe of the structure of
the DM halos within which individual galaxies, groups and galaxy
clusters are embedded. Much attention has focused on gravitational
magnification and deflection, via gravitationally lensed quasar flux
ratio anomalies, perturbations in the positions of lensed galaxies
(astrometric anomalies), and the structure of galaxy cluster cores
\cite[and references
  therein]{Dalal2002,More2009,Vegetti2023,Natarajan2024,Buckley2018}. DM
substructure can also perturb the arrival time of signals from distant
gravitationally lensed sources \cite{Keeton2009}. However,
measurements of optical lightcurves are subject to intrinsic
measurement uncertainties of $\simeq1\,\rm day$. Such uncertainties
likely swamp any arrival time perturbations induced by DM subhalos
\cite{Liao:2018}.

Joint multi-messenger probes of DM will enable a dramatic gain in the
size of halos that can be probed, thanks to the synergy between
time-delay, astrometric, and flux anomaly accuracy. Currently, optical
strong lensing can detect DM halos down to masses of
$M\gtrsim10^7-10^8\,\rm M_\odot$
\cite{Nierenberg2017,Vegetti2012,Gilman2019} via multiple separate
images whose angular separation scales with lens mass $M$ as
$\Delta\theta \approx 2\,\theta_{\rm E} \propto \sqrt{M}$, where
$\theta_{\rm E}$ is the Einstein radius. However, the angular
resolution of optical telescopes limits the minimum detectable
$\Delta\theta$, which in turn, limits the minimum detectable $M$. On
the other hand, $M$ can be obtained through the time delay. With
sub-second timing accuracy it will be possible to explore low mass DM
subhalos down to $M \simeq10^5-10^6\,\rm M_\odot$.

Gravitationally lensed GWs, GRBs and FRBs can provide arrival time
difference measurements with sub-second precision. This ultra-precise
arrival time difference regime is a promising new probe of DM
subhalos. In brief, this new probe will exploit the synergy between
optical/near-IR flux and astrometric anomalies and GW/GRB/FRB timing
anomalies for EM-bright gravitationally lensed events \cite[for
  example]{Fan2017,Liao:2018}. It is therefore well suited to, but
does not require, a \emph{golden object}.

The idea of detecting lower mass lenses through time-delays of the
lensed images has previously been explored in the context of
millilensing of GRBs. The principle is similar to strongly lensed GWs:
detecting the repeated signals in the time domain instead of being
limited by angular resolution \cite{Paczynski1987,Mao1992}. However,
since identical repeated GRB signals are difficult to identify
\cite{Levan2025} it has so far been challenging to test their lensed
nature. This could be circumvented by detecting a lensed
multimessenger signal (of both strongly lensed GWs and a lensed GRB),
where the time delays would coincide. Similar, albeit less accurate
constraints on the sub-second time precision could be achieved for
EM-dark gravitationally lensed events. This relies on probabilistic
ranking of plausible host galaxies within the joint sky localization
of candidate lensed BBH candidates (see
\cite{Hannuksela2020,Wempe2024,Uronen:2024bth}).

Further work is required to quantify the signatures in lensed GW
signals arising from realistic populations of DM subhalos and
properties such as their abundances, density profiles and radial
distributions within the main lensing halo. We need to determine the
probability of detecting such milli-lensed GW signals and whether the
lens search pipelines will be sensitive to them. Many of the
additional time-resolved, milli-lensed signals are likely to be
demagnified and thus, possibly below the typical detection thresholds
of lens search pipelines. For milli-lensed GW signals that are
overlapping, it will also be important for future GW detectors to
establish that the signals are actually lensed rather than a chance
overlap of two unrelated GW events.  Joint analysis in the optical and
GW domains may provide suitable priors not only to help discover the
milli-lensed GW signals but also to constrain the DM subhalo
properties.

\subsubsection{Microlensed GWs, the stellar mass function, and
  compact object dark matter}\label{microlensing}

\begin{figure}
  \vspace{-10pt}
  \includegraphics[width=\textwidth]{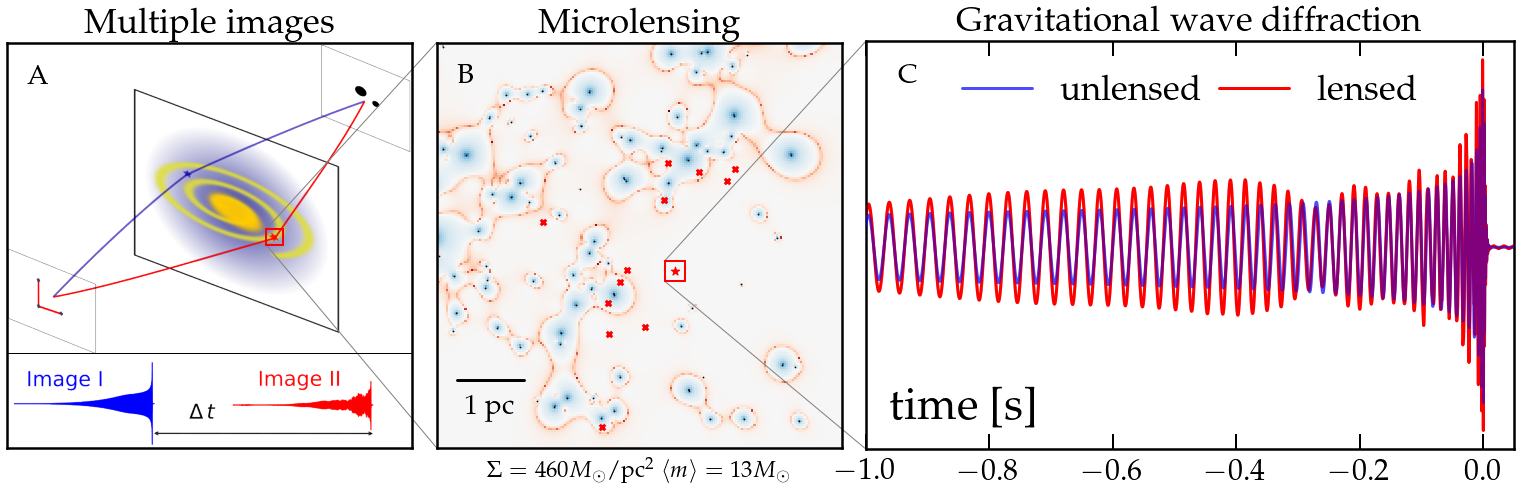} 
  \vspace*{-20pt}
  \caption{\textbf{A:} A strongly lensed system produces two images of
    a GW source (lines), separated by an arrival time difference
    $\Delta t$ (lower inset).  \textbf{B:} Image II encounters a large
    projected density of objects (dots) within the lens (with finite
    radius $10^{-4}\times$ the Einstein radius of the strong
    lens). Color shows the magnification in the lens plane. The main
    image is shown as a star, microimages appear as crosses.
    \textbf{C:} Microlensing produces a distinct modulation in the GW
    signal.
    \label{fig:gw_lensing_overview}
    \vspace{-10pt}
  }
\end{figure}

Microlensing can be used to characterize sources and the matter
distribution of gravitational lenses on small scales, including
stellar-origin objects and DM. Microlensing signatures have been
detected from a variety of EM sources, and its correct treatment is
essential to robust interpretation of a wide range of lensed
systems. Microlensing is an established probe of the structure of
quasars \cite{Vernardos2024}, the mass distribution of stars and
remnants in lens galaxies \cite{Vernardos2024,Weisenbach2024}, and the
size of supernovae \cite{Suyu2024}. Microlensing has also been used to
constrain the abundance of compact DM objects using quasars
\cite{Mediavilla:2024yhh} and caustic crossings of individual stars
\cite{Kelly2018,Venumadhav:2017pps,Oguri:2018muv}. The sensitivity of
EM microlensing is typically set by the finite size of the region
emitting the flux that is microlensed
(Figure~\ref{fig:gw_lensing_overview}; \cite{Vernardos2024}),
rendering it challenging to constrain the mass function of microlenses
\cite{Schechter:2004jg}.

Microlensing of GWs has not yet been detected
\cite{LIGOScientific:2021izm,LVKlens2023}.  This may stem from using
waveform templates in GW search pipelines that do not incorporate
signatures of lensing, resulting in a reduced detection efficiency for
microlensed GWs \cite{Chan:2024qmb}.  The signature of GW microlensing
is a modulation of the waveform (Figure~\ref{fig:gw_lensing_overview})
caused by diffraction of the GWs by objects with masses of
$M\simeq(8\pi Gf)^{-1}\simeq10\,{\rm M_\odot}(1{\rm kHz}/f)$, where
$f$ is the GW frequency \cite{Takahashi:2003ix}. However, in the
regime considered here -- microlensing of strongly lensed sources --
the effective mass of the microlenses is rescaled by $\sim\mu$, the
magnification caused by the ``macro lens'' responsible for producing
multiple images \cite{Diego:2019rzc}.  On the analysis side, whilst
microlensing by dense stellar fields within lenses is the most
plausible origin of a detection
\cite{Diego:2019lcd,Cheung_2021,Mishra:2021xzz,Meena:2022unp,Yeung_2023,Shan:2023ngi,Shan:2023qvd,Shan:2022xfx},
computational challenges in the wave optics regime have in the past
restricted analysis to isolated point lens model, however this has
begun to change\cite{Wright_2022, Janquart:2023mvf,
  Villarrubia-Rojo:2024xcj}. Microlensing of GWs is also sensitive to
the small-scale DM distribution, which can be probed by microlensing
of strongly lensed sources \cite{Dai:2019lud,Zumalacarregui:2024ocb}
and by diffraction by isolated lenses \cite{Takahashi:2003ix,
  Choi:2021jqn, Caliskan:2022hbu, Tambalo:2022wlm,
  Savastano:2023spl,Caliskan:2023zqm,Fairbairn:2022xln,GilChoi:2023ahp}.

A clear advantage of multi-messenger microlensing is the synergy
between two gravitational lensing regimes: EM microlensing in
geometric optics and GW microlensing in wave optics, because together
they provide a lever to investigate the mass of microlenses. For
example, looking at existing EM studies, the quasar emission region
sets the mass scale accessible to analysis of lightcurves without
constraining the slope of the mass function
\cite{Schechter:2004jg}. On the other hand, GW microlensing is
sensitive to heavier microlenses with $M\propto 1/f$. A
multi-messenger source, such as an AGN binary in a lensed quasar,
would constrain the stellar initial mass function (IMF) and stellar
remnants simultaneously at low and high masses, surpassing the
capacity that EM and GW have separately. On the other hand, joint
analysis of microlensing of GW and kilonova signals from lensed BNS,
even though probing essentially the same microlens population, will
have distinct observational signatures in their respective
domains. The former being frequency dependent distortions and the
latter will have time dependent evolution of the microlensed light
curve due to the increasing size of the kilonova. As a result, this
method can produce stringent and unique constraints on the microlens
population properties. In summary, while EM and GW microlensing have
been studied separately to date, joint analyses will unveil exciting
new opportunities.

Further work is required to develop the theory of microlensing to
simultaneously account for the specific effects on GWs (frequency
evolution) and EM signals (finite source size). Applications to data
will require adaptation of computational tools for parameter
estimation to include microlensing signatures, building on existing
public codes
\cite{Pagano2020,Mishra:2021xzz,Wright_2022,Cheung:2024ugg,Mishra:2023ddt,Shan:2024min,Villarrubia-Rojo:2024xcj}. Developing
tools for dedicated searches of microlensed GWs will increase the
sensitivity to events \cite{Chan:2024qmb}, especially when leveraging
information on known EM transients with a potential association
\cite{Wang:2022ryc}. Incorporating information about microlensing in
low-latency analyses can provide rapid warning on a lensed GW,
triggering follow-up searches for EM counterpart that may be otherwise
be lost.  When estimating microlensing signatures and their rates, it
is also important to revisit standard assumptions on the IMF and
remnant formation channels, motivated by EM observations (not
exclusively lensing) \cite[for
  example]{CalchiNovati:2007kkt,Oguri:2013mxl,IMF_from_SNELLS}, and
contemplate variations
\cite{Bastian:2010di,IMF_top_heavy,Mishra:2021xzz,Chan:2024qmb}.

\subsubsection{Solar mass primordial black holes}\label{sec:pbhsolar}

A wide range of empirical constraints continue to permit some of the
putative DM to be in the form of primordial black holes (PBH), whose
mass function contains structures including a prominent peak at
$M\simeq1\,\rm M_\odot$ \cite[and references therein]{Carr2024}. It is
therefore broadly accepted that convincing detection of so-called
``solar mass BHs'' would be a smoking gun for discovery of PBH,
because stellar evolution does not form BHs of this mass. Interpreting
the GW sources detected by LVK in the context of PBH has therefore
been an active field since the first direct detection \cite[for
  example]{Bird2016,AliHaimoud2017,Gow2020,Jedamzik2020,Jedamzik2021,Boehm2021}.

Multi-messenger gravitational lensing is relevant to PBH because a
gravitationally lensed merger of two solar mass PBH will occupy a
similar region of the low latency (i.e.\ based on assuming $\mu=1$)
mass-distance parameter space as gravitationally lensed BNS
(Section~\ref{sec:channels}, \cite{Smith2023}). In essence,
gravitationally lensed solar-mass PBH mergers are a ``false positive''
for gravitationally lensed BNS mergers. The main challenge in
confirming the PBH interpretation will be whether the follow-up EM
observations are sufficiently sensitive to rule all possible EM
signatures of a CBC that comprises one or more NS. Further work on
this science case is therefore needed to enhance the selection methods
of candidate gravitationally lensed CBCs and design of follow-up ToO
observations with the Vera C.\ Rubin Observatory. This will include
detailed end-to-end modelling of the expected GW and EM signatures of
gravitationally lensed EM-bright and EM-dark CBCs.

\subsection{Physics of the source populations}
\label{sec:sources}

Gravitational lensing is a well-established probe of the physics of
distant source populations, including those that are only accessible
with help from gravitational magnification. Multi-messenger
gravitational lensing unlocks new opportunities, including novel
probes of the physics of kilonovae and GRBs, the population of stellar
remnant compact objects from which CBCs emanate, the nature of FRBs
and their connection with other transient populations, the host
galaxies of CBCs across cosmic time, and the physics of core collapse
SNe.

\subsubsection{Kilonova physics}\label{sec:kilonovae}

Constraining the EoS of dense nuclear matter is a fundamental question
in nuclear physics. Observations of kilonovae provide constraints on
the EoS in a region of parameter space that cannot be replicated in
the laboratory, because the NS interiors are among the only places in
the Universe where macroscopic ``cold'' matter exists at densities at
least comparable with atomic nuclei. The observable properties of
kilonovae are driven by the outcome of NS mergers, which are all
sensitive to the structure and EoS of the component NSs. These
post-merger properties include the amount and composition of the
material that is ejected, and whether the object that remains after
the merger is a BH or short-lived NS.

Kilonovae are broadly classified as ``red'' or ``blue'', based on
their observable properties. Red kilonovae are associated with ejecta
with a low electron fraction ($Y_e$), that therefore produce
lanthanides (elements with open f-shells), and have high
opacity. Their high opacity prevents the escape of optical and UV
photons, which scatter to lower energies through fluorescence before
eventually escaping through opacity gaps in the IR
\cite{Barnes2013}. Blue kilonovae are associated with high $Y_e$,
Lanthanide-poor, low opacity ejecta. This lower opacity enables more
optical photons to escape, leading to the term ``blue''
\cite{Metzger2010}.  The distribution of $Y_e$, and therefore the
relative luminosity at blue and red wavelengths, is sensitive to the
binary mass ratio and the EoS \cite{Bauswein2013}.  However, any blue
kilonova emission is likely only detectable during the first day
post-merger. Indeed, the $\simeq12$ hour delay between detection of
GW170817 and Chilean sunset meant the rise of the optical emission was
missed in the bluer bands. The early emission that was seen could also
be explained by cooling of gas shock-heated by the GRB jet
\cite{Kasliwal2017,Piro2018}. Crucially, only observations during the
first few hours post-merger can differentiate these scenarios
\cite{Arcavi2018}.  Despite the challenges, most studies agree that
GW170817 showed evidence for multiple spatially distinct components
with different $Y_e$ \cite{Margutti2021}. More detailed discussion of
the ejecta properties inferred for GW170817 appears in
\cite{Collins2025,Nicholl2025} in this volume.

Gravitational lensing offers a unique window into the early evolution
of a kilonova at blue wavelengths. Gravitationally lensed kilonova
counterparts to gravitationally lensed BNS mergers are predicted to
reside at $z\simeq1-2$ \cite{Smith2023}. Optical searches even in red
bands therefore probe rest-frame near-UV emission. Moreover,
time-dilation increases the effective window during which the early
light curve can be detected, if searches are sufficiently
sensitive. Rapid identification of the first image associated with a
multiply-imaged kilonova / BNS with an arrival time difference of
$\gtrsim 1$\,day would enable targeted observations of the second
image during the moments after merger. A single gravitationally lensed
kilonova and BNS merger that is detected during the first days after
merger (the only time during which detection is plausible) could
therefore provide some of the best constraints on the rise and
physical origin of the early UV emission. Combined with constraints
from the GW signal, multi-messenger modelling can then be performed
\cite{Coughlin2019,Dietrich2020,Nicholl2021,Breschi2021,Gompertz2023}
to connect the pre-merger (e.g.\ mass ratio) and post merger
(e.g.\ blue ejecta) properties. The relation between these is
determined by the neutron star EoS.

Detecting the early emission will require deep and rapid ToO imaging
observations from the ground and space, that reach depths of $\rm
AB\gtrsim 25$ over multiple nights to detect the lensed kilonova,
ready for spectroscopic confirmation in the near-IR with the
\emph{James Webb Space Telescope}
\cite{Smith2023,Andreoni2024,Ryczanowski2025,Nicholl2025}.  Progress
is also needed to develop theoretical models to infer masses of
r-process material from observations and to link observational
signatures directly to the underlying EoS.  Radiative transfer
simulations can predict the expected kilonova signatures for merger
ejecta compositions resulting from employing different theoretical
EoS.  However, many uncertainties still remain in merger simulations,
r-process nucleosynthesis and atomic data, and in kilonova radiative
transfer modelling \cite{Collins2025}.  Additionally, there is the
question of whether all the early blue emission is powered by
radioactivity or if some or all of it results from the heating of
polar ejecta by a long-lived jet \cite{hamidani2024}.  With ongoing
work in this direction, kilonova simulations could predict the
timescales of the early blue component that would be measurable for
different theoretical kilonova configurations, allowing observational
constraints to be linked to the underlying EoS and r-process
compositions synthesised.

\subsubsection{Gamma-ray burst physics}\label{sec:grb}

As the most luminous explosions in nature, GRBs offer the ability to
study lensed transients across the Universe and to probe arrival time
differences as short as milliseconds (and hence lens masses down to
$M<10^4\,\rm M_\odot$). Multi-messenger detections offer a route to
rapidly confirm lensing in GW sources (via multiple co-incident GRBs)
and to test fundamental physics using the speed of light and GWs
(Section~\ref{sec:science}\ref{sec:gravity}\ref{sec:speed}). Independent
of GW detection, multi-wavelength detection of lensed GRBs holds great
promise for $H_0$ measurements, thanks to the timing accuracy of
Gamma-ray instruments
(Section~\ref{sec:science}\ref{sec:cosmology}\ref{sec:cosmography}). On
the astrophysics side, lensed GRB detections are also an opportunity
to better understand the properties of the GRBs themselves, in
particular, because the relativistic outflows from GRBs may (or may
not) have a structure on sufficiently small scales that lensed GRBs
may be chromatic, and because multiple images of a single GRB could
enable the multi-wavelength study of the emission from the earliest
times in a similar vein to that discussed in
Section~\ref{sec:science}\ref{sec:sources}\ref{sec:kilonovae}.

More than $10^4$ GRBs have been detected to date, and some of these
have likely been lensed \cite{Levan2025}. However, to date no lensed
GRBs have been confirmed, for example by the identification of the
lens, creating ambiguity over whether candidate lensed GRBs are, for
example, caused by similar pulses within a single GRB or bona fide
lensed GRBs. Multi-messenger detections of lensed GRBs are most likely
to arise from GRBs that are associated with CBCs. Traditionally this
is the short-GRB population, although recent evidence also suggests
that some long-GRBs may also arise from this channel
\cite{rastinejad2022, levan2024}.

The scientific impact of discovery of lensed GRB arrival time
differences in the seconds to hours range{ , for example as the
  counterpart to lensed GWs from a highly magnified lensed BNS merger
  \cite{Smith2023},} is enhanced by the fact that the second lensed
image is likely to occur {\em while} observations of the field
containing the first image are ongoing. Hence, rather than having only
$\gamma$-ray data, the prompt emission can be observed in the X-ray,
optical, and plausibly even radio regimes. Constructing such
broad-band SEDs of the prompt emission will be highly diagnostic, and
enhanced by, but not dependent on, a \emph{golden object} discovery.

Lensing also offers a route to probing the angular structure of GRB
jets. If GRB emission is anisotropic on small scales then we may
expect to observe multiple images which show chromatic
variations. This both poses a challenge by creating uncertainty
whether temporally and spectrally identical bursts are an accurate
route to identifying candidate lensed GRBs, and an opportunity because
it provides a direct route to determining structure in GRB jets on
scales much smaller than the opening angle of the GRB, and hence
potentially discriminating between jet structure models. In
particular, to distinguish between jets that are patchy, with hot and
cold spots \cite{meszaros1998, ioka2001} or structured, with much
stronger emission close to the axis \cite{rossi2002, lamb2022}.

Progress in this field will critically depend on the ability to
promptly recognise lensed GRBs in close to real time, rather than
identifying plausibly lensed events long after the burst, and
afterglow are gone. Since GRBs are now detected by many different
satellites (e.g.\ {\em Swift}, {\em Fermi}, {\em Einstein Probe}, {\em
  SVOM}) and the archive of old bursts is very large, the ability to
rapidly correlate locations and lightcurves from different sources
would greatly enhance the probability of correctly identifying lensed
GRBs. The recent launches of both the {\em Einstein Probe} and {\em
  SVOM} should enhance the number of well-localised bursts in the
coming years, increasing the possibility of rapid identification.
However their sky coverage is significantly less than
\emph{Fermi}/GBM, and thus rapid wide-field optical ToO follow-up from
the ground including with Rubin will have a critical role to play for
lensed GRBs that are identified in real-time \cite{Andreoni2024}.

The opportunity to probe jet structure also requires further
investigation to understand quantitatively its impact on lensed GRB
selection methods based on spectral similarity. For multi-messenger
lensing, already running searches for co-incidence between GRBs and GW
detections are highly valuable, but should be extended to new missions
to ensure events are not missed.

\subsubsection{The mass function of stellar remnant compact objects}
\label{sec:massfn}

Robust constraints on the stellar remnant mass function are central to
our understanding of stellar evolution and the physics of dense
matter, including the formation channels of CBCs, the EoS of NSs, and
SN explosion mechanisms. Direct detections of GWs from CBCs have
enabled significant progress in empirical constraints on the mass
function in recent years, with detections of sources that comprise one
or more compact objects in the putative ``mass gaps'' attracting
particular attention \cite{LVK03pop,GW230529}. The lower gap is
associated with an absence or paucity of compact objects with masses
in the range $3\lesssim M\lesssim5\,\rm M_\odot$, i.e.\ intermediate
masses between the heaviest NSs and the lightest BHs
\cite{Bailyn1998,Oezel2010,Farr2011,deSa2022}. The upper gap is
associated with an absence or paucity in the range $50\lesssim
M\lesssim120\,\rm M_\odot$, and related to the fate of massive stars
and the pair instability \cite[and references
  therein]{Mapelli:2020vfa}.

Gravitationally lensed CBCs detected via their GW emission can
masquerade as residing in one of these mass gaps, because lensed
images of distant sources are gravitationally magnified and thus the
detections appear to originate from sources that are brighter and
closer than the actual source. In particular, gravitational
magnification increases the GW strain amplitude, which leads to
under-estimating the distance to the CBC that is inferred from the
amplitude, and in turn to over-estimating the source-frame frequency
of the GW signal and hence over-estimating the mass of the system
\cite[for example]{Dai:2016igl}. Thus GW sources that are below a mass
gap can appear to be in a mass gap if lensing is not accounted in the
data analysis. Recent GW detections include sources that -- assuming
no gravitational magnification, i.e.\ $\mu=1$ -- populate both mass
gaps \cite{LVK03pop,GW230529}. Therefore, understanding the impact of
gravitational magnification on the inferred masses and distances of GW
sources is becoming critical to robust identification of real mass gap
events that can be used for formation channel studies. Methods to
break the magnification-distance degeneracy in the interpretation of
the amplitude of GW strain signals ($A\propto\mu^{0.5}D^{-1}$) are of
particular importance.

Multi-messenger gravitational lensing can break the
magnification-distance degeneracy for candidate mass-gap GW
detections, because detection of EM counterparts to GW sources is a
proven way to measure the redshift of the source independent of the GW
signal \cite{170817mm}. This is clearly relevant for GW sources that
appear to be in the lower mass gap, because the GW detector
sensitivity for low mass sources (out to $z\simeq0.2$), the
cosmological model, and the physics of gravitational lensing combine
to place the majority of gravitationally lensed BNS mergers in this
region of parameter space -- i.e.\ ``in the lower mass gap'' -- if
$\mu=1$ is assumed \cite{Smith2023}. Deep and rapid ToO observations
of GW sources that are initially placed in the mass gap are sensitive
to a wide range of kilonova physics, and can therefore probe both
lensed BNS and not lensed interpretations of mass gap sources
\cite{Bianconi2023,Andreoni2024,Ryczanowski2025}.

Multi-messenger gravitational lensing may also be relevant to the
upper mass gap, motivated among others by the detection of a candidate
AGN flare as a possible EM counterpart to GW190521 \cite{Graham2020},
and population models consistent with a fraction of BBH mergers
forming via the AGN channel \cite{Ford2022,Gayathri2023}. The dense
environment in AGN accretion disks renders this channel prone to
forming very massive BH binaries \cite{McKernan2012}. Therefore EM
follow-up observations of high mass BBH sources -- i.e.\ tuned to
search for AGN flare EM counterparts -- are also well-matched to
searching for AGN flare EM counterparts to gravitationally lensed BBH
that are magnified in to the upper mass gap \cite{Andreoni2024}.

Further work is needed to improve selection of candidate lensed GW
sources from low latency information provided by LVK, with the overall
aim of reducing the false positive rate within such selections and
guiding the design of the follow-up observations. This will maximise
the efficiency of the follow-up observations and optimise the range of
EM counterpart physics to which they are sensitive. Multi-messenger
simulations of the full range of detectable signatures of
gravitationally lensed CBCs will be crucial, including to inform which
properties of GW sources are most discriminating if released by LVK
with low latency. Current lensing-motivated ideas for expanding such
information include detector frame chirp mass, mass ratio, and the
tidal deformation parameter
\cite{Pang:2020qow,Janquart:2024ztv,Andreoni2024}.

\subsubsection{The nature of Fast Radio Bursts}\label{sec:frb}

FRBs are millisecond-duration radio transients that have intrigued the
scientific community since their discovery in 2007
\cite{Lorimer_2007}. While the exact origin of FRBs remains uncertain,
the comoving rate density of
$\simeq7\times10^{4}$\,Gpc$^{-1}$\,yr$^{-1}$ for FRBs at energies
above $10^{39}$\,erg \cite{shin_inferring_2023} far exceeds the rate
of CBCs -- 10 to 1700\,Gpc$^{-1}$\,yr$^{-1}$ for BNS mergers
\cite{abbott_population_2023}. CBCs therefore appear unlikely to
account for the majority of the FRB population. Indeed, some FRBs are
thought to be produced by magnetars, based on detections of bright
radio bursts from the Galactic magnetar SGR,1935$+$2154 in April 2020
\cite{bochenek_fast_2020, chimefrb_collaboration_bright_2020}, and
those FRBs that repeat clearly are not associated with cataclysmic
events \cite{spitler_repeating_2016,
  chimefrb_collaboration_chimefrb_2023}. Nevertheless, CBCs remain a
credible origin for some one-off FRBs -- i.e.\ the available data and
theoretical models are consistent with FRBs originating from multiple
progenitor channels \cite{piro_magnetic_2012,
  lyutikov_electromagnetic_2013, totani_cosmological_2013,
  wang_fast_2016, mcwilliams_electromagnetic_2011,
  mingarelli_fast_2015, dorazio_bright_2016}.

Multi-messenger gravitational lensing is well placed to probe whether
there is a direct connection between FRBs and CBCs. Such a connection
is currently difficult to establish due to the poor localization
constraints of GWs and the unknown time delay between the occurrence
of a CBC and associated GW emission, and the emission of any radio
burst. However, gravitational lensing provides a unique
opportunity. If both an FRB and a GW signal from the same CBC event
are gravitationally lensed, the time delays between the lensed images
would be identical, offering a strong, unambiguous association between
the two signals \cite{singh_associating_2024}. This would provide
critical insight into whether CBCs can indeed produce some of the
observed FRBs.

To date, progress in FRB observations have significantly improved our
ability to detect and study these bursts. Interferometric techniques
now allow for precise localization of FRBs to their host galaxies
\cite{shannon_commensal_2024}, opening the possibility of identifying
lensed FRBs. Moreover, many FRB surveys now store raw voltage data
when bursts are detected, preserving the phase information of radio
waves \cite{kader_high-time_2022}. These data are crucial for
identifying lensed copies of an FRB, even when propagation effects
through the interstellar and intergalactic medium complicate signal
detection. The complex spectro-temporal structures seen in many FRBs,
especially at micro and nanosecond timescales, are intrinsic to the
burst and can serve as a distinguishing feature to identify lensed
copies \cite{pastor-marazuela_fast_submitted}.

In the upcoming years, progress in several areas will be essential for
improving our prospects in multi-messenger gravitational
lensing. Increasing the FRB detection rate and improving localization
accuracy are key objectives for current FRB surveys
\cite{chimefrb_collaboration_first_2021, shannon_commensal_2024,
  rajwade_study_2024, law_deep_2024}, with the CHIME/FRB Outriggers,
currently under commission, expected to achieve subarcsecond
localization for hundreds of FRBs per year
\cite{lanman_chimefrb_2024}. In the longer term, upcoming radio
telescopes with increased sensitivity will allow for better detection
of faint, lensed FRBs \cite{macquart_fast_2015,
  vanderlinde_canadian_2019, hallinan_dsa-2000_2019}. Improving the
coordination between FRB surveys and GW observatories will become
vital for detecting lensed signals from both messengers.

\subsubsection{Studying the properties of and links between mergers
  and their hosts}\label{sec:hosts} 

There remain a great deal of unknowns about the host galaxies of GW
mergers, and they remain an active field of study in
astrophysics\cite{vijaykumar2024}. As only one GW event––the
multi-messenger GW170817 detection––from all of O1-O3 has been
confidently associated with a host galaxy, most studies about GWs and
their host galaxies rely entirely on simulations of binary formation
and galaxy evolution. However, lensing provides an opportunity to
revolutionise the field of GW host population studies: in theory as
every lensed binary merger, bright or dark, has the capacity to be
localised and its host identified, every lensed event is could become
a valuable contribution to studying the hosts of GWs.

Typically, GW binary formation is expected to correlate with certain
properties of the host galaxy, such as mass, star formation rate, and
metallicity \cite{Rauf_2023,vijaykumar2024}. However, these
conclusions are based largely on simulations combining stellar and
galaxy evolution codes. Without lensing, studies of host galaxies are
limited to GW170817-like detections, mergers confidently associated
with AGN flares, or exceptional BBHs sufficiently well localized to be
matched with a single galaxy
\cite{Chen_bbh_nonlensed_localisation_2016}. With lensing, dark GW
mergers have the possibility of being traced back to a singular host
as discussed in Section~\ref{sec:channels}\ref{sec:pathways}. And as
with GW170817, if a bright EM counterpart to a BNS is successfully
observed, host identification becomes less challenging as a
consequence of finding the BNS's exact position, typically allowing
for a confident identification of the host galaxy. However, possible
offset between the BNS/NSBH mergers and their host galaxies may make
the association more challenging \cite{Mandhai_2022, zevin2022,
  Gaspari_2024}. Thus, each lensed GW merger offers opportunities for
multi-messenger host studies\cite{Wempe2024,Uronen:2024bth}.

In addition to offering opportunities for direct host identification
for binaries, lensing has the additional benefit that the mergers, and
thus their hosts, can now originate from a variety of redshifts due to
gravitational magnification enabling discoveries beyond the redshift
frontier. This means that host population studies through lensing will
inevitably unlock information about the redshift-evolution of host
populations through the Universe.

However, as mentioned before, merger host identification does not come
without challenges, and it is likely that not all mergers can be
directly identified with their hosts. GW merger hosts may be too dim
to be observed, or the merger may be far enough offset from its host
to leave host association uncertain. Even these cases provide their
own valuable scientific applications. In the case of bright mergers,
we can directly measure the offset between the merger and candidate
hosts from optical imaging data\cite{Gaspari_2024}, which can provide
information about kicks at binary formation analogously to studies
done on ``hostless'' supernovae\cite{Zinn2011}. When the host is too
dim for identification, we can also place constraints on the maximum
luminosity––and therefore, mass––of the host galaxy for
non-observation \cite{Ryczanowski2020}. In the case of dark mergers,
the host identification may be narrowed down to a few plausible
candidate host systems that cannot be separated
\cite{Hannuksela:2019kle, Wempe2024}. While this does not offer as
direct opportunities to study the host of the GW emitter it still
allows the study of limited candidates and places constraints on
current assumptions used in simulations. Conversely, it is possible to
use information from host-GW simulations to constrain further the list
of candidates and possibly pin down the host once more in a method
similar to \cite{Artale2020}.

It is therefore clear that multi-messenger studies of the hosts of
lensed GW binaries will provide invaluable information towards
understanding the properties of the hosts of compact binaries,
understanding the evolution of this relationship with redshift,
constraining the kicks created at binary formation, and likely a
swathe of other avenues yet to be explored.

\subsubsection{The physics of core-collapse supernovae}\label{sec:ccsne}

Currently there are many competing models to describe the mechanics of
core-collapse of massive stars as they evolve in to core collapse
supernovae (CCSNe) \cite[and references
  therein]{Burrows2021}. Multi-messenger constraints on CCSNe from
neutrinos and potentially GWs are therefore central to future progress
in this field, because alongside their well known optical emission,
CCSNe are responsible for a large fraction of detected long-duration
GRBs \cite{Hjorth2003, Fruchter2006, Levesque2010, Cano2013}, and the
landmark detection of SN1987A confirmed them as sources of neutrinos
\cite{Bionta1987, Hirata1987, Alekseev1987}. Indeed, it is the ability
of neutrino and potential GW signals from CCSNe that are able to probe
beneath the optically opaque envelope to constrain the precise
timeline, geometry and thus physics of core collapse.

Multi-messenger gravitational lensing can enhance the study of core
collapse physics by taking advantage of the arrival time difference
between magnified images of gravitationally lensed CCSN. In the
optical this can probe the early phase of the CCSN lightcurve and
potentially constrain the size of the progenitor star just before core
collapse \cite{Chen2022}. Indeed, Rubin/LSST is forecast to detect
many hundreds of gravitationally lensed CCSNe \cite{Goldstein2019,
  Wojtak2019}, offering great scope to ``cherry pick'' optimal systems
for detailed further study. For example, those with a gravitationally
lensed GRB counterpart, and that are sufficiently magnified to
motivate pointed analysis of GW and neutrino datasets, are likely to
attract attention.

The main challenge that multi-messenger gravitational lensing faces
for this science case is that the event rates may be very low. This is
highlighted by the relatively small local volume within which
contemporary/imminent neutrino and GW detectors are sensitive to
signals from CCSNe. For example, pointed searches for GW signals
associated with CCSNe are limited to those located within a distance
of $D<20\,\rm Mpc$ \cite{Abbott2020ccsne}. This implies a small
redshift frontier (Section~\ref{sec:channels}\ref{sec:relative}) and
correspondingly low rate and large gravitational
magnification. Further progress therefore requires investigation of
the potential synergy between the large numbers of gravitationally
lensed CCSNe that Rubin/LSST will discover and the sensitivity of GW
and neutrino detectors to only the most highly magnified events,
including development of robust strategies for selecting and
confirming lensed CCSNe from the Rubin/LSST alert stream.

\vskip6pt

\enlargethispage{20pt}

\newpage




\aucontribute{TB, SB, CEC JME, SG, OAH, PH, MAH, JJ, DK, AJL, AM, MN, IPM, APP, GPS, HU, LEU, MW, and MZ co-led the conceiving and writing of the science cases in Section~\ref{sec:science}. M\c{C}, EC, JME, BPG, CPH, EEH, BH, GPL, RKLL, SM, QLN, HN, JSCP, ES, AJS, XS, NT contributed to writing and editing the science cases in Section~\ref{sec:science}. BPG, CPH, JJ, MN, DR, LEU, LV, MW contributed to writing and editing Section~\ref{sec:channels}. JCLC, JME, SG, OAH, PH, AL, JJ, SM, AM, SL, LEU, MW, MZ, contributed to writing and editing Section~\ref{sec:lensing}. JME, BPG, OAH, DK, GPL, and MN contributed to writing and editing Section~\ref{sec:signals}. OAH edited Section 1. OAH, PH, GPS, LEU, and MZ made the figures. JJ proof-read the entire  manuscript and provided detailed comments. GPS wrote the first draft of Sections~\ref{sec:intro}-\ref{sec:channels}. TB, FB, OAH, MAH, AM, GPS, and NT coordinated discussions at the Royal Society Theo Murphy Meeting, and subsequently conceived the manuscript. GPS, FB, MAH proposed and organised the meeting. All authors read and approved the manuscript. }

\competing{The authors declare that they have no competing interests.}

\funding{The authors acknowledge generous support from The Royal Society for the ``Multi-messenger Gravitational Lensing'' Theo Murphy Meeting in Manchester, March 2024. 
GPS, JME, SG, OAH, JJ, RL, HU, LV, and MZ also acknowledge support from the University of Vienna's Erwin Schr\"odinger Institute, when attending the ``Lensing and Wave Optics in Strong Gravity'' meeting  in December 2024.
GPS acknowledges support from The Royal Society, the Leverhulme Trust, and the Science and Technology Facilities Council (grant number ST/X001296/1).
TB is supported by ERC Starting Grant SHADE (grant no. StG 949572) and by a Royal Society University Research Fellowship (grant no. URF\textbackslash R\textbackslash 231006). 
CEC has received funding from the European Union’s Horizon Europe research and innovation programme under the Marie Skłodowska-Curie grant agreement No. 101152610 and from the European Union (ERC, HEAVYMETAL, 101071865). Views and opinions expressed are however those of the author(s) only and do not necessarily reflect those of the European Union or the European Research Council. Neither the European Union nor the granting authority can be held responsible for them.
OAH acknowledge support by grants from the Research Grants Council of Hong Kong (Project No. CUHK 14304622 and 14307923), the start-up grant from the Chinese University of Hong Kong, and the Direct Grant for Research from the Research Committee of The Chinese University of Hong Kong. 
PH acknowledges support by grants from the Research Grants Council of Hong Kong (Project No. CUHK 14304622 and 14307923), the start-up grant from the Chinese University of Hong Kong, and the Direct Grant for Research from the Research Committee of The Chinese University of Hong Kong. Acknowledgement is given to the Department of Physics, The Chinese University of Hong Kong, for the Postgraduate Studentship that facilitated this research. 
MAH acknowledges funding support from the Science and Technology Facilities Council (Ref. ST/L000946/1). 
DK was supported by the Universitat de les Illes Balears (UIB); the Spanish Agencia Estatal de Investigaci{\'o}n grants CNS2022-135440, PID2022-138626NB-I00, RED2022-134204-E, RED2022-134411-T, funded by MICIU/AEI/10.13039/501100011033, the European Union NextGenerationEU/PRTR, and the ERDF/EU; and the Comunitat Aut{\`o}noma de les Illes Balears through the Servei de Recerca i Desenvolupament and the Conselleria d'Educaci{\'o} i Universitats with funds from the Tourist Stay Tax Law (PDR2020/11 - ITS2017-006), from the European Union - NextGenerationEU/PRTR-C17.I1 (SINCO2022/6719) and from the European Union - European Regional Development Fund (ERDF) (SINCO2022/18146).
RKLL acknowledges support from the research grant no. VIL37766 and no. VIL53101 by the Villum Fonden, the DNRF Chair program grant no. DNRF162 by the Danish National Research Foundation, the European Union's Horizon 2020 research and the innovation programme under the Marie Sklodowska-Curie grant agreement No. 101131233.
MN is supported by the European Research Council (ERC) under the European Union’s Horizon 2020 research and innovation programme (grant agreement No.~948381) and by UK Space Agency Grant No.~ST/Y000692/1.
IPM acknowledges funding from an NWO Rubicon Fellowship, project number 019.221EN.019. 
APP acknowledges a PhD studentship from the Science and Technology Facilities Council and the University of Birmingham.
HU acknowledges financial support from the grants PID2021-125485NB-C22, CEX2019-000918-M funded by MCIN/AEI/10.13039/501100011033 (State Agency for Research of the Spanish Ministry of Science and Innovation), SGR-2021-01069 and FI-SDUR 2023 predoctoral grant (AGAUR, Generalitat de Catalunya). 
LEU is supported by the Hong Kong PhD Fellowship Scheme (HKPFS) from the Hong Kong Research Grants Council (RGC). LEU acknowledges support by grants from the Research Grants Council of Hong Kong (Project No. CUHK 14304622 and 14307923), the start-up grant from the Chinese University of Hong Kong, and the Direct Grant for Research from the Research Committee of The Chinese University of Hong Kong.
MW is supported by the research programme of the Netherlands Organisation for Scientific Research (NWO).
M\c{C} is supported by NSF Grants No.~AST-2307146, PHY-2207502, PHY-090003 and PHY-20043, by NASA Grant No.~21-ATP21-0010, by the John Templeton Foundation Grant 62840, by the Simons Foundation, and by the Italian Ministry of Foreign Affairs and International Cooperation grant No.~PGR01167.
JCLC acknowledges support from the Villum Investigator program supported by the VILLUM Foundation (grant no. VIL37766 and no.~VIL53101) and the DNRF Chair program (grant no. DNRF162) by the Danish National Research Foundation. 
EC is supported by ERC Starting Grant SHADE (grant no. StG 949572).
BPG acknowledges support from STFC grant No. ST/Y002253/1. 
BH acknowledges funding from the National Natural Science Foundation of China Grants No.12333001.
GPL is supported by a Royal Society Dorothy Hodgkin Fellowship (grant Nos. DHF-R1-221175 and DHF-ERE-221005). 
SM acknowledges funds from STFC grant ST/X001229/1.
HN is supported by the research programme of the Netherlands Organisation for Scientific Research~(NWO).
ES acknowledges support from the College of Science and Engineering of the University of Glasgow. 
Support for this work was provided by NASA through the NASA Hubble Fellowship grant HST-HF2-51492 awarded to AJS by the Space Telescope Science Institute, which is operated by the Association of Universities for Research in Astronomy, Inc., for NASA, under contract NAS5-26555. AJS also received support from NASA through STScI grants HST-GO-16773 and JWST-GO-2974.
NRT acknowledges funding from STFC consolidated grant ST/W000857/1. 
LV is supported by the research grant no. VIL37766 and no. VIL53101 from Villum Fonden, and the DNRF Chair program grant no. DNRF162 by the Danish National Research Foundation. This project has received funding from the European Union's Horizon 2020 research and innovation programme under the Marie Sklodowska-Curie grant agreement No 101131233. The Tycho supercomputer hosted at the SCIENCE HPC center at the University of Copenhagen was used for supporting this work.
The authors are grateful for computational resources provided by the  LIGO Laboratory and supported by National Science Foundation Grants PHY-0757058 and PHY-0823459. This material is based upon work supported by NSF's LIGO Laboratory which is a major facility fully funded by the National Science Foundation.
}

\ack{We warmly thank everyone who attended the Theo Murphy Meeting about Multi-messenger Gravitational Lensing for joining in, and for their many and varied contributions. We are also very grateful to Valentina Kostornichenko and Amy Dimmock from The Royal Society, and the staff of The Edwardian Manchester, for their organisational and practical support without which the meeting and this article would not have been possible. GPS thanks Igor Andreoni, Timo Anguita, Jocelyn Bell Burnell, Matteo Bianconi, Roger Blandford, Cl\'ement Bonnerot, Marica Branchesi, Jeff Cooke, Suhail Dhawan, Helen Eaton, Rob Fender, Mathilde Jauzac, Richard Massey, Antonella Palmese, Silvia Piranomonte, Alice Power, Stephen Smartt, Robert Stein and Glenn van de Ven for a variety of discussions and help. 
}


\newpage

\section*{Authors}

\noindent
Graham P.\ Smith\orcidlink{0000-0003-4494-8277},\!$^{1,2}$ 
Tessa Baker\orcidlink{0000-0001-5470-7616},\!$^3$ 
Simon Birrer\orcidlink{0000-0003-3195-5507},\!$^4$
Christine E.\ Collins\orcidlink{0000-0002-0313-7817},\!$^{5,6}$ 
Jose Maria Ezquiaga\orcidlink{0000-0002-7213-3211},\!$^7$ 
Srashti Goyal\orcidlink{0000-0002-4225-010X},\!$^8$ 
Otto A.\ Hannuksela\orcidlink{0000-0002-3887-7137},\!$^9$
Phurailatpam Hemantakumar\orcidlink{0000-0002-0471-3724},\!$^9$
Martin A.\ Hendry\orcidlink{0000-0001-8322-5405},\!$^{10}$ 
Justin Janquart,\!$^{11,12}$ 
David Keitel\orcidlink{0000-0002-2824-626X},\!$^{13,3}$
Andrew J.\ Levan\orcidlink{0000-0001-7821-9369},\!$^{14,15}$ 
Rico K.\ L.\ Lo\orcidlink{0000-0003-1561-6716},\!$^7$ 
Anupreeta More\orcidlink{0000-0001-7714-7076},\!$^{16,17}$ 
Matt Nicholl\orcidlink{0000-0002-2555-3192},\!$^{18}$ 
In\'es Pastor-Marazuela\orcidlink{0000-0002-4357-8027},\!$^{19}$ 
Andr\'es I.\ Ponte P\'erez\orcidlink{0009-0005-7030-8742},\!$^1$ 
Helena Ubach\orcidlink{0000-0002-0679-9074},\!$^{20,21}$ 
Laura E. Uronen\orcidlink{0009-0009-3487-5036},\!$^9$ 
Mick Wright\orcidlink{0000-0003-1829-7482},\!$^{22,23}$ 
Miguel Zumalacarregui\orcidlink{0000-0002-9943-6490},\!$^8$ 
Federica Bianco\orcidlink{0000-0003-1953-8727},\!$^{24,25,26}$
Mesut \c{C}al{\i}\c{s}kan\orcidlink{0000-0002-4906-2670},\!$^{27}$
Juno C.\ L.\ Chan\orcidlink{0000-0002-3377-4737},\!$^7$ 
Elena Colangeli,\!$^3$
Benjamin\ P.\ Gompertz\orcidlink{0000-0002-5826-0548},\!$^{1,28}$ 
Christopher P.\ Haines\orcidlink{0000-0002-8814-8960},\!$^{29}$
Erin E.\ Hayes\orcidlink{0000-0003-3847-0780},\!$^{30}$
Bin Hu\orcidlink{0000-0001-5093-8118},\!$^{31}$ 
Gavin P.\ Lamb\orcidlink{0000-0001-5169-4143},\!$^{32}$ 
Anna Liu\orcidlink{0000-0003-1081-8722},\!$^9$
Soheb Mandhai\orcidlink{0000-0001-7778-4585},\!$^{19}$
Harsh Narola\orcidlink{0000-0001-9161-7919},\!$^{22,23}$
Quynh Lan Nguyen\orcidlink{0000-0002-1828-3702},\!$^{33}$
Jason S.\ C.\ Poon,\!$^9$
Dan Ryczanowski\orcidlink{0000-0002-4429-3429},\!$^{3,1}$ 
Eungwang Seo\orcidlink{0000-0002-8588-4794},\!$^{10}$
Anowar J.\ Shajib\orcidlink{0000-0002-5558-888X},\!$^{34,35,36,37}$ 
Xikai Shan\orcidlink{0000-0003-3201-061X},\!$^{38}$
Nial Tanvir\orcidlink{0000-0003-3274-6336},\!$^{39}$ 
Luka Vujeva\orcidlink{0000-0001-7697-8361}$^7$
\bigskip

\noindent
$^1$School of Physics and Astronomy, University of Birmingham, Edgbaston, B15 2TT, United Kingdom\smallskip\\
$^2$Department of Astrophysics, University of Vienna, T\"urkenschanzstrasse 17, 1180 Vienna, Austria\smallskip\\
$^3$Institute of Cosmology and Gravitation, University of Portsmouth, Burnaby Road, Portsmouth PO1 3FX, United Kingdom\smallskip\\
$^4$Department of Physics and Astronomy, Stony Brook University, Stony Brook, NY 11794, USA\smallskip\\
$^5$School of Physics, Trinity College Dublin, The University of Dublin, Dublin 2, Ireland\smallskip\\
$^6$GSI Helmholtzzentrum f\"{u}r Schwerionenforschung, Planckstra{\ss}e 1, 64291 Darmstadt, Germany\smallskip\\
$^7$Center of Gravity, Niels Bohr Institute, Blegdamsvej 17, 2100 Copenhagen, Denmark\smallskip\\
$^8$Max Planck Institute for Gravitational Physics (Albert Einstein Institute), Am M\"uhlenberg 1, D-14476 Potsdam-Golm, Germany\smallskip\\
$^9$Department of Physics, The Chinese University of Hong Kong, Shatin, Hong Kong.\smallskip\\
$^{10}$SUPA, School of Physics and Astronomy, University of Glasgow, Glasgow G12 8QQ, United Kingdom\smallskip\\
$^{11}$Center for Cosmology, Particle Physics and Phenomenology - CP3, Universit\'e Catholique de Louvain, Louvain-La-Neuve, B-1348, Belgium\smallskip\\
$^{12}$Royal Observatory of Belgium, Avenue Circulaire, 3, 1180 Uccle, Belgium\smallskip\\
$^{13}$Departament de F\'isica, Universitat de les Illes Balears, IAC3--IEEC, Crta. Valldemossa km 7.5, E-07122 Palma, Spain\smallskip\\
$^{14}$Department of Astrophysics/IMAPP, Radboud University Nijmegen, P.O. Box 9010, Nijmegen, 6500 GL, The Netherlands\smallskip\\
$^{15}$Department of Physics, University of Warwick, Coventry, CV4 7AL, United Kingdom\smallskip\\
$^{16}$Inter-University Centre for Astronomy and Astrophysics, Post Bag 4, Ganeshkhind, Pune 411007, India\smallskip\\
$^{17}$Kavli Institute for the Physics and Mathematics of the Universe (WPI), University of Tokyo, Kashiwa, Chiba 277-8583, Japan\smallskip\\
$^{18}$Astrophysics Research Centre, School of Mathematics and Physics, Queens University Belfast, Belfast, BT7 1NN, United Kingdom\smallskip\\
$^{19}$Jodrell Bank Centre for Astrophysics, University of Manchester, Oxford Road, Manchester, M13 9PL, United Kingdom\smallskip\\
$^{20}$Institut de Ci\`{e}ncies del Cosmos (ICCUB), Universitat de Barcelona (UB), c. Mart\'{i} i Franqu\'{e}s, 1, 08028 Barcelona, Spain\smallskip\\
$^{21}$Departament de F\'{i}sica Qu\`{a}ntica i Astrof\'{i}sica (FQA), Universitat de Barcelona (UB),  c. Mart\'{i} i Franqu\'{e}s, 1, 08028 Barcelona, Spain\smallskip\\
$^{22}$Institute for Gravitational and Subatomic Physics (GRASP), Department of Physics, Utrecht University, Princetonplein 1, 3584 CC Utrecht, The Netherlands\smallskip\\
$^{23}$Nikhef – National Institute for Subatomic Physics, Science Park, 1098 XG Amsterdam, The Netherlands\smallskip\\
$^{24}$University of Delaware, Department of Physics and Astronomy, 107 The Green, Newark, DE 19716, USA\smallskip\\
$^{25}$University of Delaware, Joseph R.\ Biden School of Public Policy, Graham Hall, 184 Academy St, Newark, DE 19716, USA\smallskip\\
$^{26}$Vera C.\ Rubin Observatory, Tucson, AZ 85719, USA\smallskip\\
$^{27}$William H. Miller III Department of Physics and Astronomy, Johns Hopkins University, 3400 N Charles St, Baltimore, MD 21218, USA\smallskip\\
$^{28}$Institute for Gravitational Wave Astronomy, University of Birmingham, Edgbaston, B15 2TT, United Kingdom\smallskip\\
$^{29}$Instituto de Astronom\'ia y Ciencias Planetarias (INCT), Universidad de Atacama, Copayapu 485, Copiap\'o, Chile\smallskip\\
$^{30}$Institute of Astronomy and Kavli Institute for Cosmology, University of Cambridge, Madingley Road, Cambridge CB3 0HA, United Kingdom\smallskip\\
$^{31}$School of Physics and Astronomy, Beijing Normal University, Beijing 100875, China\smallskip\\
$^{32}$Astrophysics Research Institute, Liverpool John Moores University, IC2 Liverpool Science Park, 146 Brownlow Hill, Liverpool, L3 5RF, United Kingdom\smallskip\\
$^{33}$Phenikaa Institute for Advanced Study, Phenikaa University, Hanoi 12116, Vietnam\smallskip\\
$^{34}$Department of Astronomy and Astrophysics, University of Chicago, Chicago, IL 60637, USA\smallskip\\
$^{35}$Kavli Institute for Cosmological Physics, University of Chicago, Chicago, IL 60637, USA\smallskip\\
$^{36}$Center for Astronomy, Space Science and Astrophysics, Independent University, Bangladesh, Dhaka 1229, Bangladesh \smallskip\\
$^{37}$NHFP Einstein Fellow\smallskip\\
$^{38}$Department of Astronomy, Tsinghua University, Beijing 100084, China\smallskip\\
$^{39}$School of Physics and Astronomy, University of Leicester, University Road, Leicester, LE1 7RH, United Kingdom\smallskip\\

\newpage
\bibliographystyle{RS}


\appendix
\section{Glossary}

\begin{tabular}{ll}
AGN & Active Galactic Nucleus \cr
AT & Astronomical Telegram \cr
ATLAS & Asteroid Terrestrial-impact Last Alert System\cr
BH & Black Hole \cr
BBH & Binary Black Hole \cr
BNS & Binary Neutron Star \cr
CBC & Compact binary coalescence \cr
CCSN & Core Collapse Supernova \cr
CDM & Cold Dark Matter \cr
CE & Cosmic Explorer \cr
CHIME & Canadian Hydrogen Intensity Mapping Experiment \cr
CHORD & The Canadian Hydrogen Observatory and Radio-transient Detector\cr
DM & Dark matter \cr
EM & Electromagnetic \cr
ET & Einstein Telescope \cr
EoS & Equation of State \cr
FRB & Fast Radio Burst \cr
GBM & Gamma-ray Burst Monitor \cr
GOTO & Gravitational wave Optical Transient Observatory\cr
GR & General Relativity \cr
GRB & Gamma-ray Burst \cr
GW & Gravitational Wave\cr
GWTC & Gravitational Wave Transient Catalog \cr
$H_0$ & Hubble Constant \cr
IMF & Initial Mass Function \cr
iPTF & Intermediate Palomar Transient Factory \cr
IR & Infrared \cr
KAGRA & Kamioka Gravitational Wave Detector\cr
LIGO & Laser Interferometer Gravitational-Wave Observatory \cr
LSST & Legacy Survey of Space and Time \cr
LS4 & La Silla Schmidt Southern Survey\cr
LVK & LIGO-Virgo-KAGRA \cr
MSD & Mass Sheet Degeneracy \cr
NS & Neutron Star \cr
NSBH & Neutron Star -- Black Hole \cr
PanSTARRS & Panoramic Survey Telescope and Rapid Response System\cr
PBH & Primordial Black Hole \cr
Rubin & Vera C.\ Rubin Observatory \cr
SDSS & Sloan Digital Sky Survey \cr
SKA & Square Kilometre Array \cr
SFRD & Star Formation Rate Density \cr
SN & Supernova \cr
SNIa & Type Ia supernova \cr
SNR & Signal to Noise Ratio \cr
SLSN & Super-luminous Supernova \cr
SVOM & Space Variable Objects Monitor \cr
ToO & Target of Opportunity \cr
TDE & Tidal Disruption Event \cr
UV & Ultraviolet \cr
WFD & Legacy Survey of Space and Time, Wide Fast Deep survey \cr
XG & Next Generation GW detectors\cr
ZTF & Zwicky Transient Facility \cr
\end{tabular}

\end{document}